\documentstyle[12pt,preprint,aps,tighten,epsfig,rotate,floats]{revtex}

\def\beq{\begin{equation}}
\def\eeq{\end{equation}}
\def\beqa{\begin{eqnarray}}
\def\eeqa{\end{eqnarray}}
\def\be{\begin{equation}}
\def\ee{\end{equation}}
\def\bea{\begin{eqnarray}}
\def\eea{\end{eqnarray}}
\def\delvma{r_{V-A}}

\begin{document}
%%%%%%%%%%%%%%%%%%%%%%%%%%
\thispagestyle{empty}
\setcounter{page}{0}
%%%%%%%%%%%%%%%%%%%%%%%%%%%%%%%%
\preprint{ \vbox{\hbox{IFIC-03-20}}}
%%%%%%%%%%%%%%%%%%%%%%%%5 
\title{
QCD Condensates for the Light Quark V-A Correlator  
}
\author{Vincenzo Cirigliano}
\address{ Departament de F\'{\i}sica Te\`orica, IFIC, CSIC ---
%\affiliation{ Departament de F\'{\i}sica Te\`orica, IFIC, CSIC ---
%the alternate line is for {revtex4}
Universitat de Val\`encia \\
Apt. Correus 22085, E-46071
Val\`encia, Spain \\
Vincenzo.Cirigliano@ific.uv.es}
\author{Eugene Golowich}
\address{Physics Department, University of Massachusetts\\
%\affiliation{Physics Department, University of Massachusetts\\
Amherst, MA 01003 USA \\
golowich@physics.umass.edu \\}
\author{Kim Maltman}
\address{Department of Mathematics and Statistics, York University\\
%\affiliation{Department of Mathematics and Statistics, York University\\
4700 Keele St., Toronto ON M3J 1P3 Canada \\
and \\
CSSM, University of Adelaide \\
Adelaide, SA 5005 Australia \\
kmaltman@physics.adelaide.edu.au}
\maketitle
%%The previous line is the correct location for the old {revtex}
%%For {revtex4} it has to appear below
%%%%%%%%%%%%%%%%%%%%%%%%%%%%
\begin{abstract}
We use the procedure of pinched-weight Finite Energy 
Sum Rules (pFESR) to determine the OPE 
coefficients $a_6,\cdots ,a_{16}$ of the flavor $ud$ V-A correlator
in terms of existing hadronic $\tau$ decay data.
We show by appropriate weight choices that the error on the
dominant $d=6$ contribution, which is known to be related
to the $K\rightarrow\pi\pi$ matrix elements of the electroweak
penguin operator in the chiral limit, may be reduced to below 
the $\sim 15\%$ level. The values we obtain for OPE coefficients 
with $d>8$ are shown to naturally account for the discrepancies 
between our results for the $d=6$ and $d=8$ terms and those of 
previous analyses, which were obtained neglecting $d>8$ contributions.
\end{abstract}
%\maketitle
%%% The above line is the location needed for {revtex4}. The {revtex}
%%% location is the one above.

\vfill

\section{Introduction}
\label{Sect:intro}
In a recent work~\cite{cdgm2}, a pFESR 
analysis of the flavor $ud$ two-point V-A current correlator,
$\Delta\Pi (Q^2)$, was performed. This allowed extraction of the
dimension six V-A OPE coefficient, $a_6$, which is related by chiral 
symmetry to the $K\rightarrow\pi\pi$ matrix
element of the electroweak penguin operator, $Q_8$. 
The result for $a_6$ led directly to an improved
determination of $\epsilon^\prime /\epsilon$ in the chiral limit.

The current paper is devoted to a more detailed account of this
analysis. In it, we present the rationale for our choice of weight functions,
describe the calculation of higher dimension OPE contributions, and
discuss the relation of our work to previous treatments of the V-A
correlator. Advantages of the particular version
of the FESR formulation employed in our analysis are pointed out.
\subsection{Background}
We recall that non-strange hadronic $\tau$ decay data provide access
to the spectral functions of the flavor $ud$
vector (V) and axial vector (A) current
correlators~\cite{tauref,bnp,ledp92,pichrev}.
With $J^\mu_{V,A}$ the standard V, A currents and
the standard definitions of the spin $J=0,1$ parts of the correlators,
\begin{equation}
i \int d^4x \, e^{iq\cdot x}
\langle 0\vert T\Bigl(
J^{\mu}_{V,A}(x) J^{\nu}_{V,A}(0)^\dagger
\Bigr)\vert 0\rangle
\equiv \left( -g^{\mu\nu} q^2 + q^{\mu} q^{\nu}\right) \,
\Pi_{V,A}^{(1)}(q^2)
 + q^{\mu} q^{\nu} \, \Pi_{V,A}^{(0)}(q^2)\, ,
\end{equation}
the ratios
\begin{equation}
R_{V,A} \equiv \frac{ \Gamma [\tau^- \rightarrow \nu_\tau
\, {\rm hadrons_{V,A}}\, (\gamma)]}{ \Gamma [\tau^- \rightarrow
\nu_\tau e^- {\bar \nu}_e (\gamma)] } ,
\label{one}\end{equation}
(with $(\gamma )$ indicating additional photons or lepton pairs)
are expressible as weighted integrals over the corresponding spectral
functions $\rho^{(J)}_{V,A}\equiv {\frac{1}{\pi}}\,
{\rm Im}\, \Pi^{(J)}_{V,A}$.
Working with the combinations
$\Pi_{V,A}^{(0+1)}(s)\equiv \Pi_{V,A}^{(0)}(s)+\Pi_{V,A}^{(1)}(s)$
and $s\Pi_{V,A}^{(0)}(s)$, which have no kinematic singularities,
one has explicitly~\cite{tauref,bnp,ledp92,pichrev},
\begin{equation}
R_{V,A} =
12 \pi^2 S_{EW}\, \vert V_{ud}\vert^2
\int^{m_\tau^2}_0 {\frac{ds} {m_\tau^2 }} \,
\left( 1-{\frac{s}{m_\tau^2}}\right)^2
\left[ \left( 1 + 2 {\frac{s}{m_\tau^2}}\right)
\, \rho^{(0+1)}_{V,A}(s)
- \, {\frac{2s}{m_\tau^2}}\, \rho^{(0)}_{V,A}(s) \right] \,
\label{taufesrrho}
\end{equation}
where $S_{EW}=1.0194\pm 0.0040$
represents the leading electroweak
corrections~\cite{ms88}{\begin{footnote}{A recent 
update~\cite{erler02} yields $S_{EW}=1.0201\pm 0.0003$, compatible,
within errors, with the value $1.0194\pm 0.0040$ 
quoted above, and employed in our recent
paper~\cite{cdgm2}. Since the $\pm 0.0040$ uncertainty on $S_{EW}$
produces a negligible contribution to our total
errors below, we have chosen to retain the input employed
in Ref.~\cite{cdgm2} in what follows.}\end{footnote}}, 
and $V_{ud}$ is the $ud$ CKM matrix
element~\footnote{The additional radiative correction,
conventionally denoted $\delta^\prime_{EW}$, has been dropped
in writing this equation since it cancels in the $V-A$ difference
which is the subject of this paper.}. Since the integrals over
the $J=0$ part of the spectral function are
saturated by the pion pole contribution, up to
numerically negligible corrections of ${\cal O}(m_{u,d}^2)$,
non-strange hadronic $\tau$ decay data provide detailed
information on the sum $\rho_V^{(0+1)}(s)+\rho_A^{(0+1)}(s)$.
For states containing only pions, G-parity allows
an unambiguous separation of the V and A components of this sum.
In the range where the decay to states containing
kaon pairs is negligible (say $s \le  2~{\rm GeV}^2$),
the individual V and A terms, and hence also
the difference $\Delta\rho \equiv
\rho_V^{(0+1)}$-$\rho_A^{(0+1)}$,
are thus known very accurately from experiment.

Knowledge of the V and A spectral
functions allows access to the corresponding correlators
through the use of either dispersion relations or 
FESR's. The latter may be taken to have the form
\begin{equation}
\int_{s_{th}}^{s_0}\, ds\, \rho (s) \, w(s)\, = {\frac{-1}{2\pi i}}\,
\oint_{\vert s\vert =s_0}\, ds\, \Pi (s) w(s) \ ,
\label{basicfesr}
\end{equation}
valid for any $w(s)$ analytic in the region of the contour, and
any $\Pi (s)$ without kinematic singularities. An example is
the standard OPE representation of
$R_{V,A}$~\cite{tauref,bnp,ledp92,pichrev},
\begin{equation}
R_{V,A} = 6 \pi S_{EW}\,  \vert V_{ud}\vert^2
i\, \oint_{|s|=m_\tau^2} {\frac{ds}{ m_\tau^2}}\,
\left( 1- {\frac{s}{ m_\tau^2}}\right)^2 \left[ \left(
1 + 2 {\frac{s}{m_\tau^2}}\right) \Pi^{(0+1)}_{V,A}(s)
- 2 {\frac{s}{ m_\tau^2}}\, \Pi^{(0)}_{V,A}(s) \right] \, ,
\label{taufesr}
\end{equation}
which results from the application of Eq.~(\ref{basicfesr})
to Eq.~(\ref{taufesrrho}).

If one works at sufficiently large $s_0$ that the OPE 
representation of $\Pi(s)$ may be used reliably on the
RHS of Eq.~(\ref{basicfesr}), appropriate choices for the
weight $w(s)$ allow one to determine OPE contributions of
different dimension, $d$, in terms of experimental data for
$\rho (s)$ (analogous statements are true for the corresponding
dispersion relations and/or their Borel transforms). To reflect 
the fact that $\Pi (s)$ will differ from its OPE representation over 
at least some portion of the contour $\vert s\vert = s_0$, we recast 
Eq.~(\ref{basicfesr}) in the general form 
\begin{equation}
\int_{s_{th}}^{s_0}\, ds\, \rho (s) \, w(s)\, +  R[s_0, w]
 = {\frac{-1}{2\pi i}}\,
\oint_{\vert s\vert =s_0}\, ds\, \Pi_{\rm OPE} (s)\, w(s) \ ,
\label{basicfesr2}
\end{equation}
where
\begin{equation}
R[s_0,w]  \equiv {\frac{-1}{2\pi i}}\,
\oint_{\vert s\vert =s_0}\, ds\, \left( \Pi_{\rm OPE} (s) - \Pi (s) \right)
\, w(s)  \ \ .
\label{basicfesr3}
\end{equation}
{\it $R[s_0,w]$ then quantifies what is usually referred to as ``OPE breakdown''
or ``duality violation''.} 

Due to the intractibility of strong-coupling QCD, it is not 
possible at present to obtain an analytic expression for $R[s_0,w]$, 
and its neglect (common to all FESR analyses) therefore 
represents a key dynamical assumption. There exist several 
strategies, however, to minimize the impact of this assumption: 
\begin{itemize}
\item Work at the highest possible $s_0 \equiv s_{max}$,
where $s_{max}$ is the maximum value of $s$ for which $\rho (s)$ is
experimentally known. At high $s_0$, one has increased confidence in 
the reliability of the OPE. One can check the stability of any nominal
OPE output against changes in $s_0$ to assess theoretical uncertainties
associated with possible OPE breakdown. 

\item Work in the vicinity of certain `optimal' $s_0$ values 
$s_0^{(d)}$, called {\it duality points}~\cite{gppr}. 
The set of all such $s_0^{(d)}$ is nothing but the
zeros of $R[s_0,w]$ for certain special weight choices. 
By ``special'' is meant that the zeros can be determined
independent of the values of any unknown OPE condensates.
For example, the weights $w(s)=1$ and $w(s)=s$
are special for the V-A correlator,
since the $d=2$ and $d=4$ OPE contributions are known to be
zero in the chiral limit (this is the OPE statement
of the two Weinberg sum rules (WSR's)~\cite{sw}). 
In general, the sets of zeros of $R[s_0,w]$ for different correlators
and/or different weights are different{\begin{footnote}{See, for example, 
Figure 1 of Ref.~\cite{kmtau02}).}\end{footnote}}.
The duality point approach relies on the observation that,
for the V-A correlator, the $w(s)=1$ duality points lie close to the
corresponding $w(s)=s$ duality points (two such points 
exist in the interval $0<s_0<m_\tau^2$).
This is taken to suggest the possibility
that the zeros of $R[s_0,w]$ for {\it all}
$w(s)$ might be (approximately) the same.
If so, sum rules based on other weights would be 
reasonably satisfied at the $w(s)=1$, $w(s)=s$ 
duality points. Such sum rules, restricted to these
values of $s_0$, could then be used to extract
unknown OPE coefficients. However, the uncertainty about how close 
the true duality point for a given sum rule is
to that for the $w(s)=1, s$ sum rules will produce a corresponding 
uncertainty in the extracted OPE coefficients.
This uncertainty can be large if the $s_0$ dependence of
the corresponding spectral integrals is strong (as it is,
for example, for the weights $w(s)=s^k$ with $k\geq 2$). 

\item Work with `pinched weights', i.e.,
those satisfying $w(s_0)=0$. Such weights suppress OPE
contributions from the region of the contour near
the timelike real axis where $\left[ \Pi_{OPE}-\Pi \right]$ is
expected to be largest~\cite{pqw}. $R[s_0,w]$ will then
be small at those scales, $s_0$, for which 
the region of OPE breakdown, on the contour $\vert s\vert =s_0$,
is restricted to the vicinity of the timelike point.
If the region of scales for which this is true
extends down as far as the experimentally accessible region,
one can find a window of $s_0$ values
within which the data-based
spectral integrals admit an OPE-like representation (as given by the
RHS of Eq.~(\ref{basicfesr2})) for suitable choices of the unknown QCD
parameters (i.e. the appropriate OPE condensates). 
A successful OPE/data match implies that $R[s_0,w]$ is not detectable, 
within experimental errors, in the given analysis window. The analysis
window then represents an extended duality {\it interval}, since
every point in it, in the sense of the terminology above, is a
duality point. Moreover, working with weights which differ 
significantly in the ways they weight the experimental spectral 
data, but whose integrated OPE contributions involve the same set of 
QCD parameters, allows one to perform additional checks. 
\end{itemize}

We shall follow the last of these approaches 
and employ pFESR's to analyze the $ud$ V-A correlator.
Evidence in support of this choice
can be inferred from the results
of FESR studies of the flavor $ud$ V and A correlators,
where one knows with good accuracy both the
data and the OPE integrals for $s_0$ above $\sim 2\ {\rm GeV}^2$.
These studies show that FESR's based on the {\it unpinched}
weights, $w(s)=s^k$ ($k=0,\cdots ,3$), 
are rather poorly satisfied over the range 
$2\ {\rm GeV}^2<s_0<m_\tau^2$\cite{kmfesr,kmtau02}
({\it i.e.}, at these scales, $R[s_0,s^k]$ is typically large).
In contrast, the FESR predictions, Eq.~(\ref{taufesr}), for $R_V$ and $R_A$,
which are obtained by taking the appropriate linear combinations
of the $w(s)=1,s^2$ and $s^3$ FESR's, with
$s_0$ set equal to $m_\tau^2$ and $R[m_\tau^2,w]$ to zero,
are in extremely good agreement with experiment~\cite{ALEPH,OPAL}.
The failure of the $s^k$-weighted FESR's is a manifestation of
the breakdown of the OPE representation near the timelike
real $s$ axis for insufficiently large $s_0$, as shown by Poggio, Quinn and
Weinberg (PQW)~\cite{pqw}. The success of the OPE predictions
for $R_{V,A}$ presumably arises from the suppression of this danger
region by the (double) zeros of the kinematic weights at $s=m_\tau^2$ 
(the edge of hadronic phase space). It turns out that for
any weight of the form either $w_N(y)=(1-y)(1+Ay)$ or 
$w_D(y)=(1-y)^2(1+Ay)$ (with $A$ arbitrary and $y=s/s_0$), the
corresponding $R_{V,A}$-like pFESR is extremely well satisfied for all
$s_0$ in the range $2\ {\rm GeV}^2<s_0<m_\tau^2$~\cite{kmfesr,kmtau02}.
This indicates that for the separate V and A correlators, and
for such ``intermediate'' scales, the OPE breakdown is closely
localized to the vicinity of the timelike real axis. 
At these scales, it appears safe to neglect $R[s_0,w]$ also for other correlators
provided $w(y)$ satisfies $w(y=1)=0$, {\it but not otherwise}.
A more detailed discussion of these issues 
may be found in Section~\ref{Sect:dualityviolation}.

\subsection{Summary of Content}
In this paper we focus on the difference
$\Delta\Pi$=$\Pi_V^{(0+1)}-\Pi_A^{(0+1)}$ of the $J=0+1$ components
of the flavor $ud$ V and A correlators. In the chiral limit, its
OPE is purely non-perturbative, with
contributions beginning at dimension $d=6$. The smallness of
$m_{u,d}$ means that the physical OPE will be
dominated by $d=6$ (and higher) terms, at least until one gets to
extremely large scales. Accurate data for the associated
spectral function $\Delta\rho$ thus allow the
extraction of various vacuum condensate combinations. 
Of particular interest are the two condensates appearing in the $d=6$ part of
$\left[\Delta\Pi\right]_{\rm OPE}$, which turn out to determine the
chiral limit values of the $K\rightarrow\pi\pi$ matrix elements of the
electroweak penguin operators $Q_{7,8}$~\cite{dg}. The $d>6$ terms in
$\left[\Delta\Pi\right]_{\rm OPE}$, which enter dispersive sum rules for
these matrix elements~\cite{cdghigherd,cdgm}, are also of
phenomenological interest since a determination of their values would
allow the dispersive determination of Ref.~\cite{cdgm} to be performed
at lower scales, where uncertainties 
associated with the classical chiral sum constraints are drastically
reduced.

We will extract the higher dimension ($d>4$) terms
appearing in the OPE of $\Delta\Pi$ by constructing a
set of pFESR's designed in such a way as to minimize the impact of
experimental errors. As we will show below, it is possible to make
such determinations for $d=6,\cdots ,16$ with good accuracy using the
existing experimental $\tau$ decay data base.

In Section~\ref{Sect:input}, we detail the input required for the OPE
and data sides of the flavor $ud$ V-A $\tau$ decay sum rules and
discuss some practical considerations relevant to the choice of pFESR
weight. In Section~\ref{Sect:analysis}, we describe how, by 
appropriate pFESR choices, 
it is possible to (1) significantly improve on previous determinations
of the $d=6$ and $d=8$ OPE contributions and (2) at the same time,
extract OPE contributions with dimensions
$d=10,\cdots ,16$ not obtained in those earlier analyses. A
comparison with previous analyses is presented in
Section~\ref{Sect:comparisons}. An expanded discussion of the
issue of duality violation is given in Section~\ref{Sect:dualityviolation}.
This section also contains an outline of the techniques we have
employed to test for the presence of possible residual duality
violation in our analysis of the V-A correlator. Certain details
of these tests which are relevant to the comparison to earlier
work are deferred to the Appendix. 
%Section~\ref{Sect:disc} contains some additional 
%comments relevant to general aspects of pFESR analyses. 
Our conclusions, together with a brief discussion, are given in
Section~\ref{Sect:concl}. The implications of our results for the
chiral-limit values of the electroweak penguin matrix elements have
already been worked out in Ref.~\cite{cdgm2}.

\section{Sum Rule Analyses of the V-A Correlator $\Delta\Pi$}
\label{Sect:input}
We now describe the input required for the OPE and data sides of the
flavor $ud$ V-A sum rules. The emphasis is on
practical considerations relevant to the choice of pFESR weight
functions.

\subsection{The data side}

\begin{figure} [htb]
\unitlength1cm
\caption{The ALEPH (top panel) and OPAL 
(bottom panel) versions of the V-A spectral function. 
The errors shown are the square roots of the 
diagonal entries of the corresponding covariance matrices.}
\center{\begin{minipage}[t]{9.0cm}
\begin{picture}(8.9,8.9)
\epsfig{figure=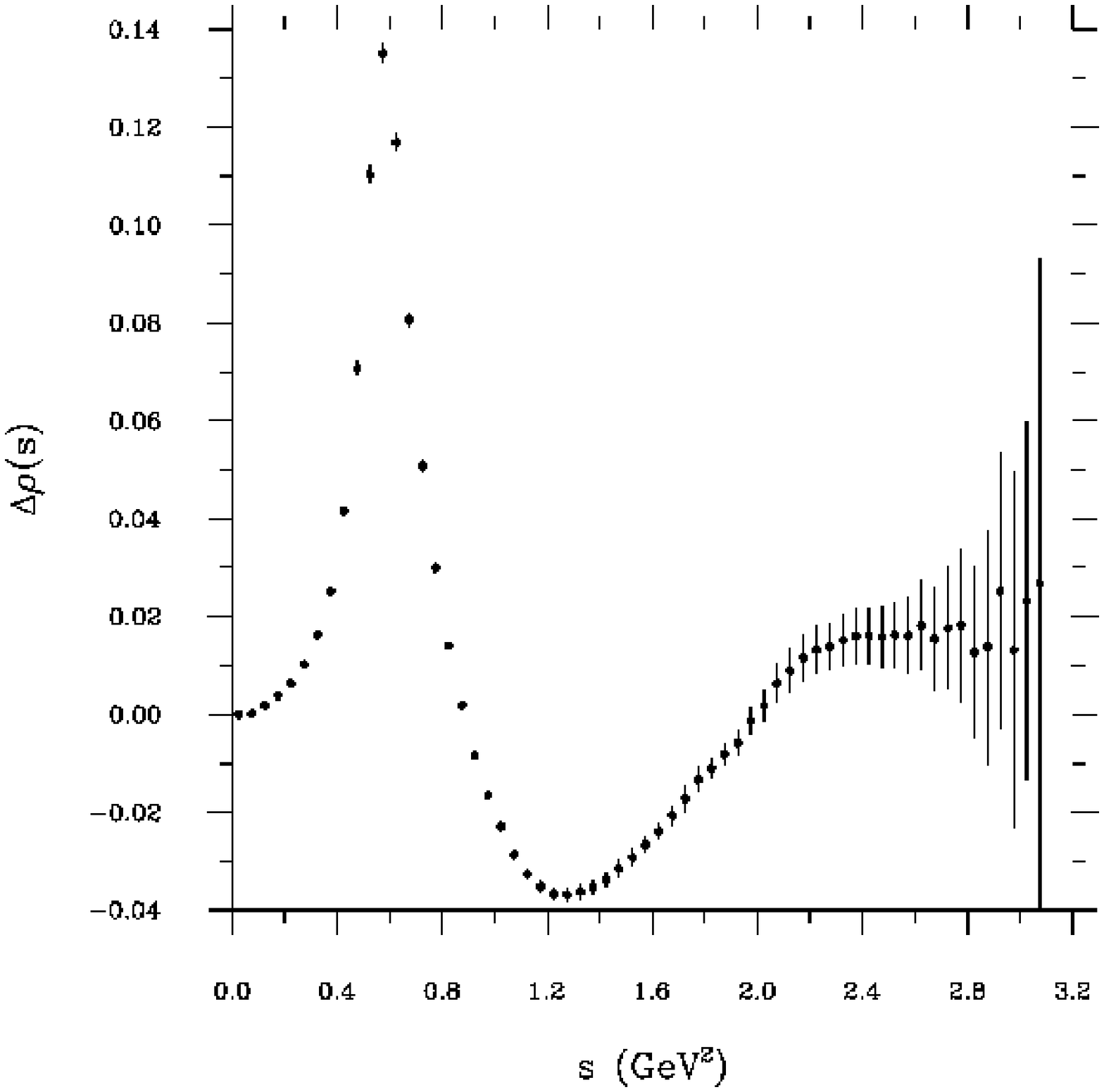,height=7.5cm,width=8.8cm}
\end{picture}
\end{minipage}
\vskip .2in
\begin{minipage}[t]{9.0cm}
\begin{picture}(8.9,8.9)
\epsfig{figure=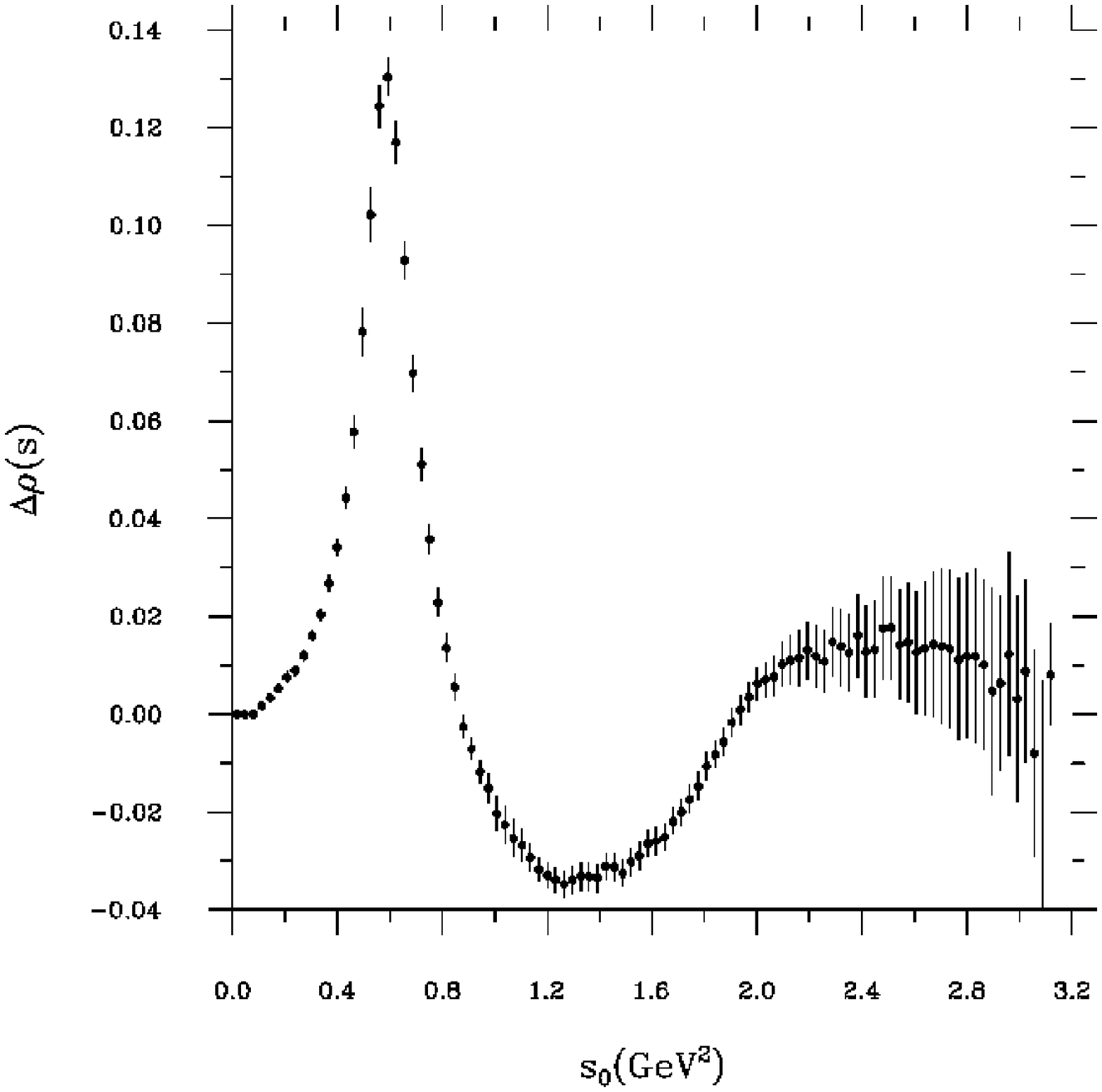,height=7.5cm,width=8.8cm}
\end{picture}
\end{minipage}}
\label{fig1}
\end{figure}

We shall employ both the ALEPH and OPAL data for
$\Delta\rho$~\cite{ALEPH,OPAL}.
The respective ALEPH and OPAL spectral functions are displayed in
Fig.~\ref{fig1}~\footnote{The points shown represent bin-averaged
$\Delta \rho (s)$ values and are plotted at the midpoints of
the ALEPH/OPAL experimental bins.}.

In the case of ALEPH, we use the publicly available data
files corresponding to the 1998 analysis, whose overall normalization
was set by the preliminary result for
the rescaled total strange hadronic branching fraction, $R_{us}\equiv
B_{us}/B_e$, $R_{us} = 0.155$, and the 1998 PDG values of $B_e$ and
$B_\mu$. The 1999 published version,
$R_{us}=0.161$~\cite{ALEPHstrange} and the recent update,
$R_{us}=0.1625$~\cite{daviertau2000,dy02} both differ slightly from the
preliminary value. This change, together with minor changes in the
values of $B_e$, $B_\mu$ and the $\tau\rightarrow\pi\nu_\tau$
branching fraction, $B_\pi$, necessitates a small global 
rescaling of the 1998 V-A data and covariance matrix~\footnote{We thank
Shaomin Chen for pointing out the necessity of this rescaling to us.
For $R_{us}=0.1625$, the PDG2002 average value~\cite{PDG2002} $B_e=0.1781$,
$B_\mu$ as implied by $\mu$-$\tau$ universality and the $\pi_{\mu 2}$
value of $F_\pi$, the rescaling turns out to be 
$1.003$, {\it i.e.}, very close to $1$.}. An input value of
$V_{ud}$ is required to convert from the
experimental number distribution provided by ALEPH
to $\Delta\rho$. We have taken this to be $V_{ud}=0.9742\pm 0.0016$, 
a value which spans both the range based on the $K_{e3}$ decay analysis and 
that based on the combination of
$0^+\rightarrow 0^+$ nuclear decays and neutron decay, as quoted in
PDG2002~\cite{PDG2002}.
In the case of OPAL, we use the publicly available data files
for $\Delta\rho$ and its correlation matrix, 
corresponding to the results of Ref.~\cite{OPAL}.
These files were constructed using a central value $V_{ud}=0.9753$.
We have, therefore, performed a small global rescaling in order 
to work with ALEPH and OPAL versions of
the spectral function which both correspond to
$V_{ud}=0.9742\pm 0.0016$.

As noted previously, the data are very accurate below $s\sim
2\ {\rm GeV}^2$. Near the kinematic endpoint
$y_\tau = 1$ ($y_\tau \equiv s/m_\tau^2$), however,
the errors on $\Delta\rho$ become large. 
This is a consequence of several factors:
\begin{enumerate}
\item The event rate becomes small in that region
due to phase space suppression.
\item There is, at present, no complete separation of V and A contributions
to the spectrum for states containing a $K\bar{K}$ pair and $\geq 1$
$\pi$'s.
\item In order to extract $\Delta\rho$ from
the experimental decay distribution ({\it c.f.} Eq.~(\ref{taufesrrho}))
one must divide by the kinematic weight factor
$w_\tau (y_\tau )=(1-y_\tau )^2(1+2y_\tau )$.
The double zero of $w_\tau$ at $s=m_\tau^2$ thus amplifies the errors
on the V-A number distributions for those $s$ near
$s=m_\tau^2$.
\end{enumerate}
In view of item $3$, pinched weights with only a single zero at
$s=s_0$ will weight the experimental number
distribution and errors with a factor which diverges
as $s\rightarrow s_0$ for $s_0$ near $m_\tau^2$. Such
behavior is to be avoided if one wishes to
keep the errors on the weighted spectral integrals under control.
For this reason we restrict our attention in the following
to pFESR's based on polynomial
weights of the form $p(y)(1-y)^2$ ($y\equiv s/s_0$).
Though this restriction is forced on us by necessity, it
has the virtue of enforcing a stronger suppression
of OPE contributions from the vicinity of
the timelike real axis, and hence of
improving the reliability of the OPE side of the pFESR's.

An important practical consideration 
in choosing pFESR weights is the non-positive definiteness
of $\Delta\rho$. Even with the very precise
data below $2\ {\rm GeV}^2$, weighted integrals which involve 
significant cancellations between contributions
from the regions of positive and negative $\Delta\rho$ will
have much larger fractional errors than
would be expected based only on the accuracy of the
spectral data alone. For some of the
weights employed previously in the literature, for example, the
V-A cancellation is at the level of a few percent of
the individual V and A integrals, leading to large errors
and significant sensitivities (as large as $20\%$)
to the exact treatment of the $\pi$ pole contribution.
Avoiding strong cancellations of this type is crucial to
reducing the errors on the final determinations
of the various vacuum condensates. To quantify this point
in our discussions below, we introduce a quantity
$\delvma$ defined as the ratio of the V-A spectral
integral to the corresponding vector spectral integral.

\subsection{The OPE side}
The OPE representation of $\Delta\Pi$ is schematically
of the form
\begin{equation}
\Delta\Pi (Q^2) = \sum_{d=2,4,\cdots} {\frac{X_d}{Q^d}}\ ,
\label{crudeOPE}
\end{equation}
where $Q^2=-s$. Perturbative corrections lead to
logarithmic dependences of the $X_d$ on $Q^2$. To NLO in
QCD one has:
\begin{equation}
X_d = a_d (\mu) + b_d \, \log \left( \frac{Q^2}{\mu^2} \right) \ .
\label{crudeOPE2}
\end{equation}
The $b_d(\mu )$ are known explicitly for
$d=2,4,6$, but not for higher $d$. For
polynomial weights, OPE contributions 
proportional to $b_d(\mu )$ involve the integrals $\int_{\vert x\vert =1}\,
dx\ x^k\, log(x)$. For the weights $w(y)=p(y)(1-y)^2$ employed in
our analysis the variations of sign in the
coefficients of $w(y)$ produce significant cancellations
(and hence additional numerical suppressions)
of these contributions relative to those of
the leading non-logarithmic terms. We thus consider
it very safe to follow earlier 
analyses in neglecting such corrections for $d \ge 8$.
The remaining non-logarithmic OPE integral contributions follow from
\begin{equation}
{\frac{-1}{2\pi i}}\,
\oint_{\vert s\vert =s_0}\, ds\, \left[ {\frac{a_d}{Q^d}}\right]
\left( {\frac{s}{s_0}}\right)^k = (-1)^k\, \delta_{k,(d/2)-1}\,
\left[ {\frac{a_d}{s_0^{k}}}\right]\ .
\label{nonlogintegrals}
\end{equation}
Weights of degree $2$ thus contain leading OPE contributions
up to $d=6$, those of degree $3$ contributions up to $d=8$, {\it etc.}.
Neglect of $d=10,\cdots , 2N+2$ contributions in pFESR's based on
weights with degree $N>3$ is therefore
dangerous unless one is working at $s_0$
large enough that such contributions may be taken to be
safely small. Typically one does not know {\it a priori}
how large an $s_0$ is ``large enough''; however, the stronger
$1/s_0$-dependence of the higher $d$ integrals allows
this question to be addressed {\it post facto},
provided one works with a range of $s_0$ large
enough to expose the presence of higher $d$ contributions
which may have been omitted when they should
not have been. If one finds that the range of
$s_0$ employed is such that the presence of such contributions is
indicated, one can use the
pFESR in question to place constraints on the relevant higher
dimension $a_d$ terms. Obviously, both
the reliability of the {\it post facto} check
and the accuracy of the higher $d$ extraction will be
enhanced for pFESR's having fewer separate $a_d$ contributions
on the OPE side of the sum rule and larger separations
between the dimensions of the contributions which do occur.

The $d=2$ term in Eq.~(\ref{crudeOPE}) is of
${\cal O}(m_{u,d}^2)$. That it can be safely neglected can be
confirmed numerically by integrating
the $J=0+1$ expression of Ref.~\cite{ck93},
which is known to ${\cal O}(\alpha_s^2)$.~\footnote{The reader might
worry that the rather bad behavior of
the integrated $J=0$, $d=2$ OPE series precludes
reliably subtracting the non-$\pi$-pole part
of the $J=0$ contribution from the data, and hence prevents us
from making such a definitive statement. While it is true
that (i) for the kinematic weight case
shown above, the last three terms in the integrated $J=0$
$d=2$ series (which is known to ${\cal O}(\alpha_s^3)$~\cite{ckp98})
are actually increasing~\cite{ckp98,taulongconvergence}, even at the
scale $s_0=m_\tau^2$ and (ii) the ${\cal O}(\alpha_s^3)$-truncated
$J=0$ $d=2$ OPE integrals corresponding
to different ``$(k,0)$ spectral weights''~\cite{ledp92}
display a significant unphysical dependence on $k$~\cite{mktaulong01},
this turns out not to be a problem.
The reason is that the behavior of the integrated $J=0$ series
has been investigated in the analogous
case involving the flavor $us$ currents, where additional
sum rule constraints were shown to allow a determination
of the corresponding spectral
integrals~\cite{mktaulong01}.
The ${\cal O}(\alpha_s^3)$-truncated $J=0$ OPE estimates were found to
represent significant {\it overestimates}~\cite{mktaulong01}.
We may thus use the (albeit poorly behaved)
OPE determinations to conclude that, apart from the
$\pi$ pole contribution, the $d=2$ $J=0$ contributions
to the measured spectral distribution
are indeed completely negligible. The $J=0+1$ part
of the spectral function can thus be reliably determined.
The integrated series in $\alpha_s$
for the $d=2$ $J=0+1$ OPE contribution converges well,
and is numerically negligible.}

The $d=4$ term in Eq.~(\ref{crudeOPE}) is given by~\cite{bnp,d4operef}
\begin{equation}
\left[\Delta\Pi (Q^2)\right]_{d=4}=\left( {\frac{8}{3}}\bar{a}
+{\frac{59}{3}}\bar{a}^2\right){\frac{\langle (m_u+m_d)\bar{u}u\rangle}
{Q^4}}\ ,
\label{D4OPE}
\end{equation}
where $\bar{a}\equiv \alpha_s(Q^2)/\pi$, with $\alpha_s(Q^2)$
the running coupling at scale $\mu^2=Q^2$ in the $\overline{MS}$
scheme. The quark condensate factor
can be evaluated using the GMOR relation~\cite{GMOR}
\begin{equation}
\langle (m_u+m_d)\bar{u}u\rangle = -F_\pi^2m_\pi^2\ ,
\label{GMOR}
\end{equation}
which is accurate to better than $6\%$~\cite{Kl4}.
We compute the weighted integrals of $\left[\Delta\Pi (Q^2)\right]_{d=4}$
using the ``contour improvement'' scheme~\cite{ledp92,cipt}, 
taking for $\bar{a}$ the version corresponding to
4-loop running~\cite{beta4} with the ALEPH determination~\cite{ALEPH}
$\alpha_s(m_\tau^2)=0.334\pm 0.022$
as input. This contribution represents only a small correction
to the dominant $d=6$ term because of the ${\cal O}(m_{u,d})$ chiral
suppression.
In the numerical analysis we have expanded to $\pm 20\%$ the errors
assigned to the GMOR evaluation of the $d=4$ OPE contributions
in order to account for the truncation of the series for the Wilson
coefficient at ${\cal O}(\bar{a}^2)$. Because the $d=4$ contribution is so
small, the resulting contribution to the total error is, however,
negligible.

Observe that in the chiral limit the  $d=2$
and  $d=4$ contributions are zero. Taking $a_2=a_4=0$ is
then the OPE implementation of the first and second WSR's.
To the extent that we use $a_2=0$ and $a_4 \sim m_\pi^2$
the WSR's are built into our procedure.

For the $d=6$ contribution, there exist several
determinations in the literature~\cite{cdgm,lsc86,ac94}, corresponding
to different schemes for the choice of
evanescent operator basis~\cite{kpdR01}.
Since one of our goals is to use our results for the $d=6$
contribution to improve the determination of the chiral limit
value of the electroweak penguin contribution to the
$K\rightarrow\pi\pi$ decay amplitudes, we employ the
most recent determination~\cite{cdgm}, which corresponds
to the same scheme as used in the calculation of the
Wilson coefficients of the effective weak
Hamiltonian~\cite{bjlw93}~\footnote{An independent determination of
$\left[\Delta\Pi\right]_{d=6}$ in this scheme was given in
Ref.~\cite{bgp01}. The results quoted in version 2 of this
reference are now in agreement with those of
Ref.~\cite{cdgm}.}.
To simplify the later application of our results
it is also convenient to work with the vacuum condensates
$\langle O_1\rangle$ and $\langle O_8\rangle$ defined
in Ref.~\cite{cdgm},
\begin{eqnarray}
\langle O_1\rangle &=& \langle  {\bar q}\gamma_\mu
{\frac{\tau_3}{2}}q~{\bar q}\gamma^\mu {\frac{\tau_3}{2}}q -
{\bar q}\gamma_\mu\gamma_5{\frac{\tau_3}{2}}q
~{\bar q} \gamma^\mu \gamma_5 {\frac{\tau_3}{2}} q \rangle\ ,
\nonumber \\
\langle O_8\rangle &=& \langle {\bar q}\gamma_\mu\lambda^a
{\frac{\tau_3}{2}}q~{\bar q}\gamma^\mu\lambda^a{\frac{\tau_3}{2}}q -
{\bar q}\gamma_\mu\gamma_5\lambda^a{\frac{\tau_3}{2}}q
~{\bar q}\gamma^\mu\gamma_5\lambda^a{\frac{\tau_3}{2}}q\rangle \ ,
\label{r6}
\end{eqnarray}
where $q = u,d,s$, $\tau_3$ is a Pauli (flavor) matrix,
and $\{ \lambda^a \}$ are the Gell~Mann color matrices.
With these choices one has
\begin{equation}
\left[\Delta\Pi (Q^2)\right]_{d=6}={\frac{1}{Q^6}}\left[
a_6 (\mu) + b_6 (\mu) \ln{Q^2 \over \mu^2} \right] \ ,
\label{d6opecoarse}
\end{equation}
with
\begin{eqnarray}
a_6 (\mu) &=& 2\left[ 2\pi\langle\alpha_s O_8 \rangle_\mu +
A_8 \langle \alpha_s^2 O_8 \rangle_\mu +
A_1 \langle \alpha_s^2 O_1 \rangle_\mu\right]\ ,\nonumber \\
b_6 (\mu) &=& 2\left[ B_8 \langle \alpha_s^2 O_8 \rangle_\mu +
B_1\langle \alpha_s^2 O_1 \rangle_\mu\right] \ ,
\label{r15a}
\end{eqnarray}
where $A_1$, $A_8$, $B_1$ and $B_8$ are
the coefficients tabulated in Ref.~\cite{cdgm}~\footnote{The
coefficients $a_6$ and $b_6$ appearing here differ
from those of Ref.~\cite{cdgm} by a factor of $2$.
This reflects the fact that in Ref.~\cite{cdgm} the coefficients
correspond to the neutral isovector current correlator, while here
they correspond to the charged isovector current correlator.
The (isospin) factor of $2$ has been made
explicit in Eqs.~(\ref{r15a}). The $A_1$, $A_8$, $B_1$ and $B_8$
of Eqs.~(\ref{r15a}) thus have the
same numerical values as in Ref.~\cite{cdgm}.}.
They depend on the number of active flavors,
the scheme employed for $\gamma_5$,
and the evanescent operator basis. For $N_f=3$,
the values for the NDR and HV $\gamma_5$ schemes are 
\begin{eqnarray}
A_1&=& 2\ ({\rm NDR}),\ -10/3\ ({\rm HV})\nonumber \\
A_8&=& 25/4\ ({\rm NDR}),\ 21/4\ ({\rm HV})\nonumber \\
B_1&=& 8/3\ ({\rm NDR\ and\ HV})\nonumber \\
B_8&=& -1\ ({\rm NDR\ and\ HV})\ .
%\begin{array}{c||c|c}
%   & {\rm NDR} & {\rm HV} \\ \hline
%A_1 & 2 & -10/3 \\
%A_8 & 25/4 & 15/4 \\
%B_1 & 8/3 & 8/3 \\
%B_8 & -1 & -1 \\
%\end{array}\ .
\label{scheme0}
\end{eqnarray}
The logarithmic ($B_{1,8}$) terms turn out to play a very
small role in the analysis, though we have kept them for completeness.
We do this by first writing $b_6=a_6(b_6/a_6)$
and then employing the existing dispersive determination
of $\langle O_1\rangle$ and $\langle O_8\rangle$~\cite{cdgm} to estimate
$r_6=b_6/a_6$. With this estimate as input,
the integrated $d=6$ OPE contribution is now, like the
non-logarithmic $d=6$ term, proportional to $a_6$. The overall
$a_6$ factor multiplying the full $d=6$ contribution
is then to be fit to data. The
central value for $r_6$ turns out to be
very small, $\simeq -0.03$. Since only the
first term in the expansion of $b_6$ in powers
of $\alpha_s$ is known, we assign a (conservative)
$50\%$ uncertainty to this estimate.

For $d=8$ and higher we take
\begin{equation}
\left[\Delta\Pi (Q^2)\right]_{d}={\frac{a_d}{Q^d}}\ .
\label{OPEhigherD}
\end{equation}
With this notation, $a_8$ is identical to $\langle O_8\rangle$ of
Ref.~\cite{DGHS98} and $O_8$ of Ref.~\cite{IZ00}. It is also
twice the negative of the integral $M_3$ 
of Ref.~\cite{bgp01,bgp02}, independent
of $s_0$, in the absence of duality violation in the $s^3$-weighted
FESR.

\section{Extraction of OPE Condensates from pFESR's}
\label{Sect:analysis}

\subsection{Choice of pFESR Weights and the $s_0$ Analysis Window}

We consider a sequence of pFESR's designed to simplify the extraction of
the OPE coefficients $a_d$ of $\Delta\Pi$. Working with pFESR's
allows us to take advantage of the freedom in the choice of weight
profile, and hence, by construction, to avoid strong V-A
cancellations. The freedom of weight choice also allows 
us to considerably simplify the task of constraining the $d>8$
contributions
and separating them from the $d=6$ and $d=8$ contributions.

Within the space of pFESR weights $w(y)=p(y)(1-y)^2$
employed in our analysis, the weight of
lowest possible degree is $w(y)=(1-y)^2$.
In a zero-error world, the corresponding pFESR would allow an extraction of
$a_6$. Unfortunately, this weight
produces a high degree of V-A cancellation, and hence is
not practical for use when employed with present
experimental data. We thus consider weights of
degree 3 (the highest degree possible involving no $a_d$
contributions with $d>8$), $w(y)=(1-y)^2(1+Ay)$. There will be
some value of $A$ for which the fractional errors on the
spectral integrals are minimized. It turns out that this
value is almost exactly equal to $-3$. The pFESR
based on
\begin{equation}
w_1(y)=(1-y)^2(1-3y)
\label{weight1}
\end{equation}
will then provide the most restrictive constraint on $a_6$, $a_8$, and
this is our first choice of weight. 
The weight $w_2(y)=y(1-y)^2$ also has reduced V-A cancellation, and
provides independent constraints on the determination of $a_6$ and
$a_8$, since degree three polynomials yield only $a_4$, $a_6$ and
$a_8$ OPE contributions, and $a_4$ is small. 
The $s_0$ dependence of both the $w_1$- and $w_2$-weighted spectral
integrals will be well-described using only two parameters,
$a_6$ and $a_8$, provided the use of the OPE representation
is justified. $w_2$ has been chosen, by construction, to weight $\Delta\rho (s)
= \Delta\rho (s_0y)$ with a profile very different from 
$w_1$ to make this test of the reliability of the
OPE representation as non-trivial as possible.

We determine our $s_0$ analysis window by fixing the upper edge at
$s_0=3.15\ {\rm GeV^2}\simeq m_\tau^2$ and decreasing the lower edge
until the fitted coefficients cease to be consistent (within
experimental errors). Since $a_6$ is the most accurately determined
coefficient we use it as our basic monitor of the onset of duality
violation. We find that duality violation for the V-A correlator
and the $w_1$, $w_2$ weight set
begins to set in below $s_0\sim 1.8\ {\rm GeV}^2$, and hence we fix the
lower edge of our analysis window at $1.95\ {\rm GeV^2}$.

To investigate $d>8$ contributions
it is convenient to construct weights for which
the only OPE contributions with $d>4$
are those proportional to $a_6$ and $a_d$, with
$d=10,12,\cdots$. The possibility of working with $s_0$ down
to $\sim 2\ {\rm GeV}^2$
is also helpful since an increased range of $s_0$ creates
an increased variation in the relative size
of the $d=10,12,\cdots$ and $d=6$ contributions over the analysis
window, and hence improves our ability to
perform the separation of contributions of different dimension.
Weights having a double zero at $y=1$, reduced V-A cancellation,
and only a single $a_d$ contribution beyond $d=6$, are
\begin{equation}
w_N(y)=y\left[ 1-\left({\frac{N}{N-1}}\right)y+
\left({\frac{1}{N-1}}\right)y^{N}\right]\
\qquad N = 2,3,4,5,6 \ .
\label{wtcaseN}
\end{equation}
The overall factor of $y$ has been introduced
in  order to reduce the level of V-A cancellation.
The case $N=2$ corresponds to the previously introduced
weight $w_2=y(1-y)^2$. For $N>2$, $w_N(y)$ produces contributions
proportional to $a_4$, $a_6$ and $a_{2N+4}$ on the OPE
side of the sum rule. We consider $N$ up to $6$, and hence
$a_d$ contributions with $d$ up to $16$.~\footnote{We
also investigated pFESR's based on the weights
$\bar{w}_N(y)=\left[ 1-\left({\frac{N}{N-1}}\right)y+
\left({\frac{1}{N-1}}\right)y^{N}\right]$
$N\geq 2$, which produce only $a_4$ and $a_{2N+2}$ OPE contributions.
The $N=2$ case is just $w_1(y)=(1-y)^2$.
For larger $N$ the smallness of the $d=4$
contributions would, in principle, make pFESR's
based on these weights good choices
for determining the higher dimension $a_d$ terms. The V-A cancellations
for the $\bar{w}_N$ family are, however, considerably stronger than
for the $w_N$ family of Eq.~(\ref{wtcaseN}), making the 
errors on the extracted $a_d$ significantly larger than those
obtained using the pFESR's based on $w_1$ through $w_{10}$.
While the results for the $a_d$ obtained using the two sets of sum
rules are in excellent agreement, the larger errors make the
analysis based on the $\bar{w}_N$ inferior to that based on the $w_N$,
at least with current experimental data as input.} 
Since each of
the resulting sum rules allows a determination of both
$a_6$ and $a_{2N+4}$, the consistency of the $a_6$
solutions obtained from the $w_1$ through $w_6$ pFESR's also provides
a strong self-consistency constraint on the reliability of the analysis.
Further constraints on $a_8$ and the higher dimension $a_d$
can be obtained by considering the weights
\begin{equation}
w_{4+N}(y)=y\left[ 1-\left({\frac{N}{N-2}}\right) y^2
+\left({\frac{2}{N-2}}\right)y^N\right]\ \qquad N = 3,4,5,6 \ .
\label{wtcaseN2}
\end{equation}
%with $N=3,4,5,6$.
These weights produce $a_4$, $a_8$ and $a_d$ contributions with
$d=10,12,14$ and $16$ for $N=3,4,5$ and $6$, respectively. The $a_8$
and $a_{d>8}$ values extracted using $w_7$ through $w_{10}$ should
be consistent with those obtained using $w_1$ through $w_6$, provided
the OPE representation of $\Delta\Pi$ is reliable for the $s_0$
employed in our analysis. We find the consistency is excellent for
all the $a_d$ with $d>4$. 

Finally, we observe that pFESRs based on the weights of
Eqs.~(\ref{wtcaseN}) and (\ref{wtcaseN2}) allow one in principle to
extract condensates of even higher dimension. With the present
experimental errors, however, higher degree pFESR's effectively work
with a smaller analysis window, localized around $s_0 = 2$ GeV$^2$
(points at higher $s_0$ suffer from much larger experimental errors,
and become irrelevant in the analysis). This feature weakens the
power of this method to detect inconsistencies through the use of an 
extended $s_0$ analysis window. We therefore quote our results for the 
condensates only up to $d=16$.

With the above choice of weights, and assuming $R[s_0,w]=0$, 
the $w_1$ through $w_{10}$ pFESR's may be 
written as
\begin{equation}
 J_{w_n}(s_0) = f_{w_n}(\{ a_d\} ;s_0)
\end{equation}
where 
%\begin{equation}
\bea
J_{w_n} (s_0) &=& \int_0^{s_0}ds\,   w_n \left(\frac{s}{s_0}\right) \,  
\Delta\rho (s)  
+{\frac{1}{2\pi i}}\oint_{\vert s\vert =s_0}ds\, 
w_n\left(\frac{s}{s_0}\right)  \, \left[ \Delta\Pi_{\rm OPE} (s)\right]_{d=4}
\\
f_{w_n}(\{ a_d\} ;s_0)
 &=&  - {\frac{1}{2\pi i}}\oint_{\vert s\vert =s_0}ds\, 
w_n\left(\frac{s}{s_0}\right)\,  \left[ \Delta\Pi_{\rm OPE} (s) \right]_{d > 4}
\ . 
\eea
%\end{equation}
The explicit form for the OPE integrals is:  
\begin{eqnarray}
f_{w_1}(\{ a_d\} ;s_0) &=& {\frac{7}{s_0^2}}\, a_6\, \left[
1+ r_6\, \log\left({\frac{s_0}{\mu^2}}\right) +{\frac{3}{14}}r_6\right]\,
+\, {\frac{3a_8}{s_0^3}}\nonumber\\
f_{w_2}(\{ a_d\} ;s_0) &=& -{\frac{2}{s_0^2}}\, a_6\, \left[
1+ r_6\, \log\left({\frac{s_0}{\mu^2}}\right) \right]\,
-\, {\frac{a_8}{s_0^3}}\nonumber\\
f_{w_3}(\{ a_d\} ;s_0) &=& -{\frac{3}{2s_0^2}}\, a_6\, \left[
1+ r_6\, \log\left({\frac{s_0}{\mu^2}}\right) +{\frac{1}{2}}r_6\right]\,
+\, {\frac{a_{10}}{2s_0^4}}\nonumber\\
f_{w_4}(\{ a_d\} ;s_0) &=& -{\frac{4}{3s_0^2}}\, a_6\, \left[
1+ r_6\, \log\left({\frac{s_0}{\mu^2}}\right) +{\frac{2}{3}}r_6\right]\,
-\, {\frac{a_{12}}{3s_0^5}}\nonumber\\
f_{w_5}(\{ a_d\} ;s_0) &=& -{\frac{5}{4s_0^2}}\, a_6\, \left[
1+ r_6\, \log\left({\frac{s_0}{\mu^2}}\right) +{\frac{3}{4}}r_6\right]\,
+\, {\frac{a_{14}}{4s_0^6}}\nonumber\\
f_{w_6}(\{ a_d\} ;s_0) &=& -{\frac{6}{5s_0^2}}\, a_6\, \left[
1+ r_6\, \log\left({\frac{s_0}{\mu^2}}\right) +{\frac{4}{5}}r_6\right]\,
-\, {\frac{a_{16}}{5 s_0^7}}\nonumber\\
f_{w_7}(\{ a_d\} ;s_0) &=& -{\frac{3}{s_0^2}} r_6 a_6 
\, +\, {\frac{3a_8}{s_0^3}}
+\, {\frac{2a_{10}}{s_0^4}}\nonumber\\
f_{w_8}(\{ a_d\} ;s_0) &=& 
-{\frac{8}{3s_0^2}}r_6 a_6 \, +\, {\frac{2a_8}{s_0^3}}
-\, {\frac{a_{12}}{s_0^5}}\nonumber\\
f_{w_9}(\{ a_d\} ;s_0) &=& 
-{\frac{5}{2s_0^2}}r_6 a_6 \, +\, {\frac{5a_8}{3s_0^3}}
+\, {\frac{2a_{14}}{3s_0^6}}\nonumber\\
f_{w_{10}}(\{ a_d\} ;s_0) &=& -{\frac{12}{5s_0^2}}r_6 a_6 \, 
+\, {\frac{3a_8}{2s_0^3}}
-\, {\frac{a_{16}}{2s_0^7}}\ .
\end{eqnarray}
Note that the small, known $d=4$ OPE contribution has been moved
to the spectral integral side in defining $J_{w_n}(s_0)$.

\subsection{pFESR Fit: Input and Results}

For our final results, we proceed in two steps. In the first step, we
extract a preferred value for $a_6$ in an analysis
employing only the weights $w_1$ and $w_2$~\cite{cdgm2}. Such an analysis is
``maximally safe'' in the sense that the numerical suppressions of the
integrated $d>8$ OPE logarithmic corrections are strongest when
$d-2n$, where $n$ is the degree of the pFESR polynomial, is as large
as possible; the neglect of such $d>8$ logarithmic terms in the OPE is
thus safest when one uses the weight(s) of the minimum possible
degree. As explained above, the accuracy of current data means that
the lowest such degree which still allows an accurate extraction of
$a_6$ is $3$. In the second step, we perform a combined least-squares
fit for the coefficients $a_6,\cdots ,a_{16}$ using, 
for each of the weights $w_1$ through $w_{10}$, defined in
Eqs.~(\ref{weight1}), (\ref{wtcaseN}), and (\ref{wtcaseN2}),
the set of 7 $s_0$ values $1.95+0.2k\ {\rm GeV}^2$, $k=0,\cdots ,6$, 
which span the range
from $s_0\sim 2\ {\rm GeV}^2$ to $3.15\ {\rm GeV}^2\simeq m_\tau^2$.

On the data side we use as input for the analyses based
on both the ALEPH and OPAL data $B_e=0.1781\pm
0.0006$~\cite{PDG2002}, 
$F_\pi =92.4\pm 0.07\pm 0.25\ {\rm MeV}$~\cite{PDG2002},
%$B_\pi = .1109\pm .0012$~\cite{PDG2000}, and
$S_{EW}=1.0194\pm 0.0040$, and $\vert V_{ud}\vert =0.9742\pm 0.0016$.
The rescaling of the 1998 ALEPH data
is determined using $R_{ud}\equiv B_{ud}/B_e =3.480\pm 0.014$~\cite{dy02}.
This value is based on the most recent update, 
$R_{us}=0.1625\pm 0.0066$~\cite{daviertau2000,dy02}, in combination with
the PDG2002 average for $B_e$ (quoted above), and the assumption of $\mu$-$e$
universality.
On the OPE side we use $\langle (m_u+m_d)\bar{u}u\rangle = -F_\pi^2m_\pi^2$,
and $r_6 \equiv b_6 / a_6 =-0.030\pm 0.015$. 

In listing final errors for the ALEPH-based analysis 
we quote separately the errors produced
by the uncertainties in the ALEPH number distribution, and those due
to all other sources, including the uncertainties on the OPE input
quantities $a_4$ and $r_6$. The former are calculated using the
rescaled ALEPH covariance matrix. The latter are combined in quadrature.

In the analysis based on the OPAL data, we again quote two uncertainties.
The first is that computed using the OPAL covariance matrix, the
second that obtained by combining in quadrature the errors
associated with uncertainties in all other input parameters
($V_{ud}$, $S_{EW}$, $a_4$, and $r_6$).

\begin{center}
{\bf Fits to the ALEPH data}
\end{center}
The results of the ``maximally safe'' analysis for $a_6$ and $a_8$
are~\footnote{Due to strong correlations between
the data integrals for different $s_0$ and different weights,
the fit values are obtained by minimizing the sum of the
squared deviations between the data and OPE integrals, weighted
by the inverse of the diagonal elements of the covariance matrix for
the set of data integrals~\cite{cdgm2}. 
With this procedure, as is well known, the
one-sigma errors and rms errors do not coincide. The former
are smaller, and underestimate the variation in the fitted $a_d$
produced by variations in the input experimental
data. All errors quoted in what follows are, therefore, the
(larger) rms errors, {\it i.e.}, the square roots of the diagonal
elements of the covariance matrix for the $\{ a_d\}$
solution set. The fitted values are, of course, also
strongly correlated, and it is crucial to employ
the full covariance matrix for the solution set
if one wishes to have accurate errors for various
sums of higher $d$ OPE contributions such as those that enter
the dispersive test of the solution set described in
the Appendix, or those required if one wishes to perform the
residual weight analysis for the $K\rightarrow\pi\pi$ EW
penguin matrix elements at lower scales~\cite{cdgm}.}
\begin{eqnarray}
a_6&=&-\left( 4.45\pm 0.61\pm 0.34\right)\times 10^{-3}
\ {\rm GeV}^6\nonumber \\
a_8&=&-\left( 6.16\pm 2.78\pm 1.40\right)\times 10^{-3}
\ {\rm GeV}^8 \ .
\label{maxsafe}\end{eqnarray}
For the ``combined fit'' analysis, we find
\begin{eqnarray}
a_6&=&-\left( 4.54\pm 0.83 \pm 0.18\right)\times 10^{-3}
\ {\rm GeV}^6\nonumber \\
a_8&=&-\left( 5.70 \pm 3.72 \pm 0.64 \right)\times 10^{-3}
\ {\rm GeV}^8\nonumber \\
a_{10}&=&\ \ \left( 4.82 \pm 1.02 \pm 0.20 \right)\times 10^{-2}
\ {\rm GeV}^{10}\nonumber \\
a_{12}&=&-\left( 1.60 \pm 0.26\pm 0.05 \right)\times 10^{-1}
\ {\rm GeV}^{12}\nonumber \\
a_{14}&=&\ \ \left( 4.26 \pm 0.62\pm 0.14\right)\times 10^{-1}
\ {\rm GeV}^{14}\nonumber \\
a_{16}&=&-\left( 1.03 \pm 0.14\pm 0.03 \right)
\ {\rm GeV}^{16}\ .
\label{finalresults}
\end{eqnarray}

\begin{center}
{\bf Fits to the OPAL data}
\end{center}
The results of the ``maximally safe'' analysis for $a_6$ and $a_8$
are
\begin{eqnarray}
a_6&=&-\left( 5.43\pm 0.72\pm 0.25\right)\times 10^{-3}
\ {\rm GeV}^6\nonumber \\
a_8&=&-\left( 1.35 \pm 3.32 \pm 1.0 \right)\times 10^{-3}
\ {\rm GeV}^8 \ .
\label{maxsafeOPAL}
\end{eqnarray}
For the combined analysis, we find
\begin{eqnarray}
a_6&=&-\left( 5.06 \pm 0.89 \pm 0.12\right)\times 10^{-3}
\ {\rm GeV}^6\nonumber \\
a_8&=&-\left( 3.12 \pm 3.82 \pm 0.45 \right)\times 10^{-3}
\ {\rm GeV}^8\nonumber \\
a_{10}&=&\ \ \left( 3.87 \pm 1.06 \pm 0.10 \right)\times 10^{-2}
\ {\rm GeV}^{10}\nonumber \\
a_{12}&=&-\left( 1.32 \pm 0.27\pm 0.03 \right)\times 10^{-1}
\ {\rm GeV}^{12}\nonumber \\
a_{14}&=&\ \ \left( 3.54 \pm 0.66 \pm 0.06\right)\times 10^{-1}
\ {\rm GeV}^{14}\nonumber \\
a_{16}&=&-\left( 0.85 \pm 0.15 \pm 0.02 \right)
\ {\rm GeV}^{16}\ .
\label{finalresultsOPAL}
\end{eqnarray}

We note that the ALEPH and OPAL determinations of
OPE coefficients are in good agreement within errors.
There is also extremely good agreement between
the combined-fit and maximally-safe-fit values for $a_6$ and $a_8$
in both the ALEPH and OPAL cases, 
providing strong {\it post facto} support for the neglect of the higher $d$
logarithmic corrections. One further point of relevance to the
self-consistency of the analysis, not evident from the results quoted
above, is the following. For each of the ten pFESR's considered above,
it is possible, because of the different $s_0$-dependence
of contributions of different dimension, to extract values for the two
unknown ($d>4$) $a_d$ coefficients occuring on the OPE side
of the sum rule in question. One can then compare 
the values of a given $a_d$ obtained using
various different individual pFESR's. It turns out that the agreement among
the results of different single-pFESR analyses is excellent 
for all the $a_d$, $d=6,\cdots ,16$. By construction six such
determinations, and hence six such consistency tests, 
exist for $a_6$ and $a_8$.
%%%%%%%%%%%%%%%%%%%%%%%%%%%%%%%%%%%%%%%%%%%%%%%%%%
%
%\subsection{Correlation matrices for the solution sets} \label{Sect:app3}
%
% Our combined fit leads to a determination of the six parameters $a_6
% \dots a_{16 }$. Because of the strong correlations between 
% data integrals corresponding to different $s_0$ and/or different
% weights, the fit parameters are strongly correlated.
%
% We report here the correlation matrices for the 
% solution sets, corresponding to both the ALEPH and OPAL fits. 
%
%\begin{equation}
%C(a_i,a_j)_{\rm ALEPH} = \left(
%\begin{tabular}{cccccc}
%1 & -0.99 & +0.98 & -0.95  & +0.92  & -0.90    \\
%-0.99  & 1  & -0.99 & +0.96  & -0.94 & +0.92   \\
%+0.98  & -0.99   & 1 & -0.99  & +0.98  &  -0.97  \\
%-0.95  & +0.96  & -0.99 & 1  & -0.99 & +0.99   \\
%+0.92  & -0.94   & +0.98 & -0.99  & 1  & -0.99    \\
%-0.90  & +0.92  & -0.97 & +0.99   & -0.99   &  1
%\end{tabular}
%\right)
%\end{equation}
%
%\begin{equation}
%C(a_i,a_j)_{\rm OPAL} = \left(
%\begin{tabular}{cccccc}
%1 & -0.99 & +0.95 & -0.90  & +0.86  & -0.82    \\
%-0.99  & 1  & -0.98 & +0.94  & -0.90 & +0.87   \\
%+0.95  & -0.98   & 1 & -0.99  & +0.97  &  -0.95  \\
%-0.90  & +0.94  & -0.99 & 1  & -0.99 & +0.98   \\
%+0.86  & -0.90   & +0.97 & -0.99  & 1  & -0.99    \\
%-0.82  & +0.87  & -0.95 & +0.98   & -0.99   &  1
%\end{tabular}
%\right)
%\end{equation}
%
% Higher precision versions of these matrices are available upon request.
%
%%%%%%%%%%%%%%%%%%%%%%%%%%%%%%%%%%%%%%%%%%%%%%%%%%%%%%%%%%%%%%%%%%%%%%%%

Our combined fit leads to a determination of the six parameters $a_6
\dots a_{16 }$. Because of the strong correlations between 
data integrals corresponding to different $s_0$ and/or different
weights, the resulting fit parameters are highly correlated.
If $C_{DD^\prime}$ is the $DD^\prime$ element of the correlation
matrix, we find that the smallest of the $\vert C_{DD^\prime}\vert$
is $0.90$ for the solution set associated with the ALEPH data
and $0.82$ for that associated with the OPAL data. The full
covariance matrices are available upon request.

\subsection{The Optimized OPE/Spectral Integral Match}
\label{ope-spec-match}

\begin{figure} [htb]
\unitlength1cm
\caption{
$J_{w_n}(s_0)$  and   $f_{w_n}(\{ a_d\} ;s_0)$ 
for the $w_1$ (top panel) and $w_2$ (bottom panel) pFESR's. 
The $J_{w_n}(s_0)$ integrals 
and errors were obtained using the ALEPH data
and covariance matrix.
Three versions of the $f_{w_n}(\{ a_d\} ;s_0)$  curve are shown.
The solid line corresponds to either the ``maximally-safe'' 
or combined fit for $a_6$ and $a_8$, as described in the text, 
the short-dashed and long-dashed lines 
to the corresponding DGHS and IZ solutions, respectively.}
\center{
\begin{minipage}[t]{9.0cm}
\begin{picture}(8.9,8.9)
\epsfig{figure=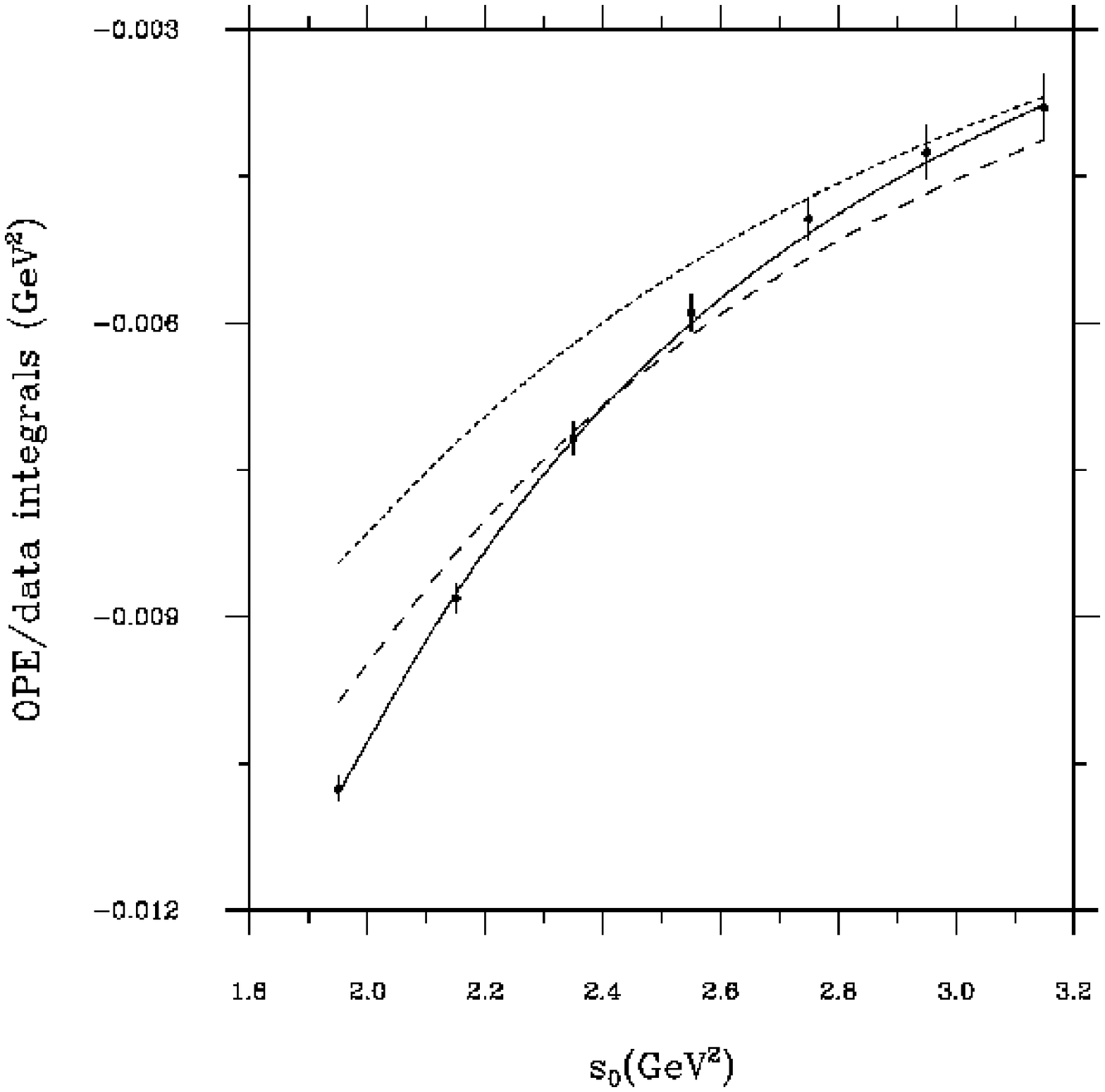,height=8.8cm,width=8.8cm}
\end{picture}
\end{minipage}
\vskip .2in
\begin{minipage}[t]{9.0cm}
\begin{picture}(8.9,8.9)
\epsfig{figure=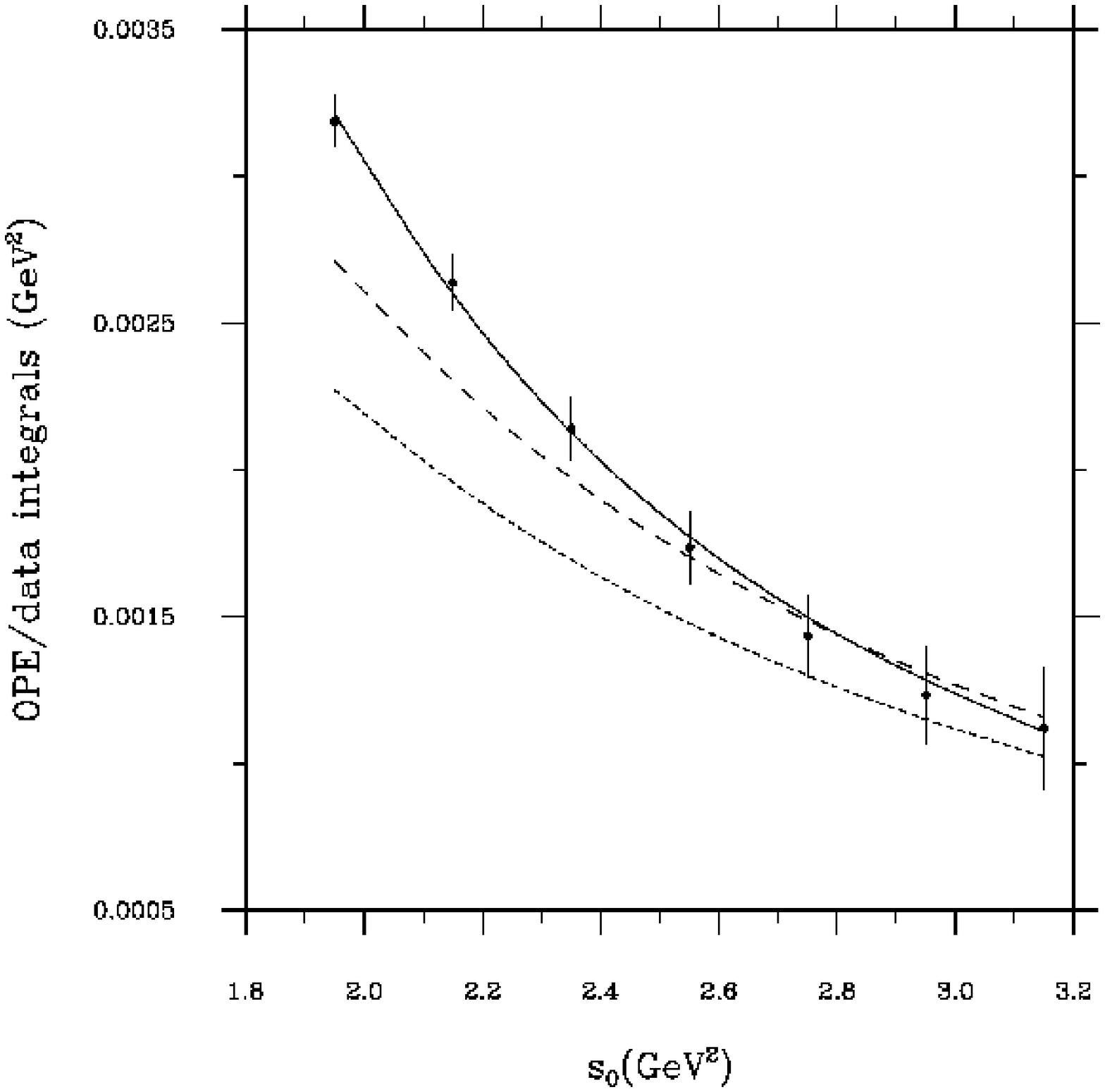,height=8.8cm,width=8.8cm}
\end{picture}
\end{minipage}}
\label{fig2}
\end{figure}

\begin{figure}[htb]
\unitlength1cm
\caption{
$J_{w_n}(s_0)$  and  $f_{w_n}(\{ a_d\} ;s_0)$ 
for the $w_3$ (top left panel), $w_4$ (top right panel),
$w_5$ (bottom left panel) and $w_6$ (bottom right panel) pFESR's. 
The $J_{w_n}(s_0)$ 
integrals and errors were obtained using the ALEPH data
and covariance matrix.
Three versions of the 
$f_{w_n}(\{ a_d\} ;s_0)$ 
curve are shown.
The solid line corresponds to the combined fit given in 
Eqs.~(\ref{finalresults}) of the text, 
the short-dashed and long-dashed lines to the DGHS and
IZ solutions (for which $a_d=0$ for $d>8$).}
\begin{minipage}[t]{8.0cm}
\begin{picture}(7.9,7.9)
\epsfig{figure=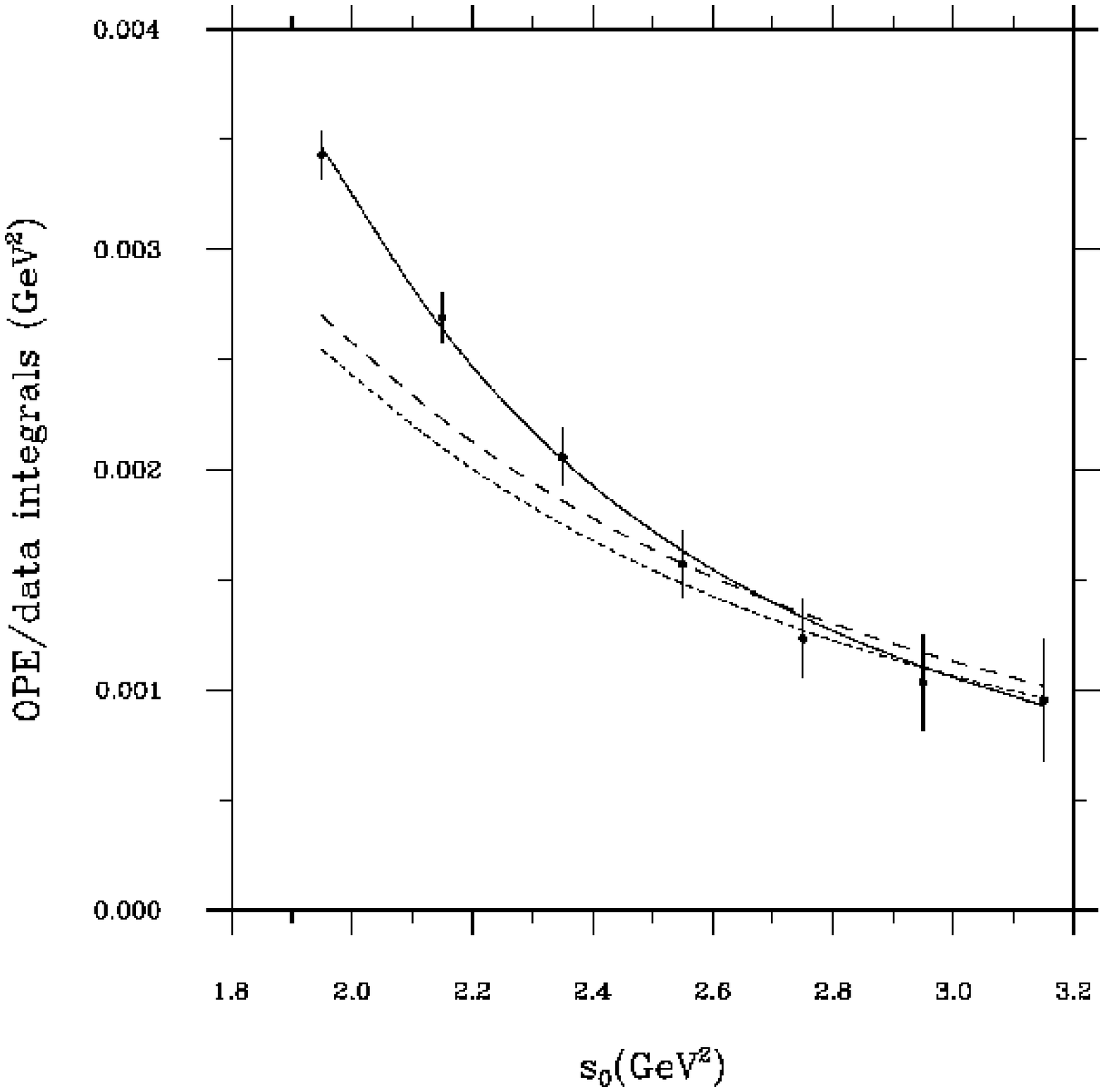,height=7.8cm,width=7.8cm}
\end{picture}
\end{minipage}
\hfill
\begin{minipage}[t]{8.0cm}
\begin{picture}(7.9,7.9)
\epsfig{figure=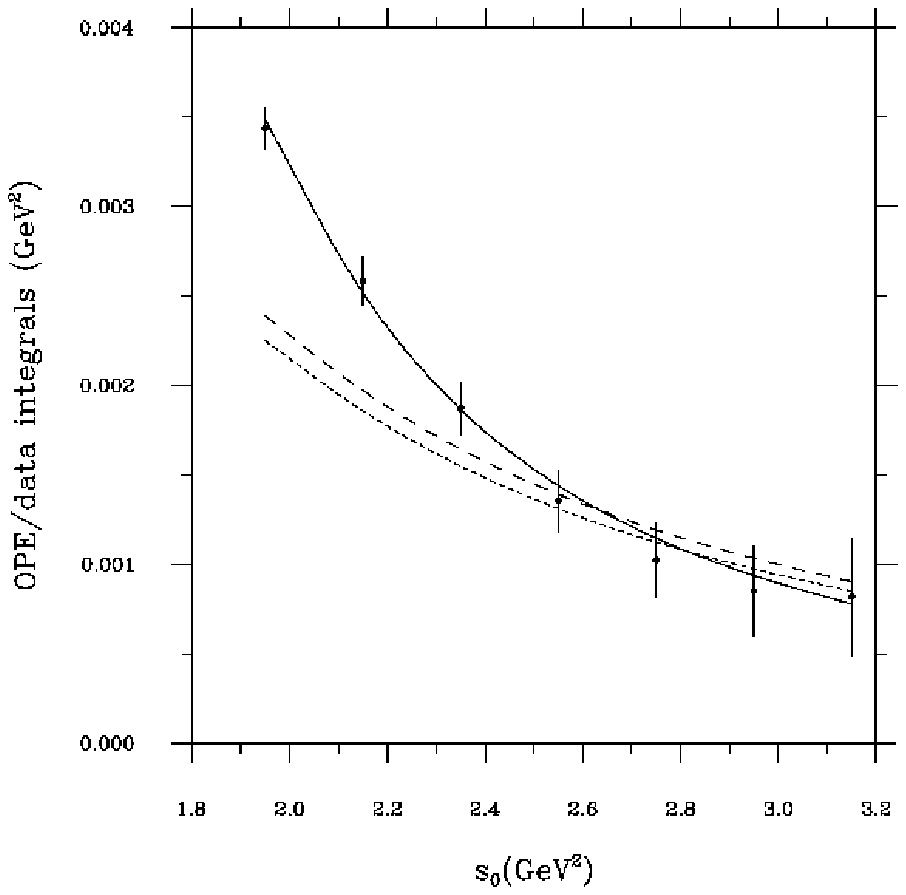,height=7.8cm,width=7.8cm}
\end{picture}
\end{minipage}
\vskip .15in\noindent
\begin{minipage}[t]{8.0cm}
\begin{picture}(7.9,7.9)
\epsfig{figure=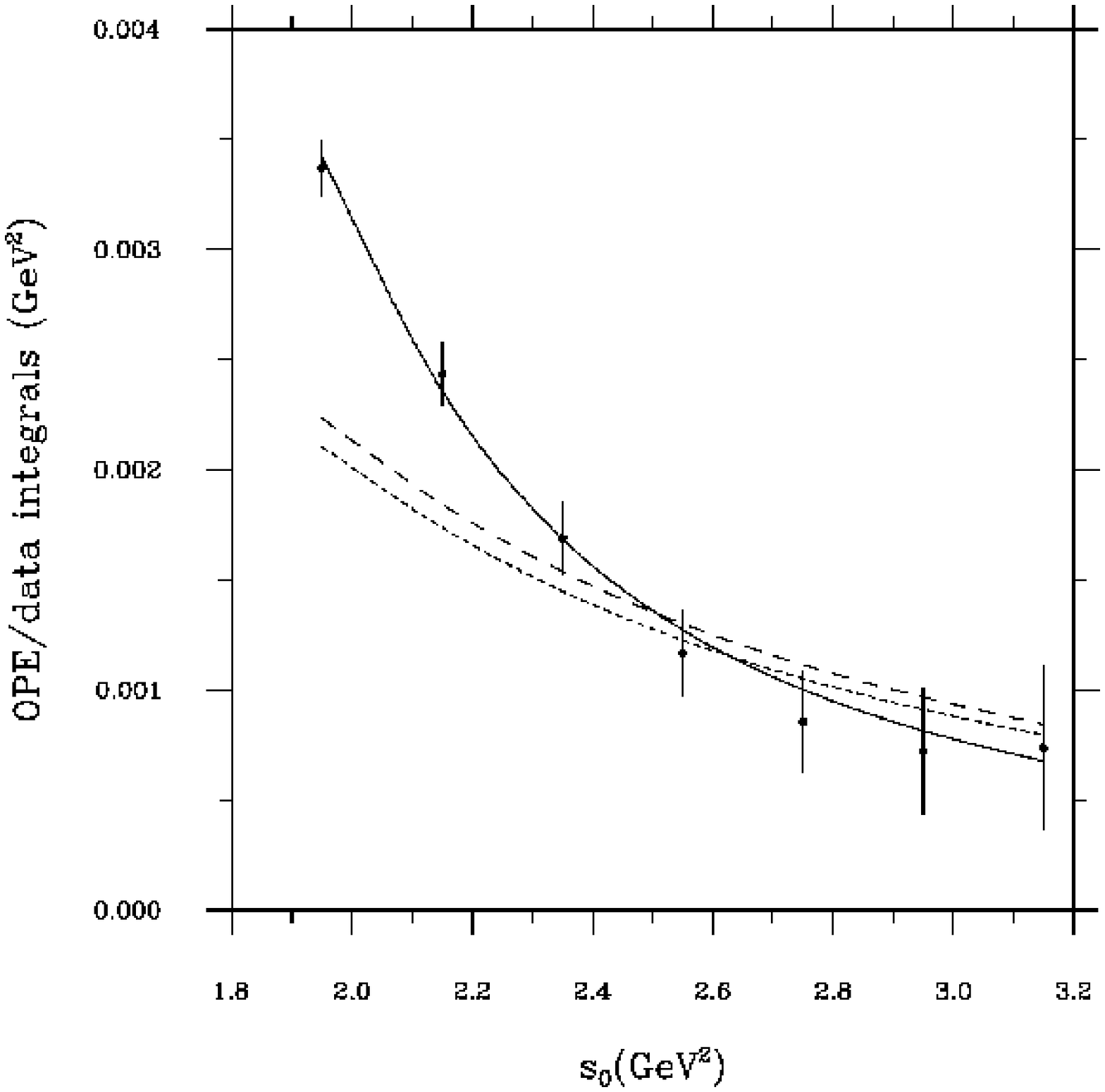,height=7.8cm,width=7.8cm}
\end{picture}
\end{minipage}
\hfill
\begin{minipage}[t]{8.0cm}
\begin{picture}(7.9,7.9)
\epsfig{figure=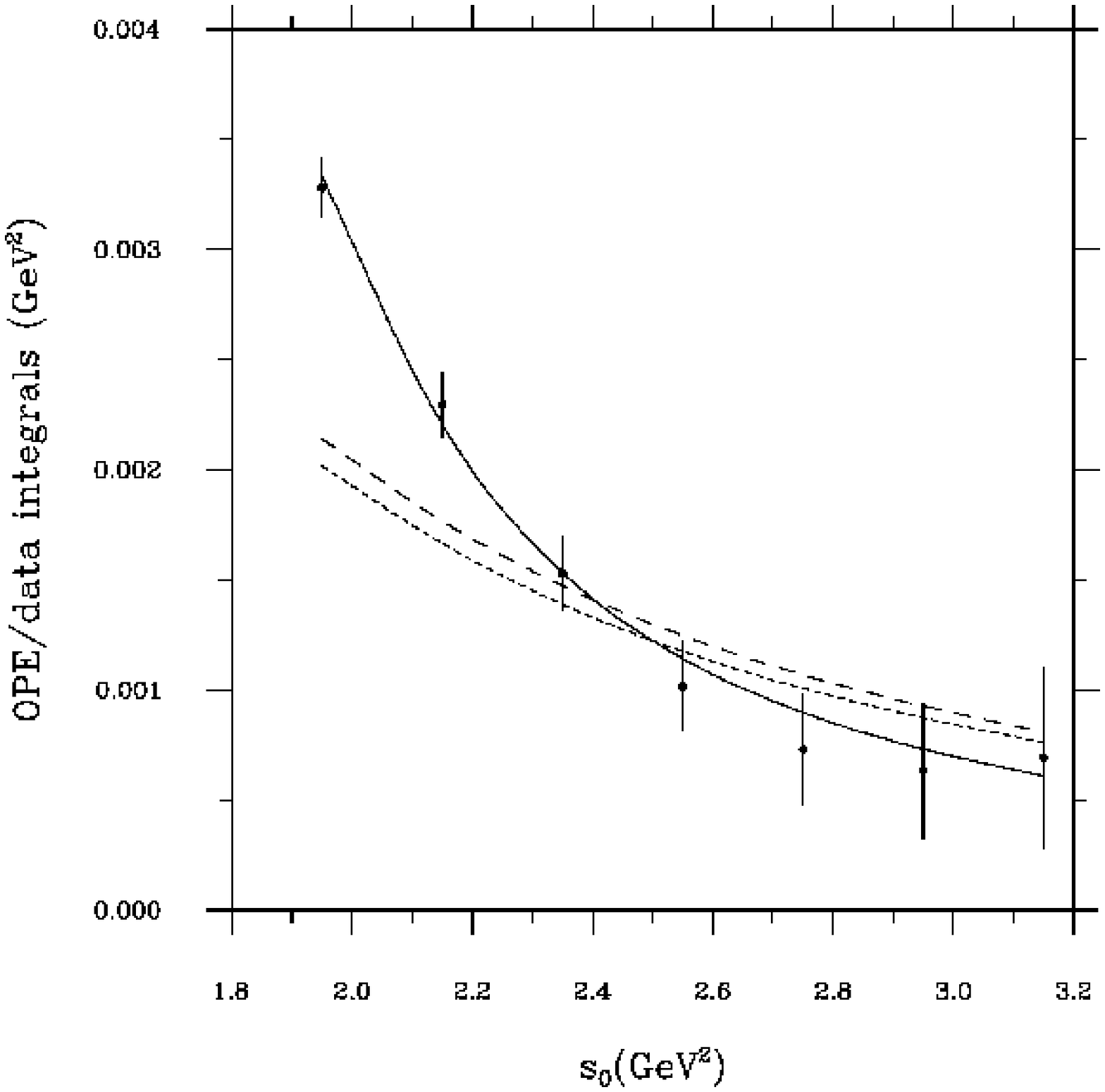,height=7.8cm,width=7.8cm}
\end{picture}
\end{minipage}
\label{fig3}\end{figure}

\begin{figure}[htb]
\unitlength1cm
\caption{
$J_{w_n}(s_0)$  and  $f_{w_n}(\{ a_d\} ;s_0)$ 
for the $w_7$ (top left panel), $w_8$ (top right panel),
$w_9$ (bottom left panel) and $w_{10}$ (bottom right panel) pFESR's. 
The 
$J_{w_n}(s_0)$ 
integrals and errors were obtained using the ALEPH data
and covariance matrix.
The notation for the three OPE curves is as in Figure 3.}
\begin{minipage}[t]{8.0cm}
\begin{picture}(7.9,7.9)
\epsfig{figure=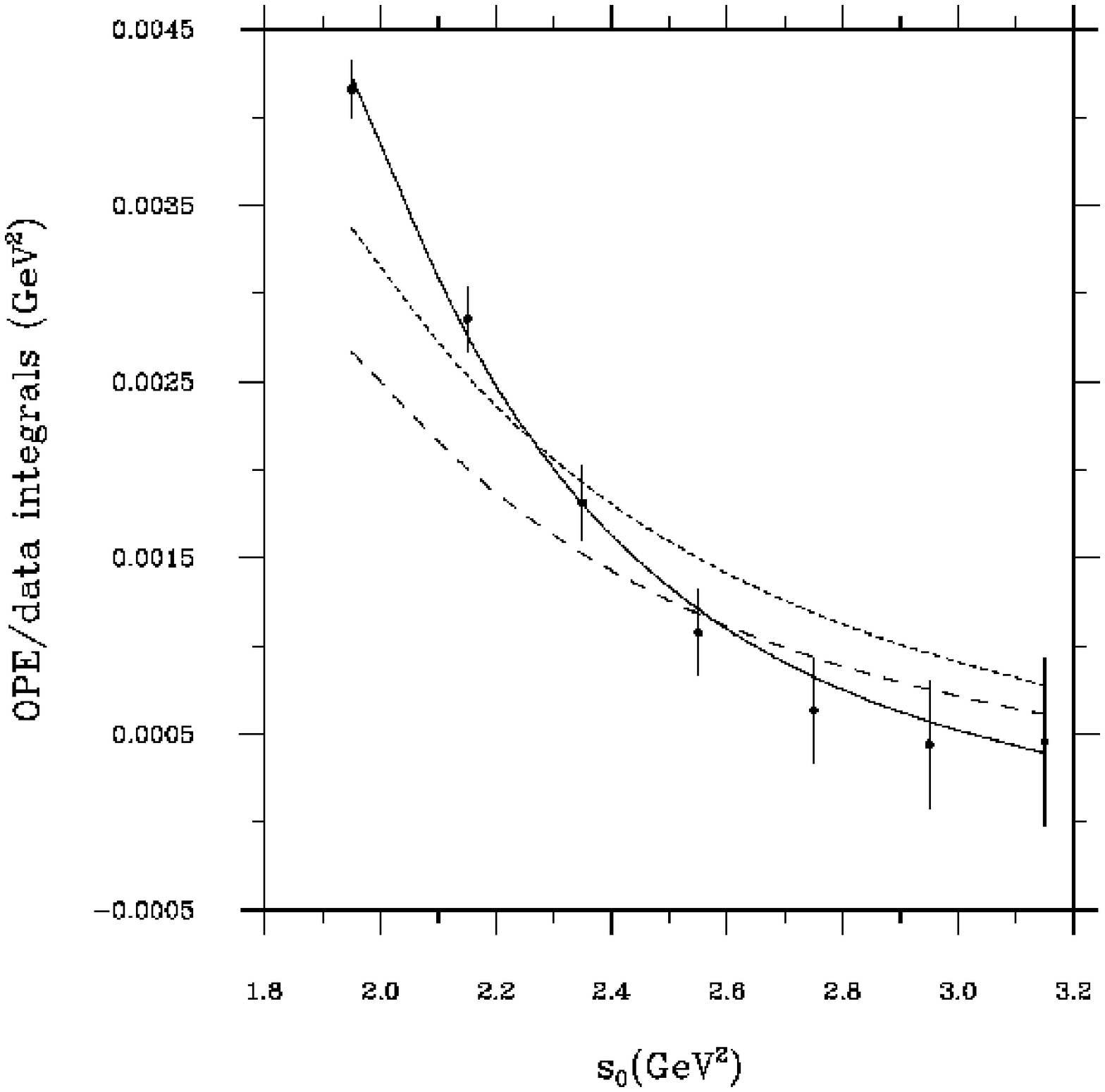,height=7.8cm,width=7.8cm}
\end{picture}
\end{minipage}
\hfill
\begin{minipage}[t]{8.0cm}
\begin{picture}(7.9,7.9)
\epsfig{figure=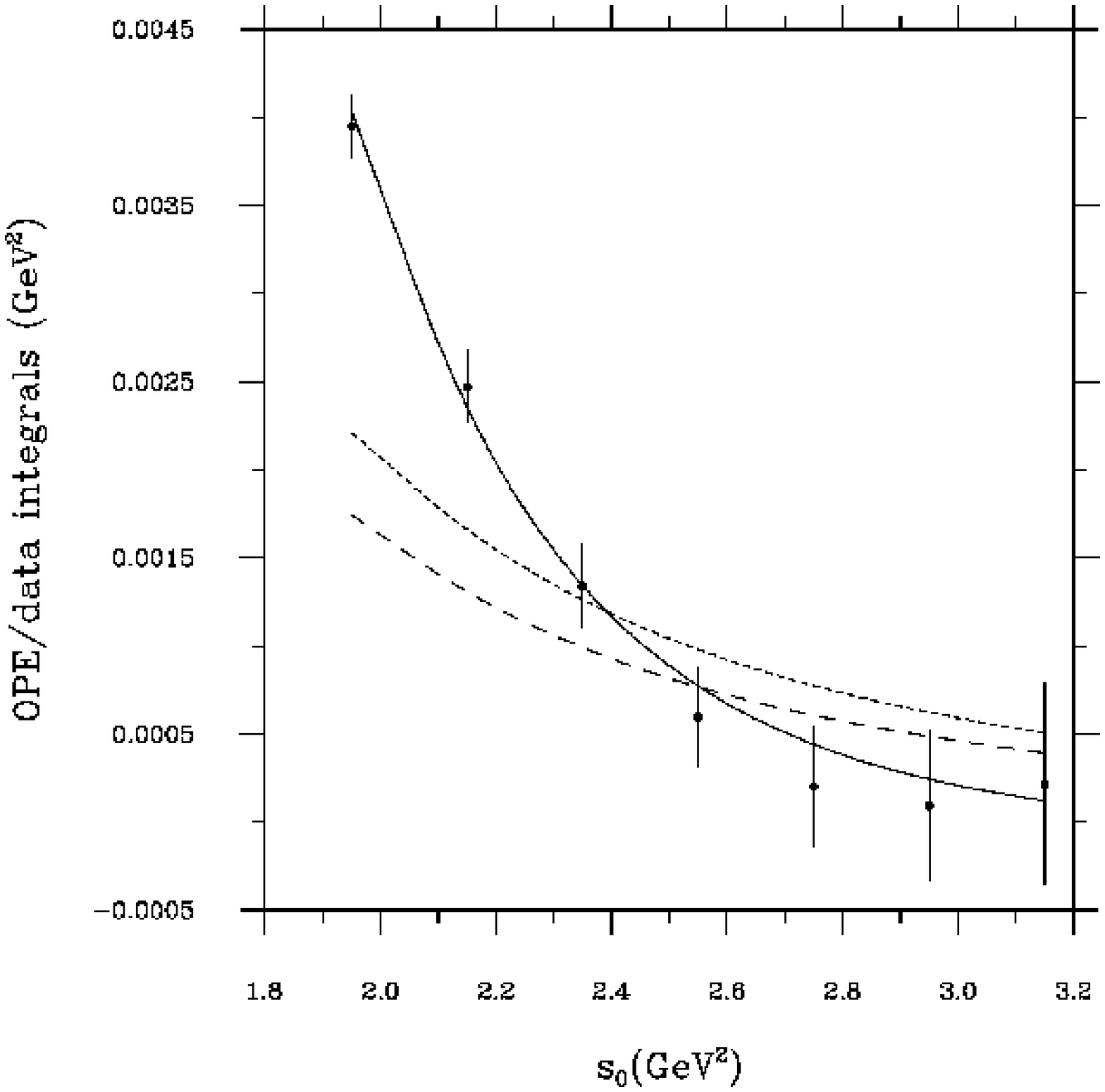,height=7.8cm,width=7.8cm}
\end{picture}
\end{minipage}
\vskip .15in\noindent
\begin{minipage}[t]{8.0cm}
\begin{picture}(7.9,7.9)
\epsfig{figure=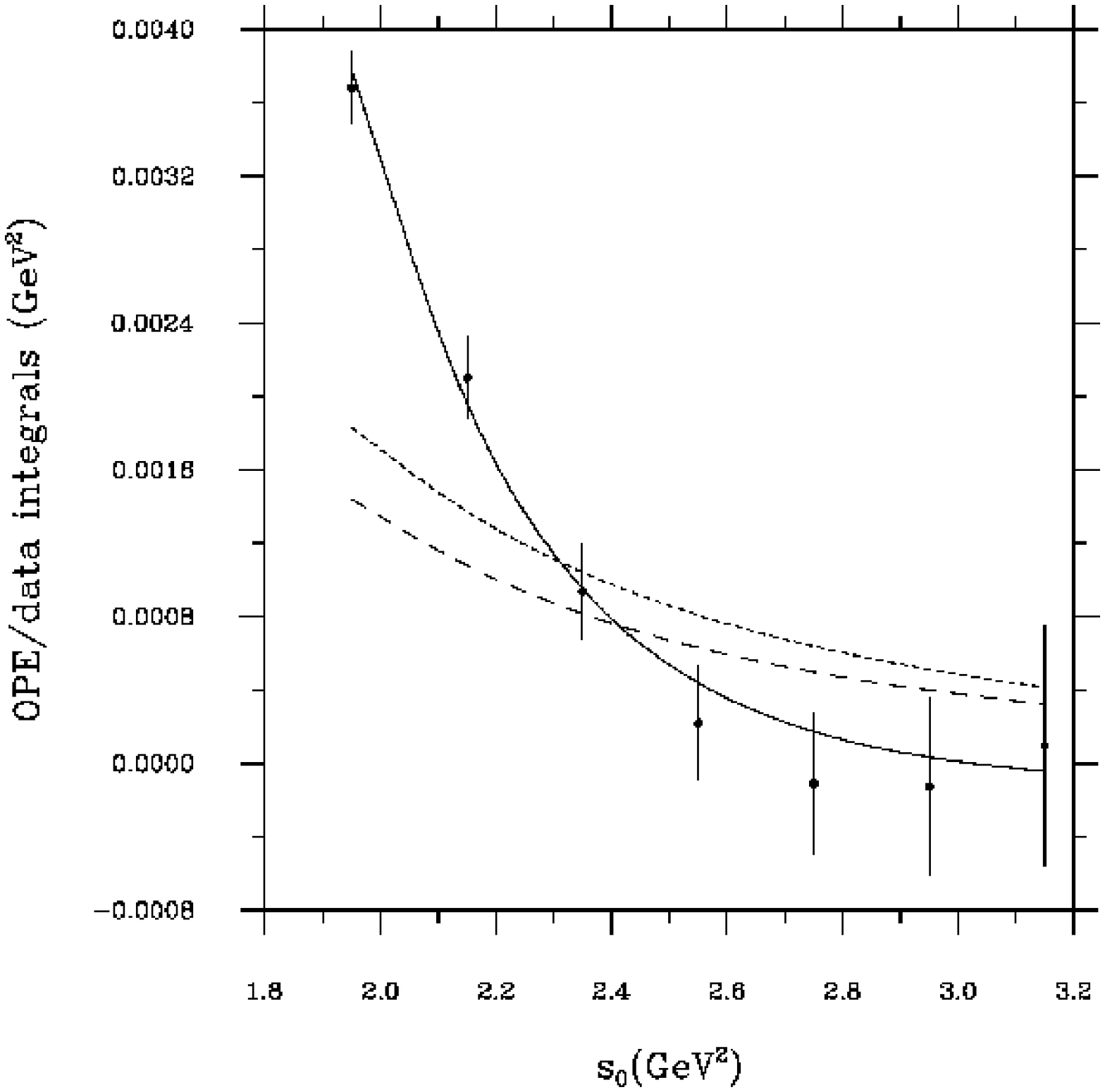,height=7.8cm,width=7.8cm}
\end{picture}
\end{minipage}
\hfill
\begin{minipage}[t]{8.0cm}
\begin{picture}(7.9,7.9)
\epsfig{figure=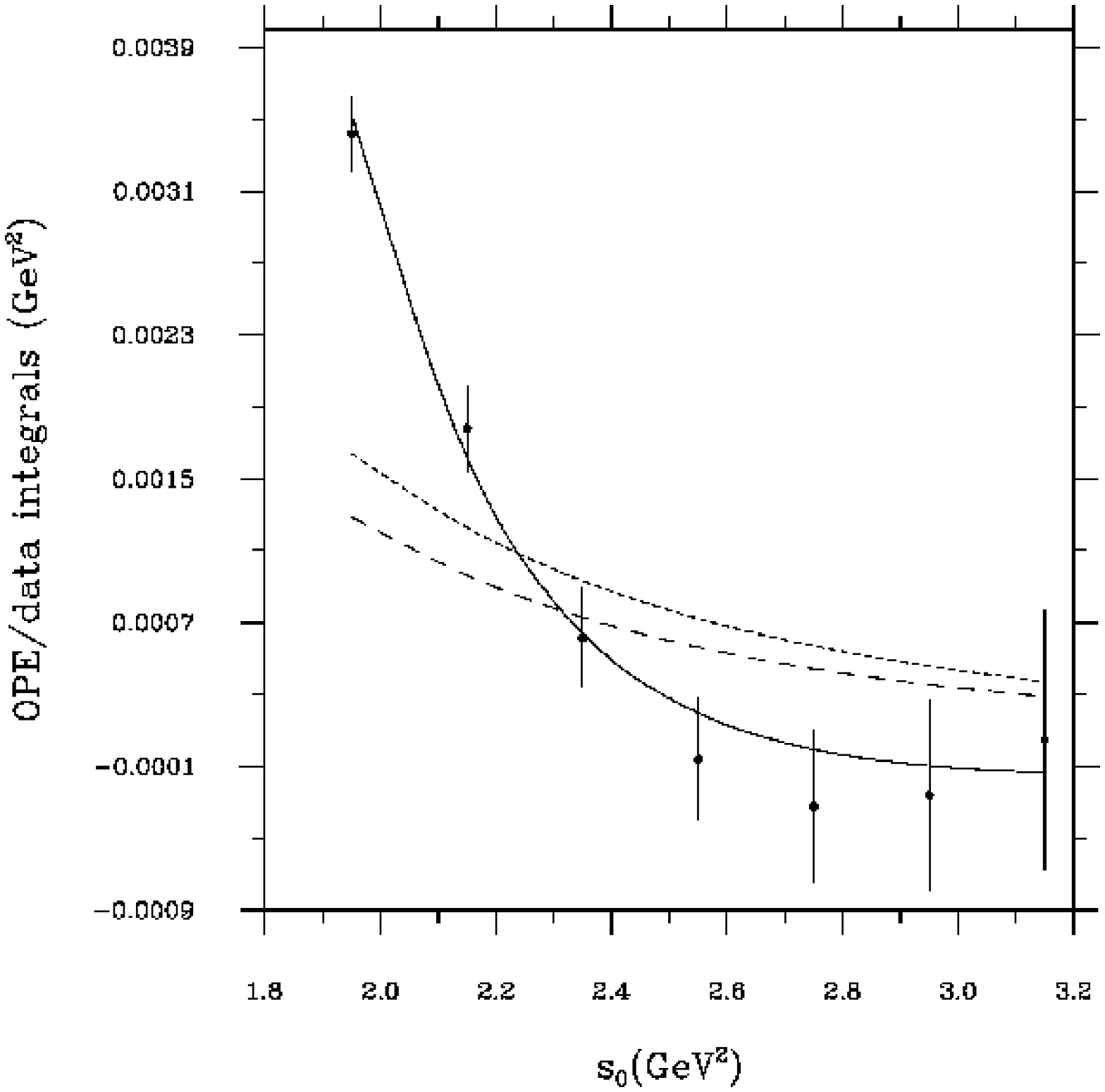,height=7.8cm,width=7.8cm}
\end{picture}
\end{minipage}
\label{fig4}\end{figure}

It is important to verify that, after fitting the OPE 
coefficients $a_d$, the resulting OPE integrals
$f_{w_n}(\{ a_d\} ;s_0)$  
provide a good match to the corresponding spectral integrals 
$J_{w_n}(s_0)$ 
over the whole of the $s_0$ analysis window. 
Failure to achieve such a match 
would represent a clear sign of duality violation. 
In Figs.~\ref{fig2}, \ref{fig3} and \ref{fig4} we display
the quality of the 
$f_{w_n}(\{ a_d\} ;s_0)$ / $J_{w_n}(s_0)$ 
%%% OPE/spectral integral 
match  for the combined fit to the ALEPH 
data. (The match for the combined fit to the OPAL data is 
of identical quality, and hence not shown separately.)
Fig.~\ref{fig2} shows the
results for the $w_1$ and $w_2$ pFESR's{\begin{footnote}
{We plot only the results of the combined fit in this case
since they are indistinguishable from those of the ``maximally-safe'' 
fit on the scale of the figure.}\end{footnote}}, 
Fig.~\ref{fig3} for the 
$w_3$ through $w_6$ pFESR's, and Fig.~\ref{fig4}
for the $w_7$ through $w_{10}$ pFESR's. 
Our results for $f_{w_n}(\{ a_d\} ;s_0)$,  
corresponding to Eqs.~(\ref{finalresults}), are given by the solid lines.
There is clearly no sign of duality violation for
any of the pFESR's employed at any of the scales, $s_0$, in our
analysis window. Improved data would reduce the errors on $J_{w_n}(s_0)$ 
and allow us to sharpen this test even 
further. Also shown for comparison in each figure are the OPE results
corresponding to the $a_6$, $a_8$ fits of Refs.~\cite{DGHS98,IZ00},
where, as in those references we take as central input values 
$a_d=0$ for $d>8$. The inclusion of $d>8$ contributions
clearly leads to a significantly improved fit to the data, as well as
a significantly reduced error on the determination, in particular, of
$a_6$. 

The excellent agreement between the optimized OPE representation 
and the corresponding data integrals displayed in Figs.~\ref{fig2}
through \ref{fig4}, while a necessary condition that
significant duality violation be absent from our analysis,
is not a sufficient one. In order to investigate this question further,
we have performed a number of additional tests on our solution sets.
Since several of these tests correspond to sum rules studied in earlier
analyses of the V-A correlator, we first discuss the relation between
our results and those of these earlier analyses. Having introduced
the relevant sum rules as part of this discussion, we
will then return to a discussion of the additional tests
which such sum rules allow us to perform on our solution sets
in Section~\ref{Sect:dualityviolation}.
%Since certain aspects of the relation between our results
%and those of earlier analyses of the V-A correlator 
%are relevant to an understanding of the efficacy of these tests,
%we expand on this relation in Section~\ref{Sect:comparisons}
%before returning to a discussion of these additional tests 
%in Section~\ref{Sect:dualityviolation}.

\section{Previous Analyses}
\label{Sect:comparisons}

Several determinations of the $d=6$ and $d=8$ contributions to the OPE
of $\Delta\Pi$ exist already in the
literature~\cite{DGHS98,ALEPH,OPAL,IZ00,bgp01,bgp02,ppr}.
In some cases the quoted results (especially for $a_8$) differ significantly 
from ours. To pin down the source of 
these discrepancies, a closer scrutiny of the previous analyses is in order.
In general, previous results have errors much
larger than those on the spectral function over most of its measured range. 
This suggests either the impact of strong cancellations or the presence of 
additional theoretical systematic uncertainties. 
One obvious possibility is the presence of $d>8$ contributions,
neglected in the analyses of Refs.~\cite{DGHS98,ALEPH,OPAL,IZ00},
in the solutions for $a_6$, $a_8$.
We will demonstrate below that, for both $a_6$ and $a_8$, 
the differences between our results and those of previous analyses
are naturally accounted for by the $d>8$ coefficients given
in Eqs.~(\ref{finalresults}), (\ref{finalresultsOPAL}).

In what follows, we shall recall the basic ingredients of the earlier
analyses and discuss possible sources of uncertainty. 
%Some more detailed
%aspects of the comparisons to our work are deferred to the Appendix.

\subsection{Spectral Weight Analyses}

In Refs.~\cite{DGHS98,ALEPH,OPAL}, the ``(k,m) spectral weights'',
\beq 
%$
w^{(k,m)}(y)=y^m(1-y)^{2+k}(1+2y) \ , 
%$
\eeq
with $(k,m)=(0,0)$ and
$(1,m)$, $m=0,\cdots ,3$, were employed to extract $a_6$ and $a_8$,
{\it under the implicit assumption that contributions with $d>8$ were
negligible in all cases}{\begin{footnote}{Ref.~\cite{DGHS98} also
employed the $(1,-1)$ spectral weight, not included in the other
analyses, in order to allow the simultaneous extraction of the 
NLO chiral LEC $L_{10}$.}\end{footnote}}. The fits for $a_6$ and $a_8$ were, 
in all cases, performed using only the highest $s_0$ available, 
$s_0=m_\tau^2$. 

Ref.~\cite{DGHS98} (DGHS) represents an update of the 
earlier ALEPH analysis~\cite{ALEPH}, and concentrates specifically 
on the V-A combination, which was not studied independently
in the original ALEPH paper. The results for $a_6$, $a_8$ thus supercede those 
inferred from the separate V, A extractions performed in Ref.~\cite{ALEPH}.
The results, in our notation, are
\begin{eqnarray}
a_6&=&\left( -6.4\pm 1.8\right)\times 10^{-3}\ {\rm GeV}^6\nonumber\\
a_8&=&\left(\ \ 8.7\pm 2.4\right)\times 10^{-3}\ {\rm GeV}^8\ .
\label{DGHSa6a8}
\end{eqnarray}
They are in good agreement with the results 
of the OPAL analysis~\cite{OPAL},
\begin{eqnarray}
a_6&=&\left( -6.0\pm 0.1\right)\times 10^{-3}\ {\rm GeV}^6\nonumber\\
a_8&=&\left(\ \ 7.6\pm 0.6\right)\times 10^{-3}\ {\rm GeV}^8\ .
\label{OPALa6a8}
\end{eqnarray}
One should bear in mind that the DGHS and OPAL analysis methods are
somewhat different: the DGHS results follow 
from a dedicated V-A analysis, while the OPAL results were generated 
by combining the $d=6,8$ contributions 
extracted for the separate V and A correlators. 
The separate V, A analyses, however, involve an additional OPE fitting 
parameter, the gluon condensate, which is absent in the V-A difference.
The fits display very strong correlations between $a_6$, $a_8$
and the gluon condensate~\cite{OPAL}. 
A dedicated V-A analysis of the OPAL data would thus, in general,
be expected to give different results for $a_6$, $a_8${\begin{footnote}{We 
thank Sven Menke for bringing this
point to our attention. No analogue of the DGHS update 
of the ALEPH analysis exists, at present, for the OPAL data.}\end{footnote}}. 
In view of this, and the
good agreement between the OPAL and DGHS results, we 
concentrate on the DGHS solution in the discussion which
follows{\begin{footnote}{A slightly different set of values,
corresponding to an average of the results of Refs.~\cite{ALEPH}
and ~\cite{OPAL}, has been used in the $(0,0)$
spectral weight analysis of Section 7 of Ref.~\cite{narison01}.
The value for $a_6$ is the same as that of DGHS,
while that for $a_8$ is $\sim 15\%$ higher. The reader interested in
the chiral limit value of the $K\rightarrow\pi\pi$ matrix
element of the electroweak penguin operator, $Q_8$, should
bear in mind, not only the difference between the $a_6$
values of Refs.~\cite{DGHS98,narison01} and our results above, but also
the fact that the extractions of the dominant,
$\langle O_8\rangle$, contribution to $a_6$ in 
Refs.~\cite{DGHS98,narison01} employ a value for the 
coefficient $A_8$ much larger than that given above. To convert 
$\langle O_8\rangle$ as determined in Refs.~\cite{DGHS98,narison01} 
to the same renormalization scheme as used for the
Wilson coefficients of the effective weak Hamiltonian
(and hence to make meaningful comparisons with the results of
Refs.~\cite{cdgm2,cdgm,bgp01,bgp02,knecht01}), one must multiply these
results by factors $1.15$ and $1.27$ for the NDR and HV $\gamma_5$ schemes,
respectively.}\end{footnote}}. 

The DGHS value for $a_6$ is consistent with ours, within errors,
but that for $a_8$ is not. 
We have studied the origin of this discrepancy, and we find that:
\begin{itemize}
\item  The discrepancy can be understood as arising 
from the neglect of the $d>8$ contributions to the spectral
weight sum rules employed by DGHS. 
\item  The ``(k,m) spectral weights'' FESR actually provide a consistency 
check on our solution set.
\end{itemize}
We first note that the $(0,0)$ and $(1,0)$ pFESR's have 
strong V-A cancellations, and hence large experimental errors on the data 
sides of the sum rules. For the $(0,0)$ case, which involves, from among 
the unknown $d>4$ $a_d$ terms, only the $a_6$ and $a_8$ contributions,
$\delvma \sim 3\%$  for $s_0=m_\tau^2$.
The $(1,0)$ case, whose OPE side in principle involves $a_6$, $a_8$ 
and $a_{10}$, also has $\delvma \sim 3\%$ for $s_0 = m_\tau^2$. 
The $(1,1)$, $(1,2)$ and $(1,3)$ weights produce much less pronounced V-A
cancellations~\footnote{For example,
for the $(1,1)$ pFESR, $\delvma =32\%$ for $s_0=m_\tau^2$.},
and hence must dominate the DGHS fit.
Note, however, that $w^{(1,m)}(y)=y^m[1-y-3y^2+5y^3-2y^4]$; these
weights thus produce numerical enhancements of the $a_d$, $d>8$ terms,
whose presence on the OPE sides of the sum rules has been assumed to
be numerically negligible (see explicit example below). 
The pFESR's dominating the fit are thus those
for which neglect of the $d>8$ terms is least safe.  

It is easy to check that the combined fit values for $a_{10},\cdots ,
a_{16}$ predict non-negligible $d>8$ contributions for all the 
$(1,m)$ pFESR's. Our results thus imply that the DGHS values for
$a_6$ and $a_8$, which are dominated by $m=1,2,3$ cases,
must contain higher dimension contamination. 
That the central DGHS $a_6$, $a_8$ values do not provide
as good a fit to the $w_1$ and $w_2$ 
pFESR's (for which $d>8$ contributions are absent) 
as does our combined fit is, presumably, a reflection
of this contamination. Further evidence is provided by the
$w_3$ through $w_{10}$ pFESR's. 

\begin{figure} [htb]
\centering{\
\epsfig{figure=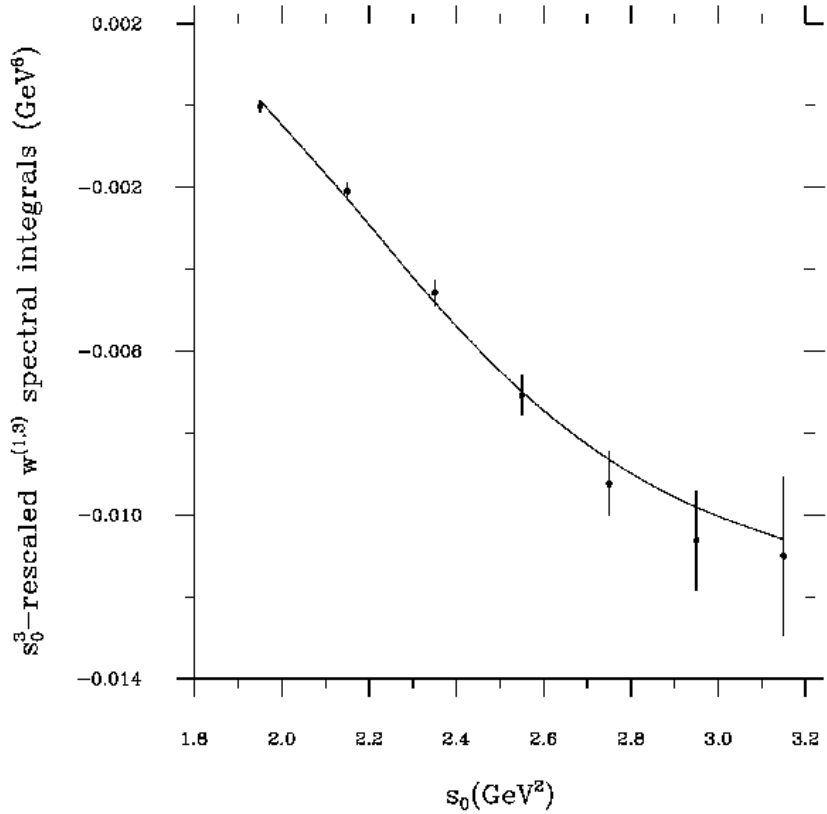,height=12.0cm}}
\caption{
The rescaled $(1,3)$ spectral weight combination,
$s_0^3 \, J_{w^{(1,3)}}(s_0)$ versus $s_0$.
%The data $-\ (d=4)$ integrals rescaled by $s_0^3$
%for the $(1,3)$ spectral weight pFESR of DGHS.
The integrals $J_{w^{(1,3)}}(s_0)$ 
and errors were obtained using the ALEPH data
and covariance matrix. 
The solid line shows the OPE prediction 
$f_{w^{(1,3)}}(\{ a_d\} ;s_0)$ 
corresponding
to the combined fit of Eqs.~(\ref{finalresults}).}
\label{fig5}
\end{figure}

One can also explicitly demonstrate that the neglected $d>8$
contributions are, indeed, important for the spectral weight
pFESR's. This demonstration is most transparent for the $(1,3)$ pFESR
since, in this case, the OPE integral is:
\beq
f_{w^{(1,3)}}(\{ a_d\} ;s_0)   = 
-  {a_8 \over s_0^3}  - { a_{10} \over s_0^4}
+3 \, {a_{12} \over s_0^5} + 5 \, {a_{14} \over s_0^6}
+2 \,  { a_{16} \over s_0^7} \ .
\eeq
If $d>8$ OPE contributions are indeed negligible
then,  rescaling  $ J_{w^{(1,3)}}(s_0) $
by $s_0^3$ should produce a result, $-a_8$, independent of 
$s_0$.~{\begin{footnote}{This is valid up to small logarithmic 
corrections. For the DGHS solution, the correction
associated with the $d=6$ logarithmic term varies 
from $2.3\%$ to $1.5\%$ as $s_0$ increases from $1.95$
to $3.15\ {\rm GeV}^2$.}\end{footnote}}
We plot $ s_0^3 J_{w^{(1,3)}}(s_0) $ for the ALEPH data
in Fig.~\ref{fig5}. The result is clearly far from constant with respect to
$s_0$, unambiguously demonstrating the presence of non-negligible
$d>8$ contributions. The solid line shows, for comparison,
the predictions corresponding to the combined fit solution of
Eqs.~(\ref{finalresults}). The good match shows that
the $d>8$ contributions produced by our solution naturally account for the
discrepancy between the DGHS predictions and the experimental results. 

\begin{figure}[htb]
\unitlength1cm
\caption{
$J_{w^{(k,m)}}(s_0)$  and  $f_{w^{(k,m)}}(\{ a_d\} ;s_0)$ 
for the $(0,0)$ (top left panel), $(1,0)$ (top right panel),
$(1,1)$ (bottom left panel) and $(1,2)$ (bottom right panel) 
spectral weight pFESR's. 
The 
$J_{w^{(k,m)}}(s_0)$ 
integrals and errors were obtained using the ALEPH data
and covariance matrix.
The dashed and solid curves correspond to the OPE fit of DGHS,
and our combined fit (Eqs.~(\ref{finalresults})), respectively.}
\begin{minipage}[t]{8.0cm}
\begin{picture}(7.9,7.9)
\epsfig{figure=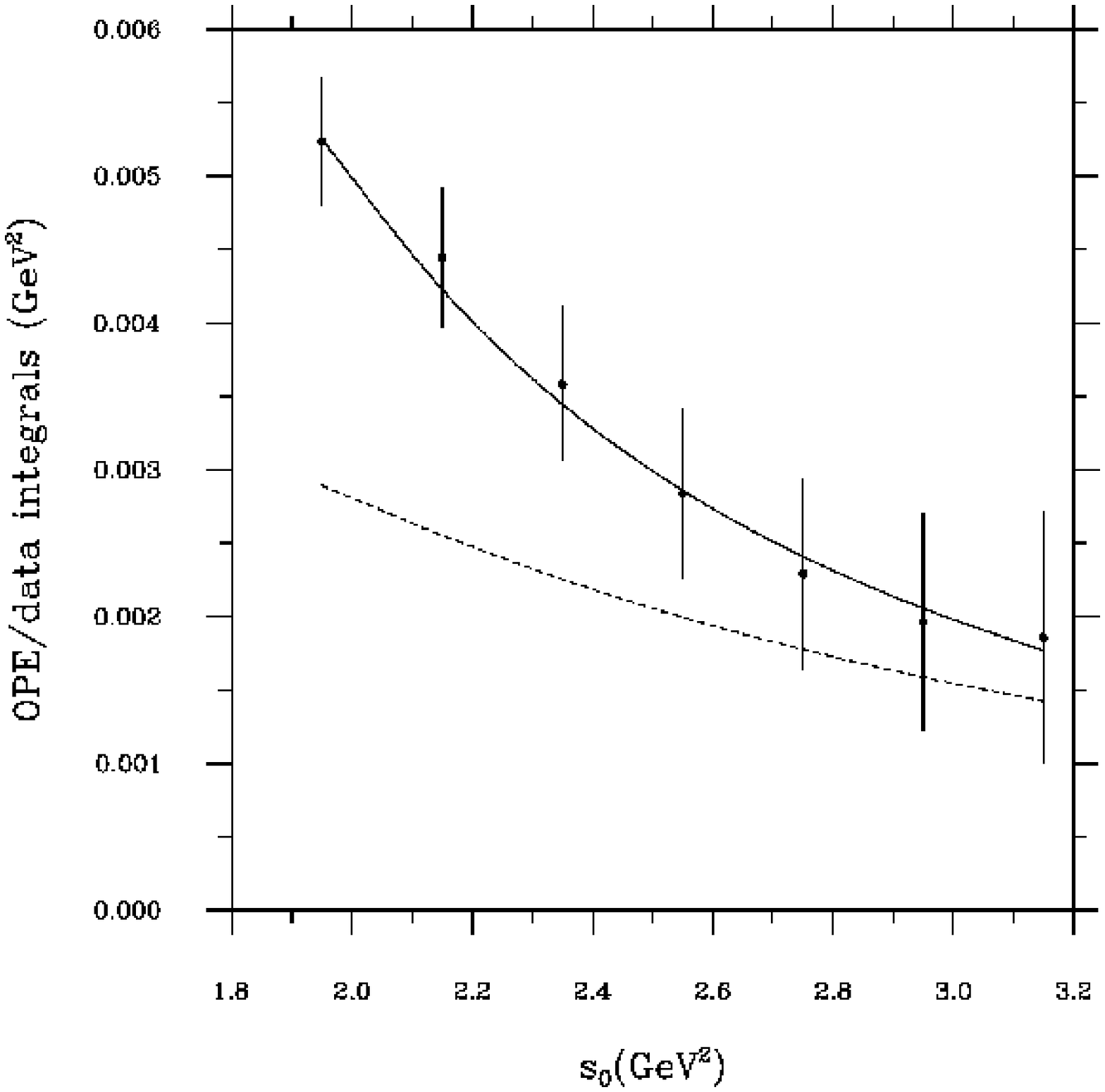,height=7.8cm,width=7.8cm}
\end{picture}
\end{minipage}
\hfill
\begin{minipage}[t]{8.0cm}
\begin{picture}(7.9,7.9)
\epsfig{figure=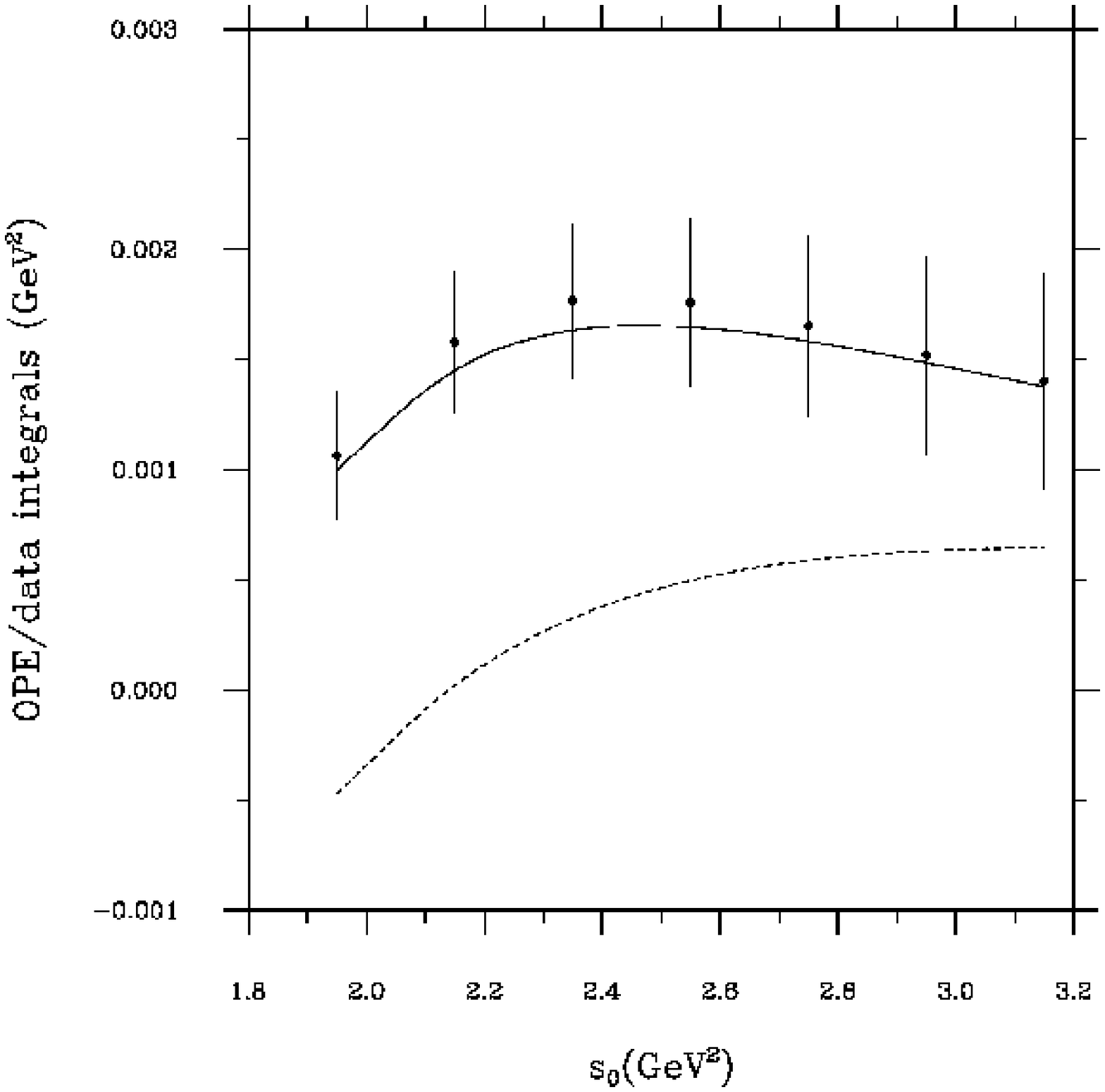,height=7.8cm,width=7.8cm}
\end{picture}
\end{minipage}
\vskip .15in\noindent
\begin{minipage}[t]{8.0cm}
\begin{picture}(7.9,7.9)
\epsfig{figure=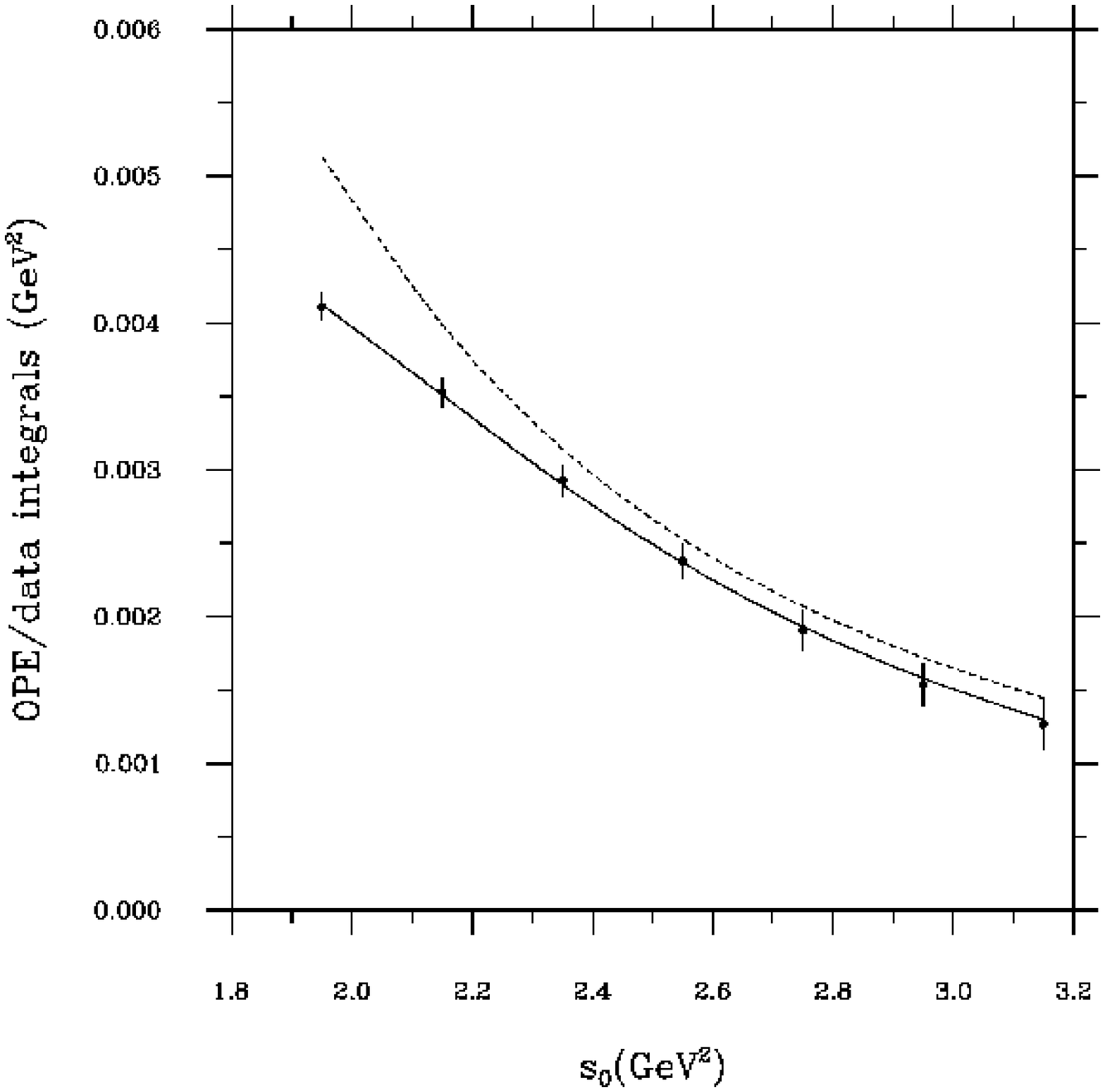,height=7.8cm,width=7.8cm}
\end{picture}
\end{minipage}
\hfill
\begin{minipage}[t]{8.0cm}
\begin{picture}(7.9,7.9)
\epsfig{figure=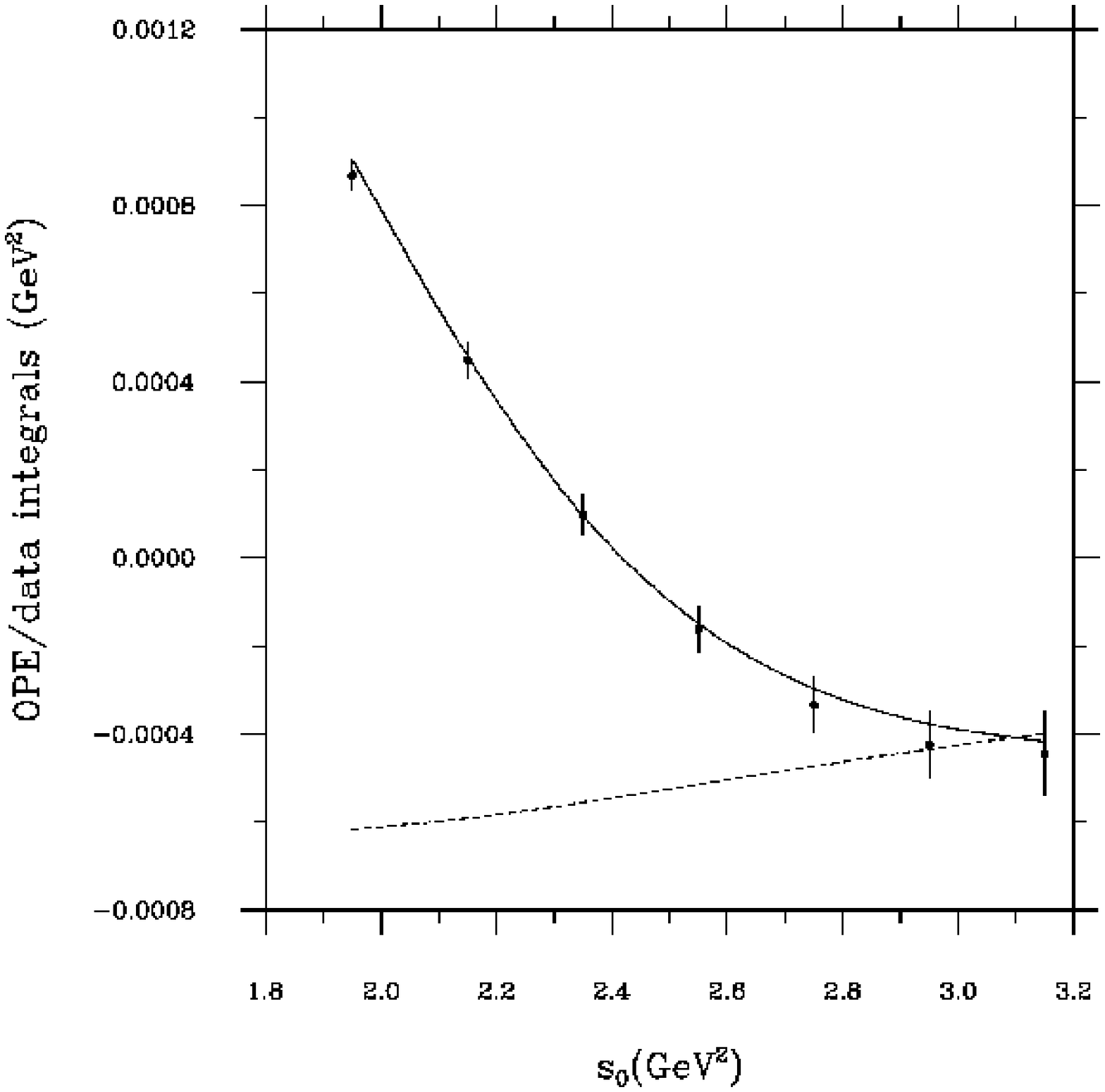,height=7.8cm,width=7.8cm}
\end{picture}
\end{minipage}
\label{fig6}\end{figure}

A similar situation holds for the other spectral weight sum rules, as
shown in Fig.~\ref{fig6}. In the figure we display the $
J_{w^{(k,m)}}(s_0)$ together with the OPE expressions
$f_{w^{(k,m)}}(\{ a_d\} ;s_0) $ corresponding to (i) the central
values of the DGHS fit, Eqs.~(\ref{DGHSa6a8}), together with $a_d=0$
for $d>8$, (shown by the dashed line), and (ii) our combined
(ALEPH-based) fit, Eqs.~(\ref{finalresults}), (shown by the solid
line). In all cases, if one takes into account the errors and
correlation for the DGHS fit parameters, the resulting OPE error bar
overlaps the spectral integral bar at $s_0=m_\tau^2$, even when the
central values are not in particularly good agreement. However, when
one goes to lower $s_0$ this is no longer the case; the shape of the
curve for the OPE integrals as a function of $s_0$ is typically rather
different from that for the spectral integrals. This is another signal
of missing higher dimension contributions on the OPE sides of the sum
rules. On the other hand when one considers the OPE contributions
implied by our combined fit, a high-quality match between the OPE and
data integrals is obtained. This is a non-trivial 
consistency test on our solution set.

\subsection{The IZ pFESR and Borel Sum Rule Analyses}

In Ref.~\cite{IZ00} (IZ), three approaches were considered: (1) pFESR's
with $w(y)=(1-y)^2$ and $y(1-y)^2$, (2) Borel transformed
dispersion relations involving $\Delta\Pi (Q^2)$ for $Q^2$ lying
along various fixed rays in the complex $Q^2$-plane, and (3) Gaussian
sum rules. The results, in this case, are
\begin{eqnarray}
a_6&=&\left( -6.8\pm 2.1\right)\times 10^{-3}\ {\rm GeV}^6\nonumber\\
a_8&=&\left(\ \ 7\pm 4\right)\times 10^{-3}\ {\rm GeV}^8\ ,
\label{IZ00a6a8}
\end{eqnarray}
and are dominated by the Borel sum rule (BSR) part of
the analysis, though
the other determinations are compatible with these, within
their (larger) errors. 

The pFESR part of the IZ analysis involves one weight, $w(y)=(1-y)^2$,
for which the V-A cancellation is extremely strong ($\delvma \sim
-4\%$ for $s_0=m_\tau^2$), and one, $w(y)=y(1-y)^2$, for which it
is considerably less so ($\delvma \sim 25\%$ for $s_0=m_\tau^2$). 
The strong cancellation for the $(1-y)^2$ case
leads to large errors on $a_6$, and to the strong
% what-at-first-appears-surprising
sensitivity to the errors on $F_\pi$ noted in Ref.~\cite{IZ00}.
The necessity of subtracting the poorly determined $a_6$ contribution
to the $y(1-y)^2$ sum rule before obtaining the residual $a_8$
contribution, then leads to large errors on $a_8$ as well.

BSR's were employed in the second part of the
analysis because of factorial suppression of high $d$
contributions ($a_d$ contributions appear in the Borel
transform of the OPE side of the sum rule multiplied by 
$1/((d-2)/2)! M^d$~\cite{svz}, where $M$ is the Borel mass). Since,
however, the spectral data are known only up to $s=m_\tau^2$, and have
significant errors above $2\ {\rm GeV}^2$, IZ are forced to work at
quite low Borel masses to suppress contributions from the
region of the spectrum where either data errors are large or
data are absent.
Explicitly, $M^2\simeq 0.8\ {\rm GeV}^2$ is used for 
sum rules dominating 
the determination of $a_6$, and $M^2\simeq 0.6\ {\rm GeV}^2$ in sum
rules used for $a_8$.
At such low $M^2$, factorial suppression of high $d$
contributions is counteracted by the enhancement
associated with the smallness of the $M^{2N+2}$ factor in the denominator,
making the sum rules potentially sensitive to higher
dimension contributions.

While the central IZ values for $a_6$ and $a_8$ are obtained
neglecting $d>8$ contributions, the quoted errors include, not only
the uncertainties due to experimental errors, but also a contribution
meant to represent a plausible bound on the magnitude of the $d>8$
terms. This bound is based on the assumptions that $\vert
a_{10}\vert$ and $\vert a_{12}\vert$ are bounded by $2\ {\rm
GeV}^4\vert a_6\vert$ and $5\ {\rm GeV}^6\vert a_6\vert$,
respectively. According to the results of our fit, these assumptions
are not sufficiently conservative: the bounds, in both cases,
lie well outside the range allowed by the errors on the combined
fit values. Recall also that, as shown in Figures~\ref{fig2},
\ref{fig3} and \ref{fig4}, the central IZ $a_6$, $a_8$ values
do not provide good fits to the $w_1$ through $w_{10}$ pFESR's.
In contrast, our combined fit implies values for the OPE sums
for the four IZ BSR's which are in excellent agreement with experiment.
A demonstration of this claim, together with a
more detailed discussion of the four IZ BSR's,
may be found in the Appendix. 

Comments similar to those on the BSR's apply to the Gaussian sum rules
studied by IZ. Since, however, the Gaussian sum rule $a_6$
and $a_8$ errors are larger than those of the BSR analysis,
and the OPE convergence even slower, we will not
comment further on that part of the IZ analysis.

\subsection{Duality Point Analyses}

A recent discussion of duality point analyses
(summarized earlier in Section~\ref{Sect:intro})
can be found in Ref.~\cite{gppr}. 
We comment here on the most recent numerical results, obtained in
Refs.~\cite{bgp01,bgp02} (BGP){\begin{footnote}{An estimate of
the four-quark vacuum matrix elements which determine $a_6$,
obtained by truncating the spectral integrals appearing
in the dispersive sum rules of Ref.~\cite{dg} at the 
duality points of the WSR's, was also
given in Section 6 of Ref.~\cite{narison01}. The assumptions underlying
this analysis are even stronger than those underlying
the duality point truncation of the WSR's,
where the corrections for the truncation can be shown to
be numerically small. In addition, the original dispersive
sum rule for the dominant $\langle O_8\rangle$
contribution suffers from potential contamination by higher
dimension effects~\cite{cdghigherd} at the scale $\mu =2 \ {\rm GeV}$
employed in Ref.~\cite{narison01}. Since, in any case, the errors
from this approach are a factor of $>3$ larger than those
obtained by averaging the results of the ALEPH and OPAL 
$(0,0)$ spectral weight analyses in Section 7 of the same reference,
we do not discuss this estimate any further.}\end{footnote}}.

BGP determine $a_6$ and $a_8$ from FESR's based on the weights $w(s)=s^2$ and
$w(s)=s^3$,{\begin{footnote}{A second determination of the 
dominant $\langle O_8\rangle$
contribution to $a_6$, which however requires an additional input
assumption, gives a compatible result.}\end{footnote}}
working at the highest duality point determined through
the second WSR. From the analysis based on ALEPH data
(for which $s_0^{(d)} = 2.53 ^{+0.13}_{-0.12}$ GeV$^2$) the
following results are quoted:
\begin{eqnarray}
a_6&=&- \left( 3.4 ^{+2.4}_{-2.0} \right)\times 10^{-3}\ {\rm GeV}^6\nonumber\\
a_8&=&- \left( 14.4 ^{+10.4}_{-8.0} \right)\times 10^{-3}\ {\rm GeV}^8\ .
\label{bgpa6a8}
\end{eqnarray}
These values are in qualitative agreement with ours, but are affected
by large uncertainties. The origin of these uncertainies is twofold. On the
one hand the analysis uses weights which emphasize the region where
the data errors are large and, on the other, the uncertainty in
the exact location of the WSR duality point gets amplified by the 
strong slope of the relevant spectral integrals with
respect to $s_0$ near $s_0 = s_0^{(d)}$. 

The errors on the second WSR duality point, $s_0^{(d)}$, quoted above are
entirely experimental in origin. The fact that the duality points for
the $s^2$, $s^3$ FESR's may not coincide exactly with those of the
second WSR, however, leads to an additional uncertainty which is not
amenable to experimental improvement. We think this uncertainty is
unlikely to be negligible, as argued below. 
%%%%%%%%%%%%%%%%%%%%%%%%%%%%%%%%%%%%%%%%%%%%%%%%%% 
If the duality points for the WSR's are universal, i.e. {\it all
FESR's} are satisfied at such $s_0^{(d)}$, then the values of the
extracted OPE parameters should not depend on the particular duality
point used in the analysis.  
Empirically, however, if one uses the lower 
of the two second WSR duality points ($s_0 = 1.47
\pm 0.02$ GeV$^2$~\cite{bgp01}), as advocated in Ref.~\cite{gppr,ppr},
one obtains~\cite{bgp01}:
\begin{eqnarray}
a_6&=&- \left( 13.2 \pm 0.4 \right)\times 10^{-3}\ {\rm GeV}^6\nonumber\\
a_8&=& \left( 24 ^{+ 2}_{-4} \right)\times 10^{-3}\ {\rm GeV}^8\ .
\label{bgpa6a8bis}
\end{eqnarray}
These results have much smaller errors 
(reflecting the better data quality at lower $s_0$) but are
not compatible with those obtained using the higher
duality point, Eq.~(\ref{bgpa6a8}).
It is thus impossible for the two duality points of the
second WSR to both be duality points of the $s^2$ and $s^3$ FESR's.
Since at least one of the two WSR duality points {\it must}
differ from the corresponding $s^2$, $s^3$ duality point, it seems
unlikely to us that either is exactly identical to its
$s^2$, $s^3$ counterpart. 

We emphasize that it is the strong slope
of the data integrals with respect to $s_0$ which is particularly
problematic for the duality point approach. The possibility of the existence
of a reasonably narrow $s_0$ region within which the actual 
duality points of a number of differently-weighted FESR's might lie 
is not itself implausible. Indeed, at those intermediate scales
suggested by the PQW argument~\cite{pqw}
(where OPE violation is small, except near the
timelike real axis), duality points for a wide range of FESR's 
would be expected to cluster in the vicinity of any $s_0$ for which
the real and imaginary parts of $\Pi (s_0)-\Pi_{OPE}(s_0)$
happened to be simultaneously small. In the case of $\Delta\Pi$,
the zeros of $Im\left[ \Pi (s_0)-\Pi_{OPE}(s_0)\right]$ on the
real axis occur at $s_0\simeq 0.9$ and $2.1\ {\rm GeV}^2$,
somewhat removed from the locations of the WSR duality points.
We would thus expect the $s^2$ and $s^3$ duality points to,
indeed, differ somewhat from the corresponding WSR duality
points. Since $\vert Im\left[ \Pi (s_0)-\Pi_{OPE}(s_0)\right]\vert$
is considerably smaller at the higher of the two WSR duality
points, we are in agreement with the authors of Refs.~\cite{bgp01,bgp02}
in expecting the higher of the two duality points to provide
the more reliable estimate of $a_6$ and $a_8$. This expectation
would appear to be borne out by comparison to our results.   
In particular, the result that $a_6$ and $a_8$ have the same sign, 
first obtained in Ref.~\cite{bgp01}, is confirmed by our analysis. We  
remind the reader that the opposite sign for $a_8$ obtained in both  
the spectral weight and BSR analyses is naturally accounted for by the  
$d>8$ contributions implied by our solution but neglected in those 
analyses.

\subsection{The MHA Analysis} 

In Ref.~\cite{ppr}, the $a_{d>4}$ are determined, not from data, but 
using a large-$N_c$-inspired, 3-pole, model approximation to $\Delta\rho$
(the so-called ``minimal hadronic ansatz'' or MHA). 
The $s^k$-weighted physical and MHA spectral 
integrals, for a given $s_0$, are in general very different. 
For all $-2\leq k\leq 4$, however, the point where the two agree 
happens to lie in the vicinity of the lower of the duality
points for the two WSR's.   
This observation is taken as evidence in support of the pattern of
long-/short-distance duality predicted by the MHA, and 
of the  reliability  of the model values of 
$a_6$, $a_8$ and $a_{10}$.{\begin{footnote}{Recall that
$a_{2k+2}=\left( -1\right)^{k}\int_0^{s_0} ds\, s^k\, \Delta\rho (s)$
if $s_0$ is a true duality point.}\end{footnote}}
The results quoted in Ref.~\cite{ppr} correspond to
\begin{eqnarray}
a_6&=&- \left( 9.5 \pm 2.0 \right)\times 10^{-3}\ {\rm GeV}^6\nonumber\\
a_8&=&\ \  \left( 16.0 \pm 4.2 \right)\times 10^{-3}\ {\rm GeV}^8\nonumber \\
a_{10}&=&- \left( 20.8 \pm 10.2 \right)\times 10^{-3}\ {\rm GeV}^{10}\ ,
\label{ppra6a8a10}
\end{eqnarray}
which are not in good agreement with our central fit values.  

One should bear in mind that the errors in Eqs.~(\ref{ppra6a8a10})
reflect only the uncertainties in the fitted values of the three
independent MHA parameters $F_0$, $m_V$, and $g_A$, and not any
possible theoretical systematic errors (due to working with the minimal
set of hadronic states, and in the large $N_c$ limit).   
The latter are not necessarily negligible.  
As a first indication of this, let us observe that MHA predicts  
the duality points for the various $s^k$-weighted FESR's 
(those $s_0$ for which the model and data integrals match) 
to be different{\begin{footnote}{This can 
be seen, for example,  by superimposing 
the plots for the $s^0$ and $s^2$
moments (the top panels of Figs. 1 and 2 of Ref.~\cite{ppr},
respectively). One finds that the band within which the 
$s^0$ matching point must lie (given the uncertainties in the
model parameters) does not overlap with the corresponding
allowed band for the $s^2$ moment. 
The situation is similar for the $s^3$ and $s^4$ moments
(the allowed matching bands in fact lies somewhat farther from the 
corresponding $s^0$ band than in the $s^2$ case).}\end{footnote}}.  
Because of the strong slope of the $s^k$ ($k=2,3,4$) spectral
integrals with respect to $s_0$, even small errors in 
the model predictions for these differences 
can correspond to large uncertainties on the $a_d$.   

\begin{figure}[htb]
\unitlength1cm
\caption{$J_w(s_0)$ and the corresponding MHA integrals
for the $w_1$ (top left panel), $(0,0)$ (top right panel),
$w_3$ (bottom left panel) and $(1,0)$ (bottom right panel) 
pFESR's. The $J_w(s_0)$ integrals and errors 
were obtained using the ALEPH data and covariance matrix.}
\begin{minipage}[t]{8.0cm}
\begin{picture}(7.9,7.9)
\epsfig{figure=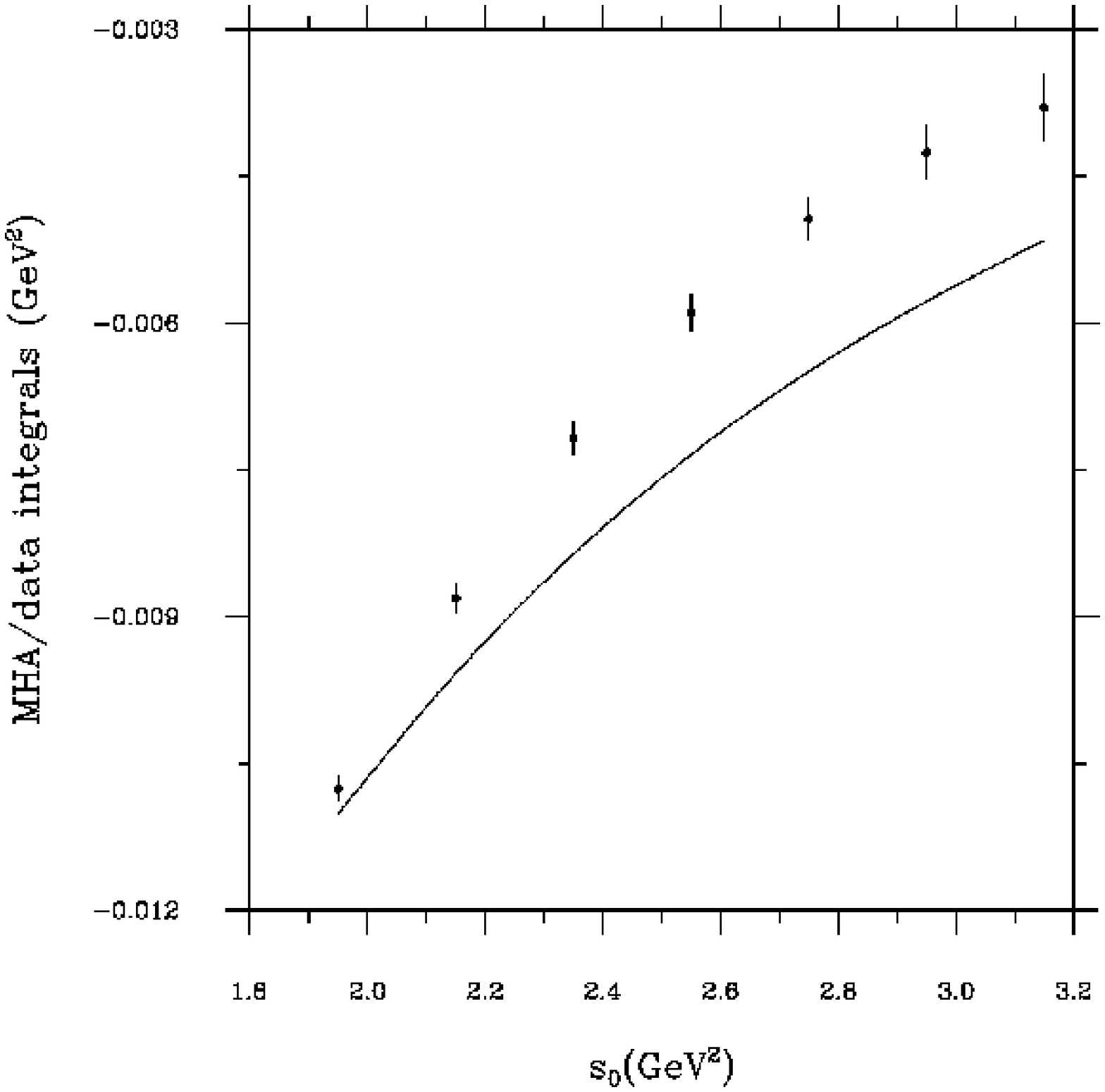,height=7.8cm,width=7.8cm}
\end{picture}
\end{minipage}
\hfill
\begin{minipage}[t]{8.0cm}
\begin{picture}(7.9,7.9)
\epsfig{figure=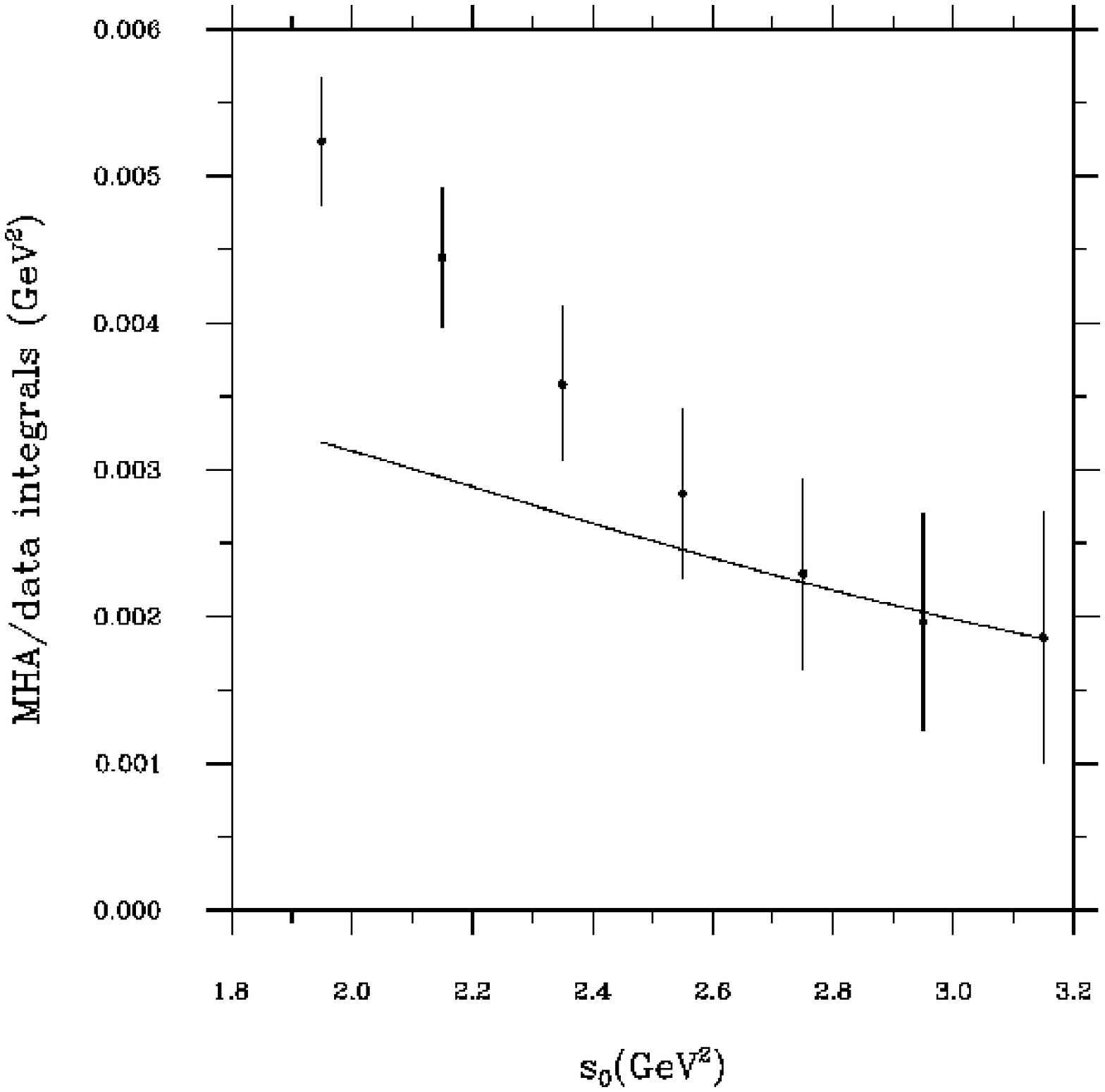,height=7.8cm,width=7.8cm}
\end{picture}
\end{minipage}
\vskip .15in\noindent
\begin{minipage}[t]{8.0cm}
\begin{picture}(7.9,7.9)
\epsfig{figure=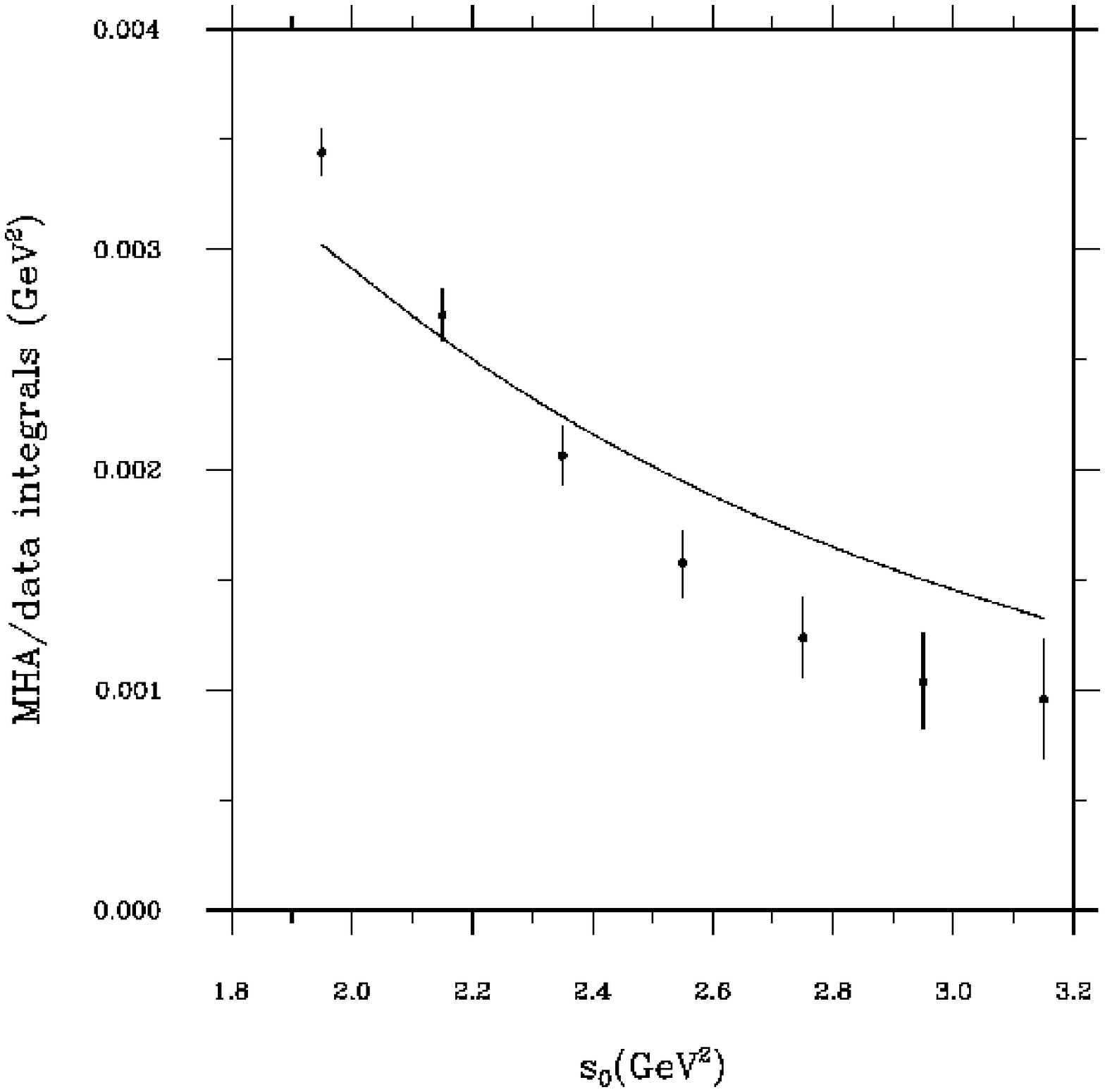,height=7.8cm,width=7.8cm}
\end{picture}
\end{minipage}
\hfill
\begin{minipage}[t]{8.0cm}
\begin{picture}(7.9,7.9)
\epsfig{figure=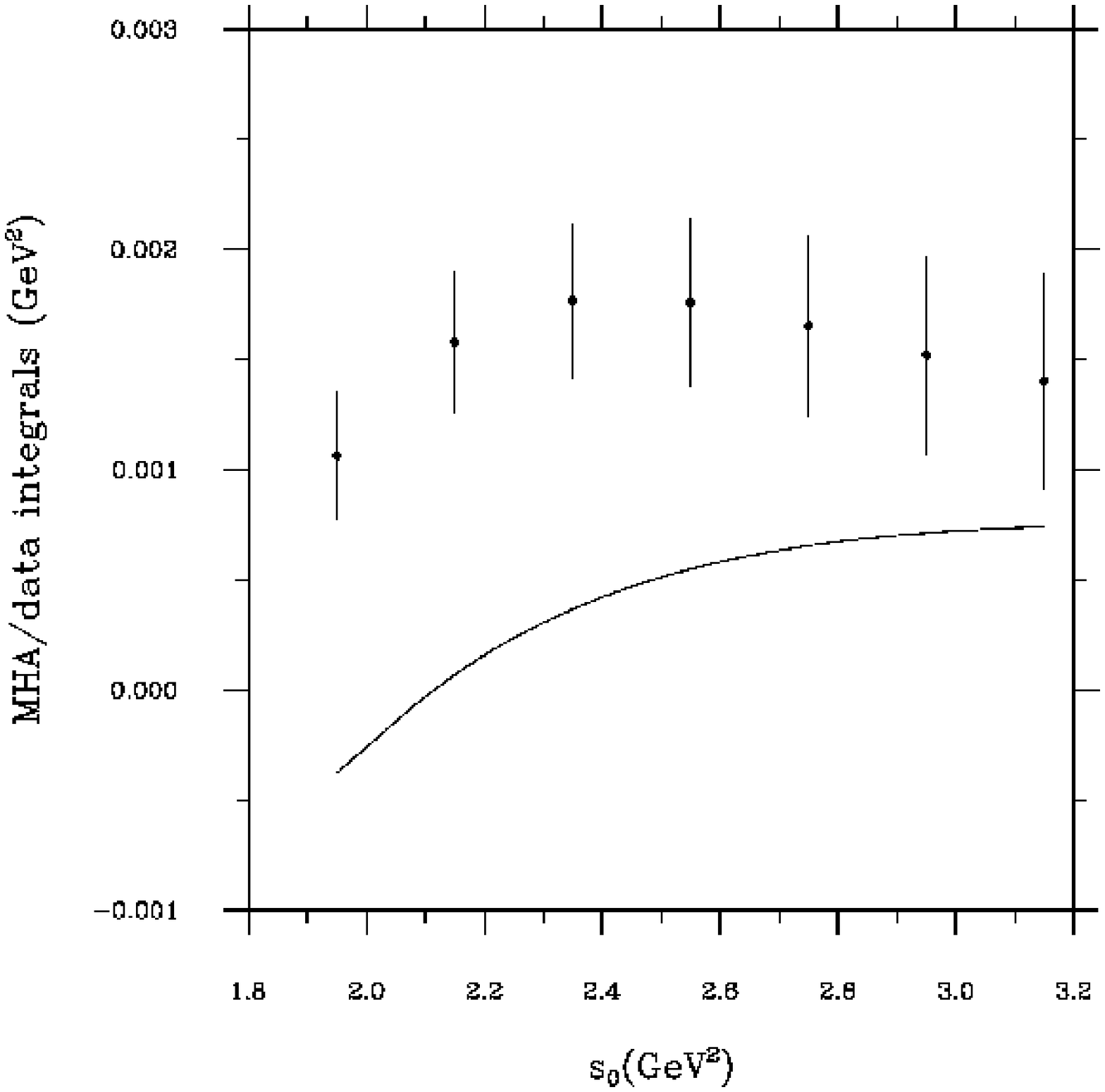,height=7.8cm,width=7.8cm}
\end{picture}
\end{minipage}
\label{fig7}\end{figure}

To further test the MHA predictions, and to get an idea of whether or  
not potential systematic uncertainties might account for the 
discepancy between the MHA predictions and our results, we may   
study those pFESR's sensitive only to $a_6$, $a_8$ and $a_{10}$.  
We display, in Fig.~\ref{fig7}, the MHA predictions for 
the $J_w(s_0)$ associated with the $w_1$, $(0,0)$, $w_3$ and $(1,0)$ pFESR's.
(The first and second sum rules are sensitive
to the MHA values of $a_6$, $a_8$, the third to $a_6$
and $a_{10}$, and the fourth to
$a_6$, $a_8$, and $a_{10}$.) We see that, although
%, given
%that it involves only three parameters, the
%model provides a not-unreasonable representation of the
%physical spectral integrals, 
the representation of the physical
spectral integrals is not-unreasonable for a three-parameter model,
the quality of this representation is not
good at the detailed level. This mismatch suggests to us the presence of  
residual theoretical systematics uncertainties in the MHA approach,  
which might be removed by going beyond the minimal ansatz and/or
incorporating $1/N_c$ corrections.   

\section{Duality Violation and V-A pFESR's}
\label{Sect:dualityviolation}
Our interpretation of the $a_d$ obtained
above as the true asymptotic OPE coefficients of
the $ud$ V-A correlator rests on the assumption that
residual duality violation (parameterized by 
$R[s_0,w]$) is small for the doubly-pinched pFESR's
and scales employed in our analysis. As noted above, the high quality of the
match between the optimized OPE representation and the corresponding
spectral integrals is a necessary, but not sufficient, 
condition for the validity of this assumption. In this section we discuss 
additional evidence in its favor. We begin by reviewing
certain relevant aspects of what is known about 
the nature of duality violation in QCD.  
 
\subsection{General Expectations for Duality Violation in QCD}
A useful review of the current status of our understanding of duality
violation in QCD is given in Ref.~\cite{Shifman:2000jv}. It is
important to bear in mind that duality violation may be small in
weighted spectral integrals even when the level of duality violation
in the spectral function itself is large over significant portions, or
even all, of the integration range{\begin{footnote}{Examples are the
spectral integrals corresponding to (i) dispersion representations of
correlators for spacelike $Q^2 \gg \Lambda_{QCD}^2$, (ii) Borel
transformed dispersion relations involving Borel masses
$M>>\Lambda_{QCD}^2$ and (iii) the $(1-y)^k(1+Ay)$-weighted ($k=1,2$)
pFESR's for the flavor $ud$ V and A correlators at scales $s_0\sim 2$
to $3$ GeV$^2$.}
\end{footnote}}. 

Two distinct types of duality violation {\it in the
spectral function} are identified in Ref.~\cite{Shifman:2000jv}.
The first is that produced by contributions to the correlator which, 
asymptotically, are exponentially suppressed relative to OPE contributions for
spacelike $Q^2$. Such terms behave asymptotically as
 $\mbox{exp} (-bQ)/Q^\kappa$,
and hence acquire oscillating imaginary
parts for timelike $Q^2=-s$. The second type of duality violation
occurs in $N_c\rightarrow\infty$ QCD, where the spectrum
consists of a tower of infinitely narrow resonances. As a result
of this spectral structure, the associated correlator 
has a different convergent Laurent expansion 
in each of the annuli lying between successive poles in the complex $s$-plane. 
In none of these annuli is the Laurent expansion equal to the
asymptotic expansion; hence duality violation exists, in this
case, in all such annuli. 

An important difference in the nature
of duality violation in these two cases lies in the structure of the
duality violating contributions to the correlator in the complex
plane. In the $N_c\rightarrow\infty$ scenario, one has a series
of different ``sub-asymptotic'' expansions, each valid in a 
different annulus. When one crosses from one annulus to the next,
all Laurent coefficients are altered, and for
no annulus are they equal to the corresponding asymptotic
OPE coefficients. Duality violation in a given annulus is
thus equally large at all points on the circle $\vert s\vert =s_0$
lying within the given annulus, and is NOT localized
to the vicinity of the timelike point on that circle for any
$s_0$, no matter how large. In contrast, for $Q^2=s_0 e^{i\phi}$
(with $\phi = -\pi \, (+\pi)$ corresponding to the top (bottom)
of the physical cut), a term of the form $\sim \mbox{exp}(-bQ)/Q^\kappa$
behaves as
\begin{equation}
{\frac{1}{s_0^\kappa}}\, \mbox{exp}\left[ -b\, s_0\, \cos(\phi /2)\right]
\, \mbox{exp}\left[ i\left( \kappa \phi /2 + \sin(\phi /2)\right)\right]
\end{equation}
and hence retains an exponential suppression, via the 
factor $\mbox{exp}[-s_0b\, \cos(\phi /2)]$, for all but the timelike
point on $\vert s\vert =s_0$. This suppression will remain
quite significant over most of the circle for scales $s_0$ 
larger than  $1/b$. 
This may be the case even if the oscillating, duality-violating
component of the spectral function is far from negligible at the same
$s_0$. Duality violation via such terms is thus PQW-like:
``intermediate'' scales exist for which duality violation in the
correlator ($\Pi_{\rm OPE}(s) - \Pi (s)$) is strongly localized to the
vicinity of the timelike real axis.

\subsection{Detecting the Presence and Nature of Duality Violation}

The two distinct patterns of duality violation for a given correlator
in the complex plane manifest themselves in readily distinguishable
different ways in sum rule analyses. In particular, these patterns
suggest different strategies and tests to explore the impact of
duality violation in a given analysis. In this section we identify
such strategies and enumerate a number of possible tests. In the next
section, we then specialize to the flavor $ud$ V-A correlator, and
we discuss the practical implementation of these tests.

In presence of a PQW-like component of duality violation, there will
exist intermediate scales $s_0$ where
\begin{itemize}
\item 
power-weighted ($w(s)=s^k$) FESR's, which fail 
to suppress contributions from the integral over $\vert s\vert =s_0$
near the timelike axis, are poorly satisfied;  
\item 
pFESR's involving weights which suppress contributions
from the vicinity of the timelike point are well satisfied.
\end{itemize} 
This observation motivates the use of ''pinched weights'' in FESR
analyses to tame PQW-like duality violation. Having adopted a set of
weights, one wants to verify that residual duality violating
contributions from the region near $s=s_0$ are not present in the
results of the analysis. In this respect, an important test is to 
\begin{itemize}
\item[1.] 
verify that the (nominally asymptotic) 
$a_d$ coefficients extracted in the pFESR analysis provide
accurate representations, not only of the spectral integrals
used in fitting the $a_d$, but also of spectral integrals
corresponding to weights with zeros of a higher order at $s=s_0$ 
(which therefore further suppress PQW-like duality violating effects). 
\end{itemize}

In the $N_c\rightarrow\infty$-like scenario, where duality
violation is not localized to the vicinity of the timelike
real axis, a rather different pattern of sum rule behavior will
be observed. Since OPE-like Laurent expansions exist in any
given annulus, so long as one restricts oneself to $s_0$
lying in a single annulus, one will obtain 
a set of coefficients, $a_d$, which provide a perfect
match between the ``OPE'' and data sides
of both power-weighted and pinch-weighted FESR's
at those scales. That set will, of course, consist of
just the coefficients of those terms in the Laurent expansion
for the given annulus which survive when integrated against
the weights employed. If one performs the same 
basic analysis (i.e., using the same set of weights),
but now for $s_0$ lying entirely in a different annulus, one
will obtain a different set of $a_d$. These $a_d$ will 
provide a perfect match between the ``OPE'' and data sides
of the sum rules employed in the new annulus. 
Schematically, this type of duality violating contribution 
implies:
\begin{itemize}
\item the existence of several sub-asymptotic regimes;
\item that pinching is not effective in removing this type of 
duality violating effect (both pinch-weighted and power-weighted FESR's
are equally well satisfied in each sub-asymptotic region). 
\end{itemize}
The existence of several such sub-asymptotic regimes 
in this type of scenario can also, in principle, be exposed in
a pFESR analysis, as follows.
Starting with some particular small range of 
$s_0$ one may gradually decrease the lower edge of the $s_0$ analysis window, 
keeping the upper edge fixed. So long as the lower edge 
lies in the same annulus as the upper edge, 
a pFESR analysis extraction of OPE-like coefficients, obtained by means of 
matching to spectral integral data, will produce an exact 
determination of the relevant coefficients in the singular part of
the Laurent expansion for that annulus. As soon as the lower edge 
of the $s_0$ analysis window leaves the single annulus, however, there 
is no longer a single set of expansion coefficients valid for all the $s_0$ 
being employed. 
The existence of such a two-annulus regime will be evident by the
sudden appearance of a poor match between the optimized OPE-like
integral and spectral integral sets. One can check these expectations
explicitly within the equal-spacing pole model of
Refs.~\cite{gppr,Shifman:2000jv}. 
This exercise shows that, in performing a pFESR analysis, it is
crucial to 
\begin{itemize}
\item[2.] demonstrate 
that there is no drift in the values of the extracted coefficients
as one decreases the lower edge of the $s_0$ analysis window and 
\item[3.] demonstrate 
that the optimized values of the fit parameters in fact produce
an accurate match to the spectral integral data used to produce the fit 
{\it over the whole of the $s_0$ analysis window employed}. 
\end{itemize}
These tests serve not only to verify the reliability of the
assumed OPE-like expansion form, but also, at least potentially, to
expose the existence of multiple sub-asymptotic expansion regimes
{\begin{footnote}{One may also test whether 
an observed deterioration in the quality
of the optimized ``OPE''/spectral integral match as $s_0$ is
lowered is, or is not, due to the existence of a new subasymptotic
regime. If it is, then the first $s_0$ for which the deterioration
appears must lie in the lower annulus. Working with $s_0$ lying
in a narrow range just below this point should then produce a new
set of fitted $a_d$ which provide a good quality representation of
the corresponding spectral integrals, when restricted to this new 
range of $s_0$. If a good quality match is not found,
then the deterioration is due to breakdown
of the OPE-like expansion form, and not to the fact that one
has entered a new subasymptotic region.}\end{footnote}}.

In the context of the $N_c\rightarrow\infty$ discussion, however, 
it is clear that,
while passing these tests is a {\it necessary} condition for the
reliability of the extraction of the OPE coefficients from data, it is
not a {\it sufficient} one: if one happened to be unlucky and perform the
pFESR analysis only for those $s_0$ lying in a single, but
sub-asymptotic, annulus, one would see a high quality match (exact in
the case of the pole model) between the spectral integrals and
optimized OPE-like integrals even though one
would have actually  extracted the coefficients relevant to the Laurent
expansion in the sub-asymptotic annulus, and not those relevant to the
asymptotic regime. A simple way to test whether or not this is the
case is to
\begin{itemize}
\item[4.] take the coefficients extracted in the pFESR analysis
and employ them as input to a dispersive analysis {\it relevant to the
asymptotic regime}. 
\end{itemize} 
If the coefficients extracted in the pFESR analysis are not those
relevant to the asymptotic regime, the resulting dispersive integrals
will be poorly approximated by the OPE-like representation generated
using the fitted coefficients. 
Again, explicit illustrations of this point can be worked out  
within the equal-spacing pole model of Refs.~\cite{gppr,Shifman:2000jv}. 
In the case of the model, performing the dispersive test is 
straightforward because the spectral function of the model is 
actually known for all $s$. The situation of interest to us, however,
is one where spectral data are available for only a limited range of
$s$. In such a situation, it would typically be difficult to construct
a dispersive test for which the errors on the dispersive integrals
were under sufficient control to make the test useful. One general
solution is to work with BSR's and restrict one's attention to Borel
masses which are both low enough that the spectral weight,
$\mbox{exp}(-s/M^2)$, is negligible in the region where spectral data
are absent and, simultaneously, high enough that the convergence with
$d$ of the Borel transformed OPE series {\it for the set of $a_d$ one
wishes to test} is acceptable. For a given BSR, such $M$ may or may
not exist. Three of the four IZ BSR's turn out to provide examples of
such tests for our solution set (details are reported in the
Appendix). Additional asymptotic tests, involving BSR's at larger
$M$, are possible for the flavor $ud$ V-A correlator. In this case,
the spectral integral uncertainties are brought under control using
the classical chiral sum rule constraints associated with the Weinberg
sum rules and the sum rule for the $\pi$ electromagnetic mass
splitting. An efficient procedure for implementing these constraints
is provided by the ``residual weight method'', which is
described in detail in Ref.~\cite{cdgm}.

\subsection{The Nature of Duality Violation in the
$ud$ V-A Channel}

In the following we argue that duality violation in the $ud$ V-A 
correlator is predominantly PQW-like. From previous work~\cite{kmtau02}, 
one has empirical evidence that duality violation in the 
individual $ud$ V and A channels is predominantly 
PQW-like. Checking for the presence or absence of 
duality violations in the $ud$ V and A correlators at intermediate
scales is straightforward because one has independent (asymptotic)
information on the value of the OPE parameter $\alpha_s$.   
Such a straightforward check is not possible for the $ud$ V-A difference. 
A qualitative argument is however available, based on the the
observation that duality violation in the flavor $ud$ V+A sum cancels,
within experimental errors, for scales above $s\sim 2\ {\rm GeV}^2$.
This can be seen from (i) the fact that the corresponding spectral
function is in agreement with the OPE prediction for such $s$ (see,
e.g., Fig. 6 of the second of Refs.~\cite{ALEPH}) and (ii) the
observation that the spectral integrals for the $s^k$-weighted FESR's,
$k=0,\cdots ,3$, are in good agreement with the corresponding OPE
integrals for $s_0$ above $\sim 1.9\ {\rm GeV}^2$~\cite{kmtau02}. This
implies that the duality violating contribution to the $ud$ V-A
correlator is, within experimental errors, twice that to 
the $ud$ V correlator.
The latter is known to be strongly localized to the vicinity of the
timelike real axis for the scales of interest to us, and hence so is
the former. 
This conclusion is compatible with the observation that,
although the $ud$ V-A $s^k$-weighted FESR data integrals are not
constant with respect to $s_0$ (i.e., not in agreement with the
behavior of the $s^k$-weighted ``OPE'' integrals), the agreement
between the data and ``OPE'' sides of our pFESR's is very good for the
optimized OPE-like fits given above. 

\begin{center}
{\bf Tests of the type 1.}
\end{center}

\begin{figure} [htb]
\unitlength1cm
\caption{
$J_{w^{(k,m)}}(s_0)$  and  $f_{w^{(k,m)}}(\{ a_d\} ;s_0)$ 
for the $(3,1)$ (top panel) and $(4,0)$ (bottom panel) 
spectral weight pFESR's. 
The 
$J_{w^{(k,m)}}(s_0)$ 
integrals and errors were obtained using the ALEPH data
and covariance matrix. 
The solid line shows the predictions based on the $d>4$ OPE 
solution represented by 
our combined fit, Eqs.~(\ref{finalresults}).}
\center{\begin{minipage}[t]{10.0cm}
\begin{picture}(9.9,9.9)
\epsfig{figure=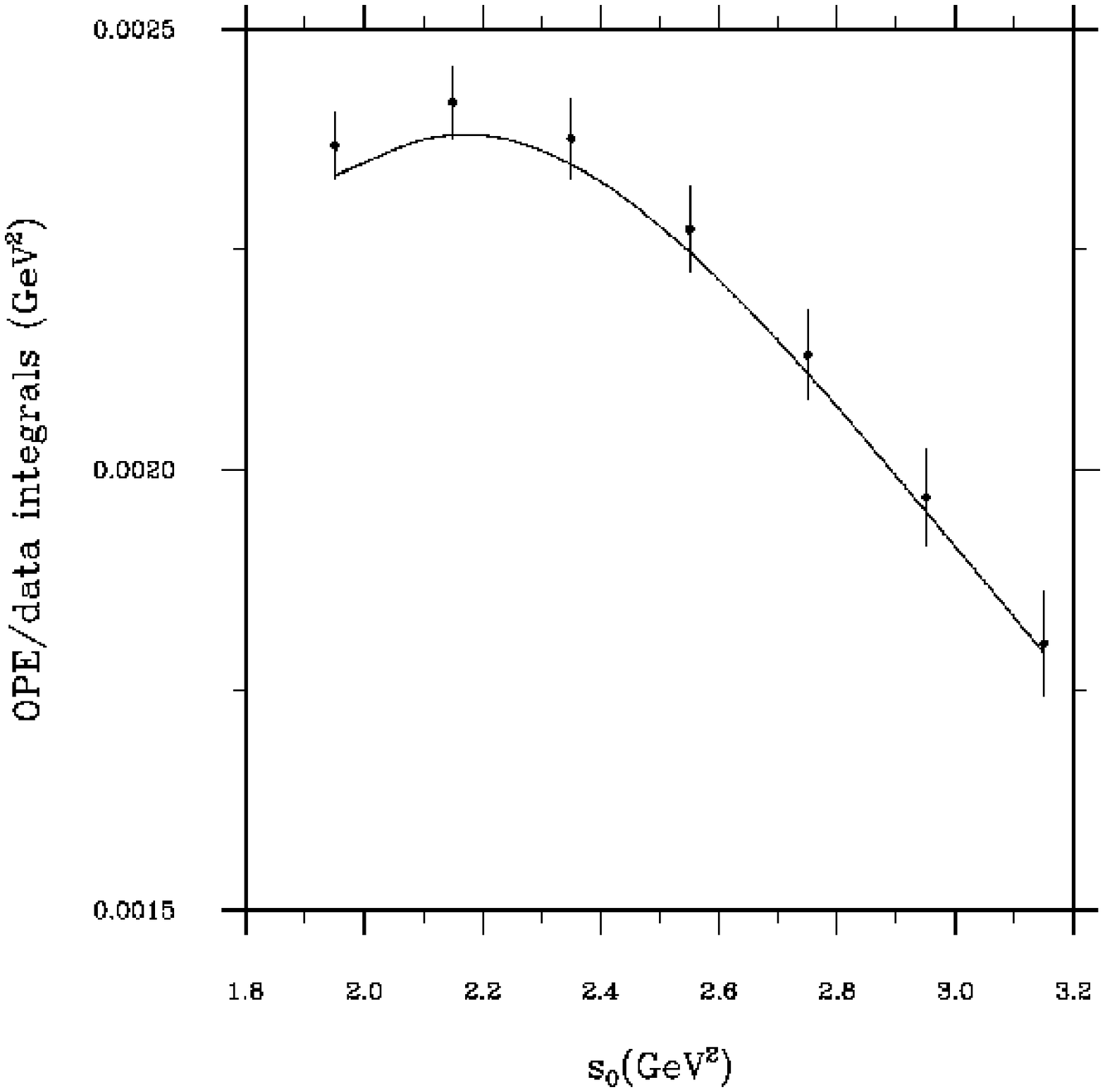,height=8.5cm,width=9.8cm}
\end{picture}
\end{minipage}
\vskip .2in
\begin{minipage}[t]{10.0cm}
\begin{picture}(9.9,9.9)
\epsfig{figure=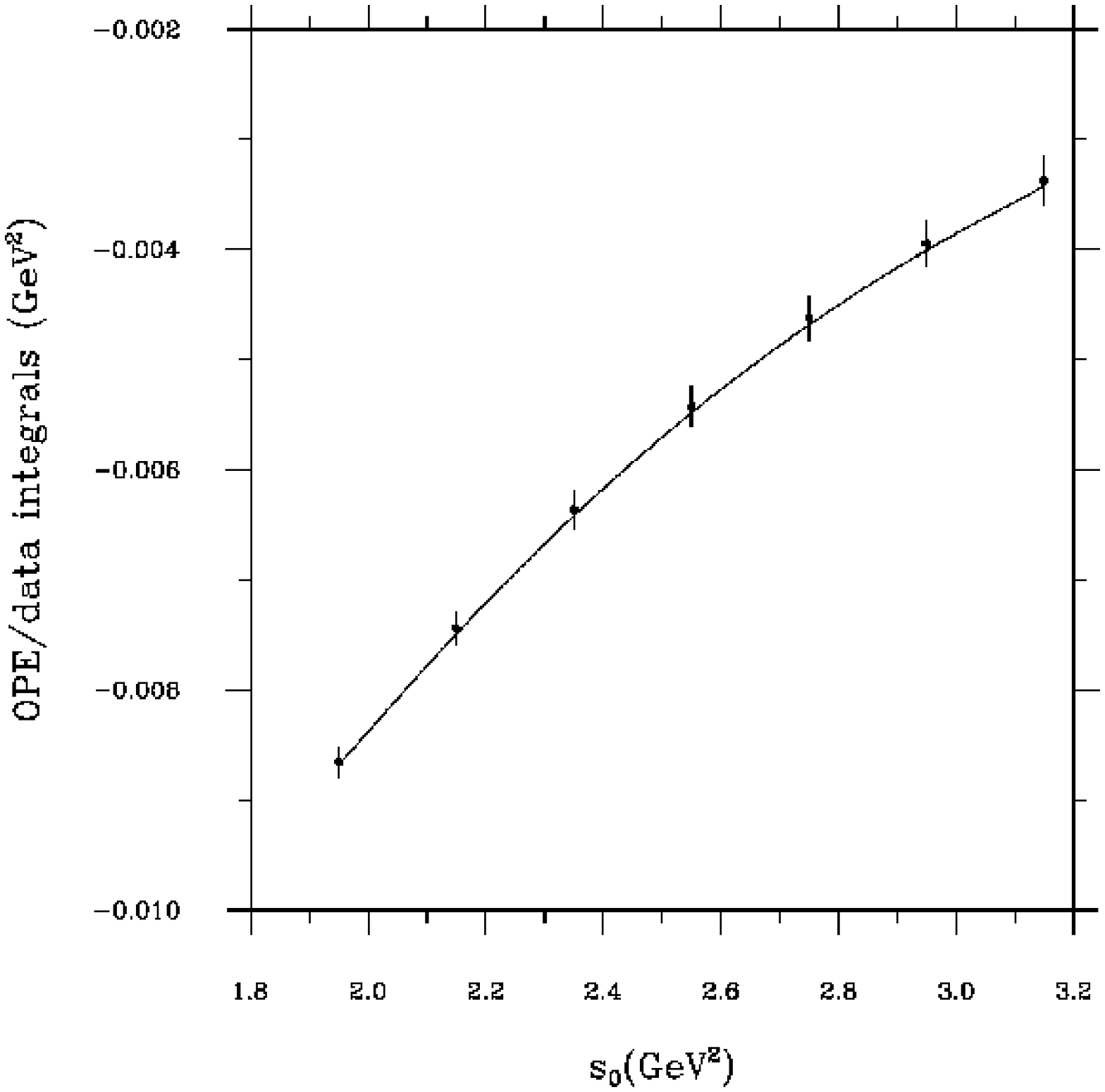,height=8.5cm,width=9.8cm}
\end{picture}
\end{minipage}}
\label{fig8}
\end{figure}

Having a suppression of duality violating contributions
which is strong enough to make such contributions negligible relative
to the $d=0$ terms in the $ud$ V and A correlators does not
necessarily mean that the same suppression is sufficient to make such
contributions small relative to the $d=6$ and higher OPE contributions
in the $ud$ V-A difference. In order to check for residual duality 
violating contributions localized to the vicinity of the timelike real
axis, we have performed tests of the type defined in the previous 
section, item 1. 
The $(1,m)$ spectral
weights discussed above (with $m=0,\cdots ,3$) have zeros of order $3$. 
As we have already seen in Figs.~\ref{fig5}
and \ref{fig6}, the results of our combined fit produce an extremely
good ``OPE''/spectral integral match for all of these weights, 
with no quality deterioration. 
We have also investigated the $(2,0)$, $(2,1)$, $(3,0)$, $(3,1)$ and
$(4,0)$ spectral weight pFESR's, which have weights with zeros of
order $4,\ 4,\ 5,\ 5$ and $6$, respectively, at $s=s_0$. Again the
quality of the match to the spectral integral sides of these sum rules
provided by our combined fit is excellent in all cases, despite the
much stronger suppression of contributions from the region on $\vert
s\vert =s_0$ near $s=s_0$. We illustrate the quality of this match
for the most extreme cases (the $(3,1)$ and $(4,0)$ pFESR's) in
Fig.~\ref{fig8}.
%%%%%%%%%%%%%%

\begin{center}
{\bf Tests of the type 2.,3.,4.}
\end{center}

The arguments given above do not completely rule out the presence of
residual duality violation of non-PQW-like nature (the 
$N_c\rightarrow\infty$-like scenario). In order to deal with this,  
we have subjected our solution set to tests of the type described 
in items 2.,3., and 4. of the previous section. 
As for test 2., we find that within the present experimental errors 
there is no drift in the extracted OPE parameters as one lowers the 
lower edge of the $s_0$ analysis window (see below for 
details and prospect of sharpening this test with improved data). 
Also tests of the type 3. are succesfully passed by our solution set 
(see Section~\ref{ope-spec-match}). 

Finally, to deal with the possibility  that our entire 
$s_0$ analysis window lies within a single sub-asymptotic region, 
we have performed a number of asymptotic dispersive
tests of the type described in the previous subsection, item 4.  
A first set of asymptotic dispersive tests is provided by the four IZ
BSR's. These are highly non-trivial since, because of the difference
in the sign of $a_8$ between our combined fit and the IZ solutions,
those IZ BSR's for which $d=6$ contributions are absent would appear
to be problematic for our combined fit. It turns out that this is not
the case; in fact, the convergence of the Borel transformed OPE series
is quite slow at the low $M$ employed by IZ and, once one extends the
sum involving our combined fit to sufficiently high $d$ to obtain
convergence, the OPE predictions are in excellent agreement with the
spectral data. Since these tests are also relevant to the comparison
to previous work, we provide a detailed demonstration of these claims
in the Appendix.

To obtain BSR's at larger Borel mass, $M$, one needs to use
``residual weight method'' improvement 
on the spectral integrals~\cite{cdgm}. 
In order to keep the errors under control, 
it is necessary to work with the product of $\Delta \Pi$ with 
appropriately chosen polynomials. We find that the combined fit 
OPE predictions are in excellent agreement with the spectral integral
sides of these BSR's for $M$ over a range sufficiently wide that,
at the upper end, the OPE integrals are completely dominated by
their $d=6$ contribution while, at the lower end, the full set
of $a_d$ obtained in the combined fit ($d=6,\cdots ,16$)
must be included before convergence of the Borel transformed OPE
sum is obtained. 

\begin{center}
{\bf Model explorations}
\end{center}

In principle, explicit models of the V-A spectral function could be
used to try and address the level of duality violation 
present in our analysis. 
One should bear in mind, however, that the only information we have about
$\Delta\rho$ in the region above $s=m_\tau^2$ is in
the form of the constraints provided by the classical chiral sum rules. 
These constraints
% these sum rules turn out to be
are far from sufficient to fully constrain the behavior of
$\Delta\rho$ above $s=m_\tau^2$ and, as a result, there exists a wide
range of model extensions of the data for $\Delta\rho$ to
$s>m_\tau^2$, all of which are compatible with these constraints. It
is easy to construct, among these, models for $\Delta\rho (s)$ for
which the asymptotic expansion coefficients are the same as those of
our combined fit. The models which have this property display
continued damped oscillations in $\Delta\rho$ as one goes to higher
$s$, and hence appear quite natural. 
It is also possible to construct models for which the asymptotic OPE
parameters differ significantly from those of our combined 
fit~\cite{jfd}. 

Because the integrated pFESR OPE contributions of dimension $d$ scale as
$1/s_0^{(d-2)/2}$, one finds that, for large $s_0$, the higher
$d$ $a_d$ contributions drop rapidly in size with increasing $d$.
With such small high $d$ contributions, a small change in the
modelling of $\Delta\rho$ in the region where it is not known experimentally
typically produces a large change in $a_d$. A very large theoretical 
systematic uncertainty for the higher dimension $a_d$ will thus be
associated with any attempts to model $\Delta\rho$ in the region above
$s=m_\tau^2$. Without being able to control
this theoretical systematic error, obtaining 
meaningful information on the level of duality
violation from such model studies is somewhat problematic.

\begin{center}
{\bf Prospects of improving the data-based tests}
\end{center}

It is worth stressing that significant improvements in the analysis
will become possible once the new hadronic $\tau$ decay data from the B-factory
experiments is available. At present both the errors on the
$a_d$ and the accuracy with which it is possible to determine
the location of the onset of duality violation in the analysis
are limited by the errors on $\Delta\rho (s)$ above $s\sim 2\ {\rm GeV}^2$.
These errors are dominated by experimental uncertainties on the 
$4\pi$, $\bar{K}K\pi$ and $\bar{K}K\pi\pi$ spectral distributions 
and uncertainties in the V/A separation for $\bar{K}K\pi$ and 
$\bar{K}K\pi\pi$ states. Major improvements should be 
forthcoming as a result of the expected $\sim 10^2$-fold 
increase in the size of the $\tau$ decay data base.
The improved spectral integral errors which result
will allow us to improve significantly on the efficiency 
of our tests for the absence of residual duality violation.
The current situation in this regard is discussed in brief below.

Recall that, by decreasing the lower edge of the analysis window,
we were able to demonstrate the presence of duality violation 
for the pFESR's used in our analysis at scales below $\sim 1.8\ {\rm GeV}^2$.
With current experimental errors there is no evidence for
duality violation in our analysis window. Ideally one would like to 
work at scales well above $1.8\ {\rm GeV}^2$, in order to suppress, 
as much as possible, any residual duality violating contributions 
which might be present, but masked by current experimental errors.
While current errors are small enough that $a_6$ may still
be determined, even if one
works with only a small portion of our present analysis 
window{\begin{footnote}{For example, using only $s_0=2.75, 2.95$
and $3.15\ {\rm GeV}^2$, one finds, from the maximally-safe
analysis of the ALEPH data, $a_6=-0.0049\pm 0.0022$, where the error quoted is
that associated with the ALEPH covariance matrix. The error is, of course,
significantly larger than that obtained from the larger analysis
window, but still less than $50\%$ of the signal.}\end{footnote}},
this is not true for the $a_d$ with $d\geq 8$. In fact, with current
experimental errors, the uncertainties on the extracted $d\geq 8$ $a_d$ 
do not become smaller than $\vert a_d\vert$, until the lower edge of 
the analysis window has been reduced to below $s_0\sim 2.5\ {\rm GeV}^2$.
If, as an example, we perform our analysis of the ALEPH data using the
$s_0=2.35 \rightarrow 
%, 2.55, 2.75, 2.95$ and $
3.15\ {\rm GeV}^2$ sub-window,
then, with central values for all non-spectral input,
the maximally-safe output for $a_6$ and $a_8$ is:
\bea
a_6 &=& - ( 3.88 \pm 1.21 ) \times 10^{-3} \, {\rm GeV}^6  \nonumber \\
a_8 &=& - ( 9.32 \pm 6.58 ) \times 10^{-3}\,  {\rm GeV}^8 \ . 
\eea
Within the quoted errors, these results
are compatible with those of the full-window 
analysis. The situation is similar for
the results of the combined analysis: the results of the sub-window 
analysis for $a_{10}$ through $a_{16}$ are:
\bea
a_{10} &=& (6.62 \pm 2.83) \times 10^{-2} \,{\rm GeV}^{10}
\nonumber \\
a_{12} &=& - (2.16\pm 0.86) \times 10^{-1}\, {\rm GeV}^{12} 
\nonumber \\
a_{14} &=&  (5.88\pm 2.48) \times 10^{-1}\,  {\rm GeV}^{14}
\nonumber \\
a_{16} &=& - (1.47\pm 0.69) \,  {\rm GeV}^{16} \ , 
\eea
again compatible with the full-window analysis within the sub-window
analysis errors. Were the errors to be $1/3$ as large, however, the
full-window and sub-window results would no longer be compatible and
we would be forced to conclude that residual duality violation was
present for those $s_0$ in the lower part of the full analysis
window. We stress that there is no reason for reaching such a
conclusion at present. In fact, there are strong reasons for trusting
 the results of the full-window analysis:
\begin{itemize}
\item where the existence of duality violation can be
explicitly demonstrated, the OPE-like expansion is known {\it not} to
provide a good representation of the spectral integrals;
\item in the lower part of our full analysis window the OPE-like
form provides an excellent representation of the
spectral integrals;
\item the combined fit from the full analysis window provides
an excellent representation of the spectral integrals
not only in the lower part, but also the upper part, of the analysis window;
\item the combined fit results obtained from {\it the sub-window version
of the analysis} turn out to provide a poor representation
of the spectral integrals in the lower part of the full window.
\end{itemize}
Nonetheless, the size of the sub-window errors are such
that much stronger tests of the absence of 
residual duality violation, using various sub-windows, 
will become possible once the
errors on $\Delta\rho (s)$ above $2\ {\rm GeV}^2$ are 
reduced{\begin{footnote}{If such 
reduced errors were to expose residual duality violation in
the lower part of the current full-window analysis, one would of
course be forced to raise the lower edge of the analysis window.}
\end{footnote}}. 
While it seems unlikely to us, for the reasons given above,
it is not at present possible to conclusively rule out
additional uncertainties, associated with residual duality
violating, at the level of the difference of the full-window
and sub-window analysis central $a_d$ values.

\section{Conclusions}
\label{Sect:concl}

In this paper we have used Finite Energy Sum Rules with
``pinched-weights'' (pFESR's) to determine the OPE
coefficients $a_6,\cdots ,a_{16}$ of the flavor $ud$ V-A correlator
with good accuracy using existing hadronic $\tau$ decay data. 
While it is not possible at present to either prove or disprove 
on {\it rigorous} analytic grounds that this approach (or any other) yields 
a valid approximation to the actual dynamics of QCD, 
we have carefully demonstrated the advantages of pFESR's among the 
class of sum rule techniques and have described a large number of 
checks on our own work and that of others.

At a technical level, we have employed a set of ten polynomial
weights carefully chosen to minimize the impact of
experimental errors and of duality violating effects, as well as to
optimally separate the contributions from condensate
combinations of different
dimension. Our analysis shows that the OPE contributions with $d>8$
are typically not negligible at scales $\sim 2-3\ {\rm GeV}^2$.

We have performed a number of tests to explore the presence 
of duality violating effects in our analysis. 
These support the conclusion that our combined fit values 
are not affected by duality violation within the existing experimentally 
induced errors. 
We recall the main observations in support of this statement:
\begin{itemize}
\item[(i)] independent
determinations of the $a_d$ using pFESR's based
on different (independent) weights
are in excellent agreement; 
\item[(ii)] the results of the combined fit
for the $a_d$ lead to an extremely good match between
the OPE and spectral integral sides of all the pFESR's employed
in the fitting procedure; 
\item[(iii)] the combined fit values also 
lead to extremely good matches for the $(k,m)$ spectral weight pFESR's, 
where $d>8$ contributions are much larger relative to $d=6,8$ contributions
than is the case for the $w_3$ through $w_{10}$ pFESR's; 
\item[(iv)] there is 
no deterioration in the quality of the combined fit prediction for the pFESR
spectral integrals even for those spectral weights
with zeros at $s=s_0$ of much higher order than those
used in obtaining the combined fit ;
\item[(v)]  the dispersive tests, described above, 
and in the Appendix, are successfully passed by our solution set. 
This  provides additional support for the reliability of
the extracted values, and our interpretation of them
as asymptotic OPE coefficients of the V-A correlator. 
\end{itemize} 
Improved experimental data would allow
one to significantly sharpen some of the tests reported above.

Some general observations also follow from the results and discussion
above. First, the OPE representation of $\Delta \Pi$, with
the $a_d$ given by the combined fit values of either Eq.~(\ref{finalresults})
or Eqs.~(\ref{finalresultsOPAL}),
provides a very accurate representation of the corresponding
spectral integrals down to scales as low as $s_0=2\ {\rm GeV}^2$,
at least for pFESR's based on weights with a double zero at $s=s_0$.
This suggests that the OPE remains reliable at intermediate
scales, $Q^2\sim 2\to 3\ {\rm GeV}^2$, apart perhaps from a
region near the timelike real axis. In contrast, if one considers
weights which do not suppress contributions from this region,
one sees clear evidence for the breakdown of the OPE.
The situation is similar to that
for the flavor $ud$ V and A correlators. 
The double zeros of the pFESR weights at $s=s_0$ in the V-A
case evidently again provide sufficiently strong suppression
in the vicinity of the timelike real axis to efficiently remove
contributions from the region of OPE breakdown on the circle
$\vert s\vert =s_0$.

A second point concerns the relative sizes of the various $a_d$.
The results of the combined 
fit indicate that $a_{d+2}/a_d$ is typically of order
$2 \to 3\ {\rm GeV}^2$ for the V-A correlator. This means
that, at intermediate ($2\to 3\ {\rm GeV}^2$) scales, there is
no `natural' ordering of contributions with different $d$,
in the sense that pFESR weights with comparable coefficients
for the $y^N$ and $y^{M}$ terms in $w(y)$ will produce
comparable $d=2N+2$ and $d=2M+2$ contributions
to the pFESR OPE integrals. This makes explicit the danger
of neglecting terms with $d>8$ for pFESR's based on
$w(y)$ with degree greater than $3$. This observation
also raises the possibility that the analogous neglect of
higher $d$ terms in other pFESR analyses, such as those
used to extract $m_s$ from the flavor-breaking difference
of $ud$ and $us$ V+A correlators~\cite{ckp98,mstaurefs}, may suffer from
similar problems.

\vspace{1.0cm}

\acknowledgments
The work of E.G. was supported in part by the National
Science Foundation under Grant PHY-9801875.
The work of V.C. was supported in part by MCYT, Spain (Grant No.
FPA-2001-3031), by ERDF funds from the European Commission, 
and by the EU RTN Network EURIDICE, Grant No. HPRN-CT2002-00311. 
K.M. would like to thank A. H\"ocker and S. Chen
for providing detailed information on the ALEPH
data, S. Chen for pointing out the need for
the normalization correction to the 1998
nonstrange data necessitated by the results of the 1999
strange data analysis, and to acknowledge
the ongoing support of the Natural Sciences and
Engineering Research Council of Canada, and the hospitality of the
Special Research Centre for the Subatomic Structure of Matter at the
University of Adelaide and the Theory Group at TRIUMF.
We are happy to acknowledge useful input from M. Eidemuller, 
S. Menke, S. Peris, A. Pich, J. Prades and M. Roney and 
especially from J. Donoghue for his many stimulating and 
instructive discussions. 

\vspace{1.0cm}
%\vfill\eject
{
\begin{appendix}
\section{The IZ Low-Scale Asymptotic Dispersive Tests}

In this Appendix we provide details of the four IZ BSR's and
complete the comparison of our results to those of early work
by subjecting our combined fit to the asymptotic dispersive
tests provided by these sum rules.

Incorporating the small $d=6$ logarithmic contribution, the BSR's
employed by IZ may be cast into the
form~\footnote{In writing Eqs.~(\ref{realbsr}), the factor
$F_\pi^2$ from the RHS of Eq. (21) of IZ has been moved to the LHS
of Eq.~(\ref{realbsr}), and absorbed into the spectral function via the shift
$\Delta\rho^{(1)}\rightarrow \Delta\rho^{(0+1)}$.
Eq.~(\ref{imaginarybsr}) is just $M^2$ times Eq. (22) of IZ, up
to the logarithmic correction term (proportional to $b_6$).}
\begin{eqnarray}
\int_0^\infty\, ds\,&& exp[s\, cos(\phi )/M^2]\, cos[s\, sin(\phi )/M^2]
\, \Delta\rho^{(0+1)} (s)\ =\ \sum_{k=1} (-1)^k
{\frac{cos(k\phi )\, a_{2k+2}}{k!M^{2k}}}\nonumber \\
&&\qquad -{\frac{b_6}{2M^4}}\left[ (\phi -\pi )sin(2\phi )
+\left(\ell n(M^2/\mu^2)-\gamma_E+3/2\right) cos(2\phi )\right]
\label{realbsr} \\
\int_0^\infty\, ds\,&& exp[s\, cos(\phi )/M^2]\, sin[s\, sin(\phi )/M^2]
\, \Delta\rho^{(0+1)} (s)\ =\ \sum_{k=1} (-1)^k
{\frac{sin(k\phi )\, a_{2k+2}}{k!M^{2k}}}\nonumber \\
&&\qquad -{\frac{b_6}{2M^4}}\left[ (\phi -\pi )cos(2\phi )
-\left(\ell n(M^2/\mu^2)-\gamma_E+3/2\right) sin(2\phi )\right]\ ,
\label{imaginarybsr}
\end{eqnarray}
where $\phi$ is the angle fixing the ray in the complex $Q^2$
plane along which the Borel transform was performed ($\phi =0$
corresponds to the top of the physical cut).
The four cases considered by IZ correspond to
\begin{enumerate}
\item  Eq.~(\ref{realbsr})
with $\phi =5\pi /6$ and  $M^2=0.8\ {\rm GeV}^2$,
\item  Eq.~(\ref{imaginarybsr}) with $\phi =2\pi /3$ and
$M^2=0.85\ {\rm GeV}^2$,
\item  Eq.~(\ref{realbsr}) with $\phi =3\pi /4$ and $M^2=0.6\ {\rm GeV}^2$,
\item  Eq.~(\ref{imaginarybsr}) with $\phi =3\pi/ 4$ and
$M^2=0.65\ {\rm GeV}^2$.
\end{enumerate}
The first two cases have no $d=8$ contribution, the third no $d=6$
contribution. We test the combined fit by employing the fitted 
$a_d$ values as input on the OPE side of the IZ BSR's. 
This leads to a prediction
for the value of the corresponding spectral integral (less the known
$d=4$ contribution) for each such sum rule.

Since the $a_6$ or $a_8$ values obtained by IZ
reflect the values of the spectral integrals, the change in sign of
$a_8$ between the IZ fit and our combined fit would seem to
represent a problem for the combined fit, especially in
the case of the third IZ sum rule. This is, however, not the case.
It is easy to check that, with the combined fit values as input, 
the convergence of the Borel transformed OPE series,
at the low values of $M^2$ employed by IZ, is rather slow. Not only
is the first of the $d>8$ contributions neglected by IZ in 
obtaining their central values, in all cases,
larger in magnitude than the corresponding sum of
$d=6$ and/or $d=8$ contributions, but also, 
in order to be certain that we have reached the region of convergence,
we have had to extend the extraction of the $a_d$ to higher $d$.
This is done using the extensions to higher $N$ of the families of
weights of which $w_6$ and $w_{10}$ are members (recall that
these weights produce $d>4$ OPE contributions
proportional to either $a_6$ and $a_{2N+4}$ or
$a_8$ and $a_{2N+4}$). We are able to extract
terms with $d$ up to $24$, albeit with larger errors
than for the $a_d$ given by the combined fit. The
$a_d$ $d=18,\cdots 24$ values obtained from the two different
weight families are in good agreement, and the $a_6$ and
$a_8$ values obtained using each of the new sum rules separately
are also in good agreement with those corresponding combined
fit values. With these values in hand
one finds that the OPE sides of
the BSR's may be safely truncated, as can be seen 
from Table~\ref{tablenew}. As in turns out, the additional
($d>16$) terms play a role in the full OPE sum only for the
third of the IZ sum rules. In none of the four cases is the
$d=6$ or $d=8$ contribution the dominant one.

In Table~\ref{table2} we show the combined-fit predictions, together with
the actual values (inferred from IZ) for the IZ spectral integrals
(less the known $d=4$ OPE terms). Shown for comparison are
the predictions corresponding to the IZ solution. Note that
the second of the four sum rules has been used by IZ to obtain
their quoted value for $a_6$. The errors quoted for the combined
fit $d=6$ through $d=24$ sum are obtained using the
covariance matrix for the solution set, generated from
covariance matrix of the ALEPH data{\begin{footnote}{Because
there are significant cancellations amongst the $d>4$
contributions, and strong correlations among
the combined fit values for the $a_d$, it is crucial to employ the 
full covariance matrix for the solution in 
determining the uncertainties on the combined fit predictions
shown in Table~\ref{table2}. The fact that, at the low
values of $M$ for which the spectral integral errors
are under control, the OPE sides of the IZ BSR's turn out to involve
a complicated cancellation between a large number
of terms of different dimension, in fact means that it would
not be possible to disentangle such contributions
using a BSR analysis.}\end{footnote}}. 
As is evident from the table, 
the combined fit predictions are in excellent agreement
with the data, providing further support for the asymptotic
nature of the coefficients extracted in the combined fit.
Table~\ref{tablenew}, in addition, shows that the four sum
rules weight the different $a_d$ contributions in very
different ways, demonstrating that the results represent
four independent, highly non-trivial tests of the combined
fit results.

\begin{table}[ht]
\begin{tabular}{cccrrrrrrrr}
IZ case&$d=6$&$d=8$&$d=10$&$d=12$&$d=14$&$d=16$&$d=18$&$d=20$&$d=22$&$d=24$\\
\tableline\\
1&$1$&$0$&$1.33$&$-1.93$&$1.23$&$-0.46$&$0.09$&$0.00$&$-0.01$
&$0.00$\\
2&$1$&$0$&$1.18$&$-0.93$&$0.00$&$0.20$&$-0.07$&$0$&$0.00$&$0.00$\\
3&$0$&$1$&$-5.19$&$4.09$&$0.00$&$-1.73$&$1.16$&$-0.34$&$0.00$&$0.04$\\
4&$1$&$0.42$&$0.00$&$-1.46$&$1.41$&$-0.53$&$0$&$0.09$&$-0.04$&$0.01$\\
\end{tabular}
\vskip .05in\noindent
\caption{The relative size of $d>4$ OPE contributions
to the four IZ BSR's for the extended version
of the combined fit described in the Appendix. In all cases,
the entries have been normalized to the lowest dimension
($d=6$ or $d=8$) contribution.
IZ cases 1 through 4 label the four IZ BSR's according
to the enumeration scheme given in the Appendix.}
\label{tablenew}
\end{table}

\begin{table}[ht]
\begin{tabular}{crrr}
IZ case&Combined fit\ \ \ &IZ fit\ \ \ \ \ \ \ \ \ &Data $-\ (d=4)$\\
\tableline\\
1&$-0.0023\pm 0.0002$&$-0.0027\pm 0.0008$&$-0.0023\pm 0.0006$ \\
2&$0.0038\pm 0.0002$&$0.0041\pm 0.0013$&$0.0041\pm 0.0009$\\
3&$-0.0030\pm 0.0004$&$-0.0038\pm 0.0022$&$-0.0032\pm 0.0009$\\
4&$0.0050\pm 0.0001$&$0.0050\pm 0.0033$&$0.0051\pm 0.0004$\\
\end{tabular}
\vskip .05in\noindent
\caption{The combined fit $d>4$ predictions for the four IZ BSR's.
IZ cases 1 through 4 label the four IZ BSR's according to the enumeration
scheme given in the Appendix. Column 2 gives our prediction 
for the $d>4$ OPE sum, obtained using extended version of the combined fit
corresponding to the ALEPH data. The results obtained using the 
OPAL data are the same, except in the fourth case, where our
prediction becomes $.0051\pm .0001$. 
Column 3 gives the $d>4$ OPE sum corresponding to the central values of the
IZ fit. Column 4 gives
the spectral integrals, less the known $d=4$ terms, corresponding
to the results quoted by IZ. The entries in columns 2, 3, and 4
are in GeV$^2$.}
\label{table2}
\end{table}
\end{appendix}}

%\vfill\eject
%%%%%%%%  %VCC% %%%%%%%%%%

%%%%%%%%%%%%%%%%%%%%%
%\end{references}
%\vfill\eject

\noindent
%\vfill\eject

\noindent
%\vfill\eject

\noindent
\noindent
%\vfill\eject

\noindent
%\vfill\eject

\noindent
%\vfill\eject

\noindent
\vfill\eject
\noindent

\end{document}